\documentclass[%
 reprint,
superscriptaddress,
nofootinbib,
 amsmath,amssymb,
 aps,
prl,
floatfix,
]{revtex4-2}

\usepackage{graphicx}% Include figure files
\usepackage{dcolumn}% Align table columns on decimal point
\usepackage{bm}% bold math

\usepackage{aas_macros}
\usepackage{xspace}
\usepackage{color}
\usepackage{hyperref}% add hypertext capabilities
\newcommand{\eg}{\mbox{e.\,g.,}\xspace}

\newcommand{\ie}{\mbox{i.\,e.,}\xspace}

\renewcommand{\vec}[1]{\bm{#1}}

\definecolor{darkgreen}{rgb}{0,0.5,0}
\definecolor{mypink1}{rgb}{0.858, 0.188, 0.478}
\definecolor{red}{rgb}{1,0,0}

\begin{document}

\preprint{APS/123-QED}

\title{New limits on light dark matter -- proton cross section from the cosmic large-scale structure}% Force line breaks with \\

\author{Keir K. Rogers}
 \email{keir.rogers@utoronto.ca}

\affiliation{%
Dunlap Institute for Astronomy and Astrophysics, University of Toronto, 50 St.\,George Street, Toronto, ON M5S 3H4, Canada}%

\author{Cora Dvorkin}
 \email{cdvorkin@g.harvard.edu}
 
 \affiliation{Department of Physics, Harvard University, 17 Oxford Street, Cambridge, MA 02138, USA}

\author{Hiranya V. Peiris}
 \email{h.peiris@ucl.ac.uk}

\affiliation{
Department of Physics \& Astronomy, University College London, Gower Street, London WC1E 6BT, UK}%

\affiliation{%
Oskar Klein Centre for Cosmoparticle Physics, Department of Physics, Stockholm University,\\
AlbaNova University Center, Stockholm 10691, Sweden}%
%\affiliation{
%Oskar Klein Centre for Cosmoparticle Physics, Department of Physics, Stockholm University,\\
%AlbaNova University Center, Stockholm 10691, Sweden
%\affiliation{%

\date{\today}% It is always \today, today,

\begin{abstract}
We set the strongest limits to-date on the velocity-independent dark matter (DM) -- proton cross section \(\sigma\) for DM masses \(m = 10\,\mathrm{keV}\) to 100 GeV, using large-scale structure traced by the Lyman-alpha forest: \eg a 95\% lower limit \(\sigma < 6 \times 10^{-30}\,\mathrm{cm}^2\), for \(m = 100\,\mathrm{keV}\). Our results complement direct detection, which has limited sensitivity to sub-GeV DM. We use an emulator of cosmological simulations, combined with data from the smallest cosmological scales used to-date, to model and search for the imprint of primordial DM--proton collisions. Cosmological bounds are improved by up to a factor of 25. \end{abstract}

\maketitle

\textbf{Introduction} -- The evidence for dark matter comes from its gravitational effect on cosmic structure at many scales, from galaxies to the cosmic microwave background. A determination of its interaction with Standard Model particles would open a new window on understanding dark matter microphysics. The dark matter particle or particles could theoretically have any mass, bounded only by the Planck scale and the cosmic horizon. Searches for weakly interacting massive particles by direct detection experiments have set strong bounds at GeV to TeV mass scales \citep[\eg][]{PhysRevD.97.115047}. Lighter dark matter particle candidates with masses from keV to GeV are well-motivated in many particle physics scenarios \citep{2008PhRvL.101w1301F,2014arXiv1402.5143H,2015PhRvL.115b1301H,2016PhRvL.116v1302K,2017PhRvD..96k5021K}. However, the search for lighter, sub-GeV dark matter is hampered in traditional direct detection analyses by limited sensitivity to nuclear recoil kinematics. Many novel experiments are proposed to search for sub-GeV dark matter, \eg using atomic targets \citep{PhysRevD.85.076007}, scintillators \citep{PhysRevD.96.016026}, semicondictors \citep{PhysRevD.85.076007,GRAHAM201232}, superconductors \citep{2016JHEP...08..057H}, two-dimensional targets \citep{HOCHBERG2017239}, although these mostly rely on electron interactions. Directly detecting nuclear interactions for dark matter lighter than \(\sim 100\,\mathrm{MeV}\) remains difficult even in the near future, although new technologies are proposed \citep{2017arXiv170704591B}.

Cosmological searches for light dark matter are highly complementary to direct detection, extending down to keV masses \citep{2002astro.ph..2496C,2014PhRvD..89b3519D,2018PhRvD..97j3530X,2018PhRvD..98h3510B,2018PhRvD..98b3013S,2021PhRvL.126i1101N,2021ApJ...907L..46M,Dvorkin:2022bsc}. Thermal coupling between dark matter and ordinary ``baryonic'' particles collisionally dampens structure growth in the early universe. Here we consider dark matter with a velocity-independent cross section, elastically interacting with hydrogen nuclei in the non-relativistic limit. This is a model-independent approach to setting dark matter limits, but our results map to specific models, \eg dark matter with a magnetic dipole moment \citep{2004PhRvD..70h3501S,2012JCAP...08..010D,2014PhRvD..89b3519D}. For velocity-independent cross sections, at current bounds, the coupling is strong in the very early universe (redshift \(z \gg 10^6\)). The rate of momentum transfer between dark matter and protons, relative to the Hubble expansion rate, decays rapidly such that interactions at late times are exceedingly rare. Nonetheless, the dark matter -- proton coupling in the early universe strongly suppresses the growth of late-time small-scale structure (sub-Mpc at current bounds). The suppression scale is a function of cross section and dark matter particle mass. Smaller cross sections correspond to smaller suppression scales (larger wavenumbers \(k\)); for a given cross section, the lighter the dark matter, the larger the suppression scale (see Fig.~\ref{fig:scale}).

In order to search for dark matter with smaller cross sections, we must look for suppression on the smallest scales currently accessible in the linear matter power spectrum. We do so using the Lyman-alpha forest \citep[\eg][]{1998MNRAS.296...44G,1998ApJ...495...44C}. This is neutral hydrogen absorption observed in high-redshift quasar spectra (\(2 \lesssim z \lesssim 6\)). The absorption traces fluctuations in the intergalactic medium (IGM). This is mostly primordial gas at about mean cosmic density in-between galaxies. Therefore, the flux power spectrum (correlations in the Lyman-alpha forest flux transmission) is a powerful tracer of the linear matter power spectrum \citep{2005PhRvD..71j3515S,2005ApJ...635..761M,2006ApJS..163...80M}. Using the highest resolution spectra, we probe the matter power spectrum on sub-Mpc scales \citep{2019ApJ...872..101B} and so the signal from smaller cross sections, improving existing upper limits.

In this work, we improve upon previous cosmological bounds on the dark matter -- proton cross section by, for the first time, simulating a full forward model for the effect of interactions in the early universe on the cosmic large-scale structure, as traced by the Lyman-alpha forest \citep{yu_feng_2018_1451799,2001NewA....6...79S,2005MNRAS.364.1105S,Anderson:2018zkm}. This allows us to exploit small-scale information in the Lyman-alpha forest and to avoid conservative assumptions in translating bounds from other dark matter models \citep{2021PhRvL.126i1101N,2021ApJ...907L..46M}. We are able to compare computationally-expensive simulations with data in a robust statistical inference by using the dark matter emulator we introduced in Ref.~\cite{2020RogersPRD}. This emulator is a computationally-cheap but accurate machine-learning model for the power spectrum, trained on cosmological hydrodynamical simulations of the IGM \citep{2020RogersPRD,2019JCAP...02..031R,2019JCAP...02..050B,2021JCAP...05..033P}. By using an active learning technique called Bayesian optimization, we test the robustness of our bounds with respect to the fidelity of the theoretical modeling \citep{2019JCAP...02..031R}.

\textbf{Model} -- We model the effect of dark matter -- proton scattering on the IGM with modified initial conditions in our hydrodynamical simulations at \(z = 99\). The initial conditions are given by a transfer function \(T(k) \equiv \left[P_\mathrm{pDM}(k)/P_\mathrm{CDM}(k)\right]^\frac{1}{2} = [1 + (\alpha (m, \sigma) k)^{\beta (m)}]^{\gamma}\) \citep{2017JCAP...11..046M}. \(P_\mathrm{pDM}(k)\) and \(P_\mathrm{CDM}(k)\) are respectively the linear matter power spectra for dark matter with proton scattering and cold, collisionless dark matter as a function of wavenumber \(k\). \(\alpha (m, \sigma)\) and \(\beta (m)\) are free functions and \(\gamma\) is a free parameter, where \(m\) is the dark matter particle mass and \(\sigma\) is the dark matter -- proton cross section. These are fit using a polynomial form to power spectrum calculations from a modified Boltzmann code, which accounts for dark matter -- proton interactions and the drag force exerted on the dark matter fluid in the early universe \citep{2014PhRvD..89b3519D}. The best fit is given in the Supplemental Material, which includes Refs.~\cite{2019JCAP...02..031R,2020RogersPRD,2020RogersPRL,2017PhRvL.119c1302I,2021PhRvL.126i1101N,2014PhRvD..89b3519D,rasmussen2003gaussian}. This parametric model accurately captures the key feature of a small-scale suppression in the linear matter power spectrum.

For reliable bounds on the dark matter -- proton cross section, we marginalize over the uncertainty in the thermal and ionization state of the IGM. The vast majority of the IGM gas at \(z \sim 5\) is accurately described by a power-law relation between its temperature \(T(z)\) and over-density \(\Delta\): \(T(z) = T_0(z) \Delta^{\widetilde{\gamma}(z) - 1}\) \citep{1997MNRAS.292...27H}. The two free parameters are the temperature at mean density \(T_0 (z)\) and the index \(\widetilde{\gamma}(z)\). We marginalize over the integrated energy injected from the ionizing background of photons per unit mass at the mean density \(u_0 (z)\) \citep{2016MNRAS.463.2335N}. This correlates with the filtering scale in the IGM, which is the relevant pressure smoothing scale for an evolving thermal state in an expanding universe \citep{1997MNRAS.292...27H,2015ApJ...812...30K}. We also marginalize over the effective optical depth \(\tau_\mathrm{eff} = - \ln \langle\mathcal{F}\rangle\), where \(\langle\mathcal{F}\rangle\) is the mean transmitted flux fraction, by a multiplicative correction \(\tau_0 (z)\) to the fiducial model given by Ref.~\cite{2019ApJ...872..101B}. This accounts for uncertainty in the photo-ionization rate as the two are degenerate in their effect on the flux power spectrum.

We further marginalize over the primordial power spectrum by its power-law index \(n_\mathrm{s}\) and amplitude \(A_\mathrm{s}\) at a pivot scale \(k_\mathrm{p} = 2\,\mathrm{Mpc}^{-1}\). For the other cosmological parameters, we fix to the baseline \textit{Planck} 2018 parameters \citep{refId0}: in particular, physical baryon energy density \(\Omega_\mathrm{b} h^2 = 0.0221\), physical dark matter energy density \(\Omega_\mathrm{c} h^2 = 0.121\) and dimensionless Hubble parameter \(h = 0.669\).

\textbf{Simulations} -- We use the cosmological hydrodynamical simulations and mock quasar spectra presented in Ref.~\cite{2020RogersPRD}, augmented with additional simulations by active learning as described below. Here, we summarize the main details. We evolve \(512^3\) particles each of dark matter and gas in a \((10\,h^{-1}\,\mathrm{Mpc})^3\) box from \(z = 99\) to \(z = 4.2\) using the publicly-available code \texttt{MP-Gadget}\footnote{\url{https://github.com/MP-Gadget/MP-Gadget}.} \citep{yu_feng_2018_1451799, 2001NewA....6...79S, 2005MNRAS.364.1105S,Anderson:2018zkm}. At each redshift bin of the data \(z = [4.2, 4.6, 5.0]\), we generate mock spectra containing only Lyman-alpha absorption and then measure the 1D flux power spectrum using \texttt{fake\_spectra} \citep{2017ascl.soft10012B}. The 1D flux power spectrum only includes correlations in the flux along the line of sight (\ie integrated over transverse directions).

Our simulations assume that the dark matter is collisionless from \(z = 99\). This is a very good approximation as, at existing bounds, the rate of momentum transfer between protons and dark matter (normalized to the expansion rate) is negligible at late times \citep{2014PhRvD..89b3519D}. Our constraining power comes from the imprint on cosmological structure from small-scale collisional damping in the early universe. Our simulations are optically thin, and heated and ionized by a spatially-uniform ultra-violet background (UVB) \citep{2012ApJ...746..125H}. Refs.~\cite{2019MNRAS.490.3177W, 2021arXiv210906897M} find this is a good approximation at the precision of current data. In order to vary the thermal IGM parameters \([T_0 (z), \widetilde{\gamma} (z), u_0 (z)]\) (see above), we vary the input ionization and heating rates by the model of Ref.~\cite{2017ApJ...837..106O}. This model modifies fiducial ionization and heating rates \citep{2012ApJ...746..125H} to account for variations in the timing of hydrogen reionization and its heat injection, as well as uncertainties in the strength and density-dependence of UVB heating. This in particular removes spurious heating before reionization arising in previous models. The effective optical depth \(\tau_0 (z)\) is varied in a computationally-cheap post-processing of the mock spectra. The other dark matter and cosmological parameters are varied in the simulation initial conditions according to the model presented above.

\textbf{Data} -- We use the 1D flux power spectrum presented in Ref.~\cite{2019ApJ...872..101B}, measured from fifteen high-resolution Keck-HIRES \citep{1994SPIE.2198..362V} and VLT-UVES \citep{2000SPIE.4008..534D} quasar spectra. It has three redshift bins at \(z = [4.2, 4.6, 5.0]\) and, at each redshift, has sixteen logarithmically-spaced wavenumber bins from \(\mathrm{log}(\mathrm{k}_\mathrm{f} [\mathrm{s}\,\mathrm{km}^{-1}]) = -2.2\) to \(\mathrm{log}(\mathrm{k}_\mathrm{f} [\mathrm{s}\,\mathrm{km}^{-1}]) = -0.7\). These are the smallest scales in the Lyman-alpha forest ever used to set dark matter -- proton cross section bounds and we therefore anticipate a strengthening of constraints. Since the current number of high-redshift, high-resolution spectra is relatively low \citep[\eg compared to lower-resolution surveys like the Baryon Oscillation Spectroscopic Survey][]{2019JCAP...07..017C}, the small-scale flux power spectrum is measured to a statistical precision of \(\sim 10 \%\) to \(25 \%\). This means that current dark matter bounds are limited by statistics and we anticipate improved bounds from upcoming larger sets of spectra, \eg ESPRESSO \citep{2014arXiv1401.5918P,2021A&A...645A..96P}. Metal line contamination is estimated from data that do not contain Lyman-alpha forest and with supplemental spectra, and is subtracted at the power spectrum level. The data would be insensitive to a \(100 \%\) mis-estimation of the metal flux power \citep{2019ApJ...872..101B}.

\textbf{Emulation and inference} -- We compare simulated flux power spectra to data using the dark matter emulator we presented in Ref.~\cite{2020RogersPRD}. This allows us to sample the parameter space in a computationally feasible way. The emulator is a computationally cheap but accurate model for the flux power spectrum called a Gaussian process, which is trained on the simulations discussed above. We discuss improvements made to the emulator model and training in the Supplemental Material. We use active learning (Bayesian optimization) to expand the emulator training set \citep{2019JCAP...02..031R} with 39 more simulations, giving 89 in total. Active learning lets us test the convergence of our bounds with respect to the fidelity of the emulator model. We present tests of convergence and cross-validation in the Supplemental Material.

We sample the posterior distribution for parameters \(\vec{\theta} = [\log(m [\mathrm{eV}]), \log(\sigma [\mathrm{cm}^2]), \tau_0 (z = z_i), T_0 (z = z_i), \widetilde{\gamma} (z = z_i), u_0 (z = z_i), n_\mathrm{s}, A_\mathrm{s}]\), for \(z_i = [4.2, 4.6, 5.0]\), using the Markov chain Monte Carlo ensemble sampler \texttt{emcee} \citep{2013PASP..125..306F}. The likelihood function is Gaussian with covariance \(C(\vec{\theta}) = C_\mathrm{data} + C_\mathrm{emu}(\vec{\theta})\). The data covariance between power spectrum bins \(C_\mathrm{data}\) is estimated by data bootstrapping and regularized by simulations \citep[see][]{2019ApJ...872..101B}. In order to propagate the modeling uncertainty, we include \(C_\mathrm{emu}(\vec{\theta})\), the (diagonal) covariance of the emulator-predicted theory power spectrum (this covariance is modeled by the emulator). In practice, \(C_\mathrm{emu}(\vec{\theta}) \ll C_\mathrm{data}\), for \(\vec{\theta}\) that we sample in the converged posterior distribution.

In the prior distribution, we exclude the edges of the \(T_0 (z = z_i)\) --- \(u_0 (z = z_i)\) and \(T_0 (z = z_i)\) --- \(\widetilde{\gamma} (z = z_i)\) planes at each redshift not spanned by the training set. This excludes unphysical IGMs, while preventing the emulator from extrapolating beyond the training set. We prevent, in neighbouring redshift bins, unphysical changes in \(T_0\) greater than 5000 K and changes in \(u_0\) greater than 10 \(\mathrm{eV}\,m_\mathrm{p}^{-1}\) (\(m_\mathrm{p}\) being the proton mass). There are \textit{Planck} 2018-motivated \citep{refId0} priors on \(n_\mathrm{s}\) and \(A_\mathrm{s}\) (translated to the pivot scale we use): Gaussian distributions with means \(0.9635\) and \(1.8296 \times 10^{-9}\) respectively and standard deviations \(0.0057\) and \(0.030 \times 10^{-9}\) respectively. There is a Gaussian prior on \(\tau_0 (z = z_i)\) with mean \(1\) and standard deviation \(0.05\), as these parameters are poorly constrained by the data. Following the tests in Ref.~\cite{2020RogersPRD}, which demonstrated the insensitivity of our previous bounds on ultralight axion dark matter \citep{2020RogersPRL} to a prior on \(T_0 (z = z_i)\), we instead now use a uniform prior.

\begin{figure}
\includegraphics[width=\columnwidth]{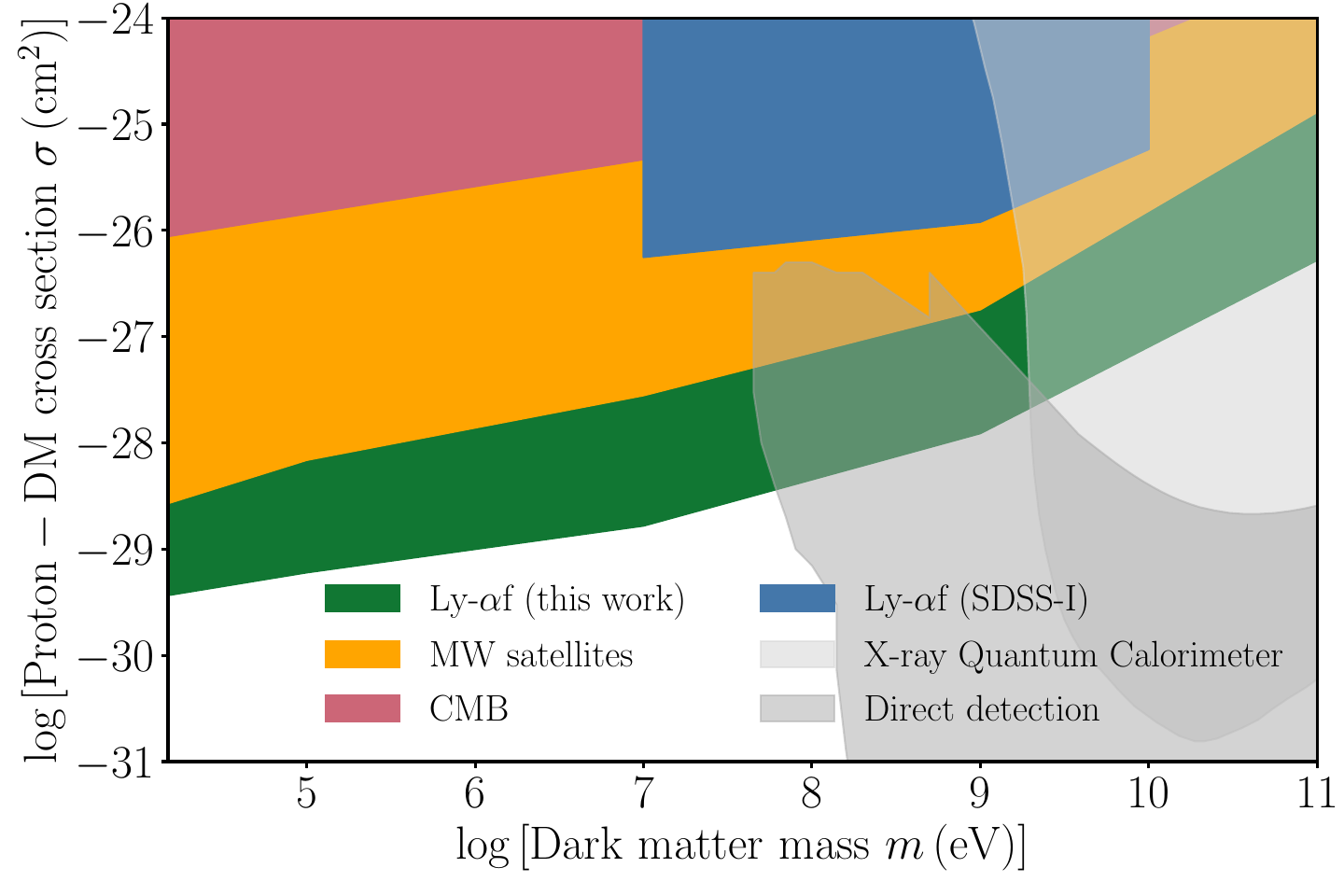}
\caption{\label{fig:comparison}Exclusion plot comparing our proton -- dark matter cross section bound (\textit{in green}) to some other competitive cosmological and direct detection bounds. Cosmological bounds are from: (\textit{in red}) \textit{Planck} 2015 cosmic microwave background (CMB) \citep{2016A&A...594A..11P} temperature, polarization and lensing data \citep{2018PhRvD..98h3510B}; (\textit{in blue}) the Lyman-alpha forest (Ly-\(\alpha\)f) from the Sloan Digital Sky Survey \citep[SDSS-I;][]{2005ApJ...635..761M} in combination with \textit{Planck} CMB temperature and polarization data \citep{2018PhRvD..97j3530X}; (\textit{in orange}) the Milky Way (MW) satellite population \citep{2020ApJ...893...47D} as inferred from the Dark Energy Survey and PanSTARRS-1 \citep{2021ApJ...907L..46M}. In the parameter space we consider, direct detection bounds are from CRESST-surface \citep{2017EPJC...77..637A,PhysRevD.97.115047} and EDELWEISS-Surf \citep{2019PhRvD..99h2003A}. The X-ray Quantum Calorimeter (XQC) bound is from Ref.~\cite{2018JCAP...10..007M}.}
%\kkr{Email E. Nadler for XQC.}}
\end{figure}

\begin{table}
\begin{ruledtabular}
\begin{tabular}{lcc}
&\multicolumn{2}{c}{\(95 \%\) credible interval}\\
\hline
\(\log[\sigma (m = 100\,\mathrm{GeV})\,[\mathrm{cm}^2]]\)&\multicolumn{2}{c}{\(< -26.28\)}\\ %-26.27977397048768
\(\log[\sigma (m = 1\,\mathrm{GeV})\,[\mathrm{cm}^2]]\)&\multicolumn{2}{c}{\(< -27.91\)}\\ %-27.91406950983393
\(\log[\sigma (m = 10\,\mathrm{MeV})\,[\mathrm{cm}^2]]\)&\multicolumn{2}{c}{\(< -28.78\)}\\ %-28.78461055880187
\(\log[\sigma (m = 100\,\mathrm{keV})\,[\mathrm{cm}^2]]\)&\multicolumn{2}{c}{\(< -29.22\)}\\ %-29.21786085176801
\(\log[\sigma (m = 10\,\mathrm{keV})\,[\mathrm{cm}^2]]\)&\multicolumn{2}{c}{\(< -29.48\)}\\ %-29.475489472511562
\hline
\(n_\mathrm{s}\)&0.953&0.975\\
\(A_\mathrm{s}\)&\(1.77 \times 10^{-9}\)&\(1.89 \times 10^{-9}\)\\
\hline
\(\tau_0 (z = 4.2)\)&0.917&1.075\\
\(T_0 (z = 4.2)\) [K]&9117&12285\\
\(\widetilde{\gamma} (z = 4.2)\)&1.02&1.67\\
\(u_0 (z = 4.2)\) \([\mathrm{eV}\,m_\mathrm{p}^{-1}]\)&6.42&17.4\\
\hline
\(\tau_0 (z = 4.6)\)&0.954&1.062\\
\(T_0 (z = 4.6)\) [K]&9553&12972\\
\(\widetilde{\gamma} (z = 4.6)\)&1.16&1.57\\
\(u_0 (z = 4.6)\) \([\mathrm{eV}\,m_\mathrm{p}^{-1}]\)&6.60&16.9\\
\hline
\(\tau_0 (z = 5.0)\)&0.893&1.008\\
\(T_0 (z = 5.0)\) [K]&8463&13880\\
\(\widetilde{\gamma} (z = 5.0)\)&1.09&1.60\\
\(u_0 (z = 5.0)\) \([\mathrm{eV}\,m_\mathrm{p}^{-1}]\)&3.09&17.4
\end{tabular}
\end{ruledtabular}
\caption{\label{tab:bounds}1D marginalized \(95 \%\) credible intervals.}
\end{table}
\textbf{Results} -- The main result is summarized by a 95\% credible marginalized upper limit on the logarithm of the dark matter -- proton cross section as a function of dark matter particle mass from 10 keV to 100 GeV. This is the edge of the green contour in Fig.~\ref{fig:comparison}, which shows how the cross-section is excluded at the 95\% credible interval. Table \ref{tab:bounds} gives the cross section bound for some fixed dark matter masses, as well as marginalized nuisance IGM and cosmological parameter constraints: \eg a 95\% credible lower limit \(\sigma < 6 \times 10^{-30}\,\mathrm{cm}^2\), for \(m = 100\,\mathrm{keV}\). The thermal and ionization state of the IGM over which we marginalize is statistically consistent with previous results \citep{2020RogersPRL,2019ApJ...872..101B,2019ApJ...872...13W,2021arXiv211100019V} and fiducial models \citep{1997MNRAS.292...27H,2012ApJ...746..125H,2019MNRAS.485...47P}. We find that the IGM at \(z = 4.2\) to \(z = 5\) is at a plateau between the end of hydrogen reionization and the start of helium reionization (although IGM bounds are weak since we marginalize dark matter parameters). Even after marginalizing small-scale smoothing in the IGM from temperature and pressure effects, we rule out strong suppression from the largest cross sections we consider. The marginalized posterior distribution of the cosmological parameters is not significantly different from the prior, indicating as expected no constraining power on these parameters from our dataset. We show marginalized 2D posterior corner plots in the Supplemental Material.

\begin{figure}
\includegraphics[width=\columnwidth]{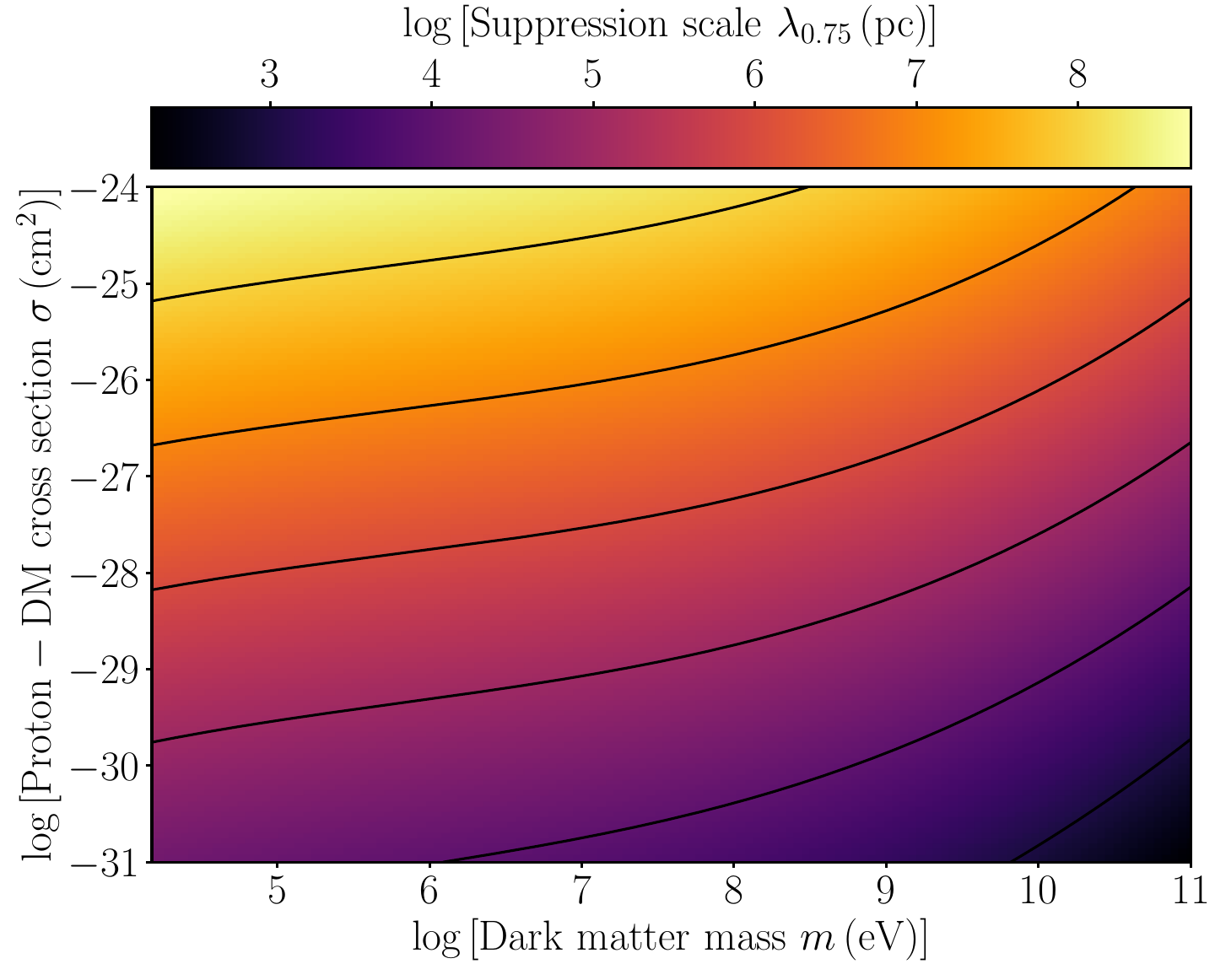}
\caption{\label{fig:scale}The suppression scale \(\lambda_{0.75}\) in the growth of cosmic structure arising from dark matter -- proton interactions (relative to standard cold, collisionless dark matter; CDM), as a function of dark matter particle mass \(m\) and interaction cross section \(\sigma\). The black lines are contours of constant suppression scale; these are parallel to the cosmological bounds shown in Fig.~\ref{fig:comparison}, indicating how, to first order, cosmological limits are driven by the non-detection of suppressed growth on the smallest scale that a given dataset probes. We show the length \(\lambda_{0.75}\) (at which the linear matter power spectrum is suppressed by 25\% relative to the CDM case) as the scale to which our analysis is sensitive (see Supplemental Material), but, in general, the exact scale will depend on a particular dataset.}
\end{figure}
\textbf{Discussion} -- Figure \ref{fig:comparison} compares our bound to some other competitive published bounds from cosmology and direct detection (see caption for details)\footnote{There are two different existing approaches to setting cross-section limits using Milky Way satellites data, in the absence of dedicated cosmological simulations that include the imprint of primordial dark matter -- proton interactions. Both entail translating bounds on the mass of a warm dark matter (WDM) particle. In one approach, the bound is translated by finding the cross section at which the corresponding transfer function is always more suppressed than the WDM equivalent \citep{2021ApJ...907L..46M}. In the other, the bound is translated by matching the interacting dark matter transfer function to the WDM transfer function at the scale at which the power spectrum is suppressed by 75\% relative to cold dark matter \citep{2021PhRvL.126i1101N}; both approaches are considered in Ref.~\cite{2021arXiv210712377B}. In the preprint \cite{2021arXiv210712377B}, there are also limits using Lyman-alpha forest data and an approximate method called the ``area criterion'' to translate bounds from other dark matter models \citep{2017JCAP...11..046M}. Dark matter and neutrino direct detection limits have been reinterpreted in the context of dark matter upscattering by cosmic rays \citep{2019PhRvL.122q1801B,2019PhRvD.100j3011C}; we do not show these limits here as they constrain the cross section at relativistic energies, which precludes a simple mapping to the velocity-independent cross section bounds we consider here. In Fig.~\ref{fig:comparison}, cosmological limits are given at the 95\% C.I., while direct detection and XQC limits are given at the 90\% C.I. as is the convention in experimental particle physics; this motivates the establishment of a common statistical language in setting dark matter bounds.}. We set the strongest cosmological upper limit to-date on the dark matter -- proton cross section for dark matter particle masses from 10 keV to 100 GeV. Our bounds are highly complementary to direct detection searches for sub-GeV dark matter, where traditional nuclear recoil techniques have limited sensitivity to light dark matter. We improve over previous cosmological cross section bounds by accessing smaller scales in the matter power spectrum at high redshift (\(z \sim 5\)), as traced by the Lyman-alpha forest flux power spectrum. Fig.~\ref{fig:scale} shows, as a function of mass and cross section, the relevant suppression scale in the clustering of matter arising from dark matter -- proton interactions in the early universe. It follows that by probing smaller scales using the Lyman-alpha forest, we rule out the presence of a suppression arising from a smaller cross section and reduce the allowed upper limit.
% There are also limits using the same Milky Way satellites data as in the figure, but with a less conservative approach to translating bounds from other dark matter models \cite{2021PhRvL.126i1101N,2021arXiv210712377B}.

In the low-density IGM that the Lyman-alpha forest traces at \(z \sim 5\), there is less contamination of the dark matter suppression scale by non-linear gravitational evolution than, \eg in the distribution of Milky Way satellites at \(z = 0\). Nonetheless, these two methods are themselves highly complementary, providing independent limits in different physical environments and with different nuisance parameter models. This motivates future joint analyses, also in combination with strong gravitational lensing observations \citep{2020MNRAS.491.6077G,2021ApJ...917....7N}.

We are able to set strong limits and access small scales by, for the first time, modeling the data using cosmological simulations that include the effect of dark matter -- proton interactions in the early universe. This means that we do not need to translate from bounds on other dark matter models using conservative assumptions \citep{2021PhRvL.126i1101N,2021ApJ...907L..46M}. Instead, we are able to exploit the full constraining power of the data. We compare simulations to data in a robust statistical inference using the dark matter emulator we presented in Ref.~\cite{2020RogersPRD}. This employs active learning \citep{2019JCAP...02..031R} to test the robustness of our bounds with respect to the fidelity of the theoretical modeling (see Supplemental Material). This work demonstrates the power of combining active learning with an emulator in order to test cosmological models in an accurate and efficient manner \citep[see also previous bounds on ultralight axion dark matter;][]{2020RogersPRL}. We anticipate that other dark matter models can be tested using our method, \eg dark matter -- proton cross sections with velocity dependence \citep{2014PhRvD..89b3519D,2018PhRvD..97j3530X,2018PhRvD..98h3510B,2021ApJ...907L..46M}, warm dark matter \citep{2005PhRvD..71f3534V,ENQVIST1990531,2016JCAP...11..038K,2009JCAP...05..012B} or freeze-in dark matter scenarios \citep{2019PhRvD..99k5009D,2021PhRvL.127k1301D}.

\textbf{Conclusions} -- We present the strongest cosmological limits to-date on a velocity-independent cross section between dark matter and protons, for dark matter particles from 10 keV to 100 GeV: \eg a 95\% credible lower limit \(\sigma < 6 \times 10^{-30}\,\mathrm{cm}^2\), for \(m = 100\,\mathrm{keV}\). These results are highly complementary to direct detection searches, where traditional nuclear recoil techniques have limited sensitivity to sub-GeV masses. Our bounds can inform future experimental design. We set these cross section limits using, for the first time, cosmological simulations that account for the imprint of dark matter -- proton interactions in the early universe. This uses the dark matter emulator we present in Ref.~\cite{2020RogersPRD}. We anticipate future improvements to these bounds from: external constraints on the thermal and ionization state of the IGM \citep[\eg high-redshift quasar transmission spikes;][]{2006AJ....132..117F,2021arXiv210906295Z}; larger sets of Lyman-alpha forest observations (\eg Dark Energy Spectroscopic Instrument \citep{2016arXiv161100036D,2016arXiv161100037D}, ESPRESSO \citep{2014arXiv1401.5918P,2021A&A...645A..96P}); and joint analyses with other astrophysical data (\eg Milky Way satellites \citep{2020ApJ...893...47D,2021PhRvL.126i1101N,2021ApJ...907L..46M,2021arXiv210712377B}, stellar streams \citep{2018JCAP...07..061B}, strong gravitational lensing \citep{2020MNRAS.491.6077G}).

\textbf{Acknowledgments} -- KKR thanks Jose O{\~n}orbe for sharing his code for the reionization model of Ref.~\cite{2017ApJ...837..106O} and for valuable discussions. The authors also thank Kimberly Boddy, Daniel Gilman, Vera Gluscevic, Ethan Nadler and Linda Xu for their assistance. The Dunlap Institute is funded through an endowment established by the David Dunlap family and the University of Toronto. CD was partially supported by the National Science Foundation (NSF) under Cooperative Agreement PHY - 2019786 (the NSF AI Institute for Artificial Intelligence and Fundamental Interactions). HVP was supported by the Science and Technology Facilities Council (STFC) Consolidated Grant number ST/R000476/1 and the research project grant ``Fundamental Physics from Cosmological Surveys'' funded by the Swedish Research Council (VR) under Dnr 2017-04212. This project has received funding from the European Research Council (ERC) under the European Union’s Horizon 2020 research and innovation programme (grant agreement no. 101018897 CosmicExplorer). This work used computing facilities provided by the UCL Cosmoparticle Initiative; and we thank the HPC systems manager Edd Edmondson for his indefatigable support. This work used computing equipment funded by the Research Capital Investment Fund (RCIF) provided by UK Research and Innovation (UKRI), and partially funded by the UCL Cosmoparticle Initiative.

\subsection{\label{sec:supplement}Supplemental Material}

\begin{figure*}
\includegraphics[width=\textwidth]{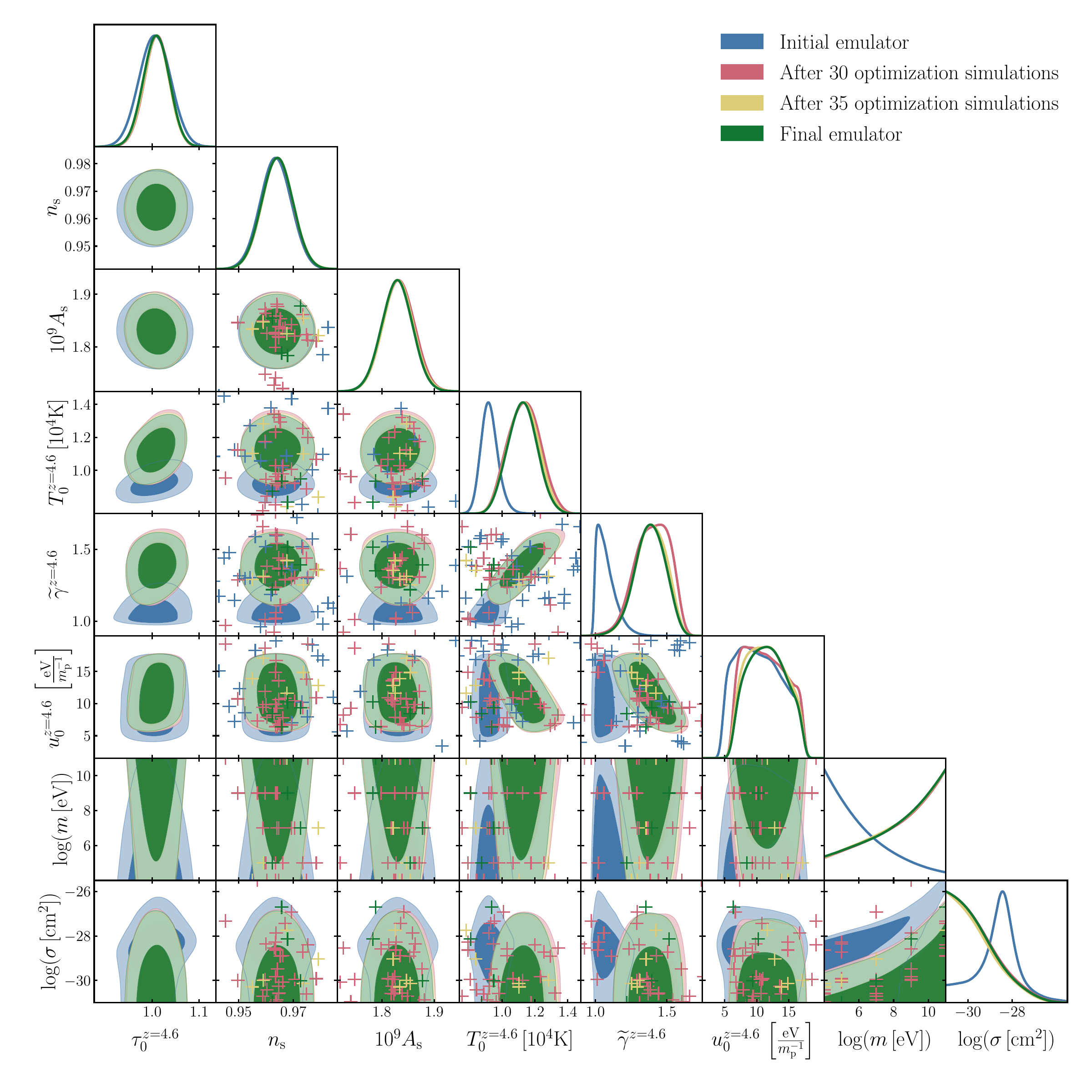}
\caption{\label{fig:posterior}The convergence of the posterior distribution (marginalized 1D and 2D distributions) with respect to the construction of the flux power spectrum emulator. Each set of colored contours shows the estimate of the posterior at different stages of building the emulator. The green contours are our final estimate of the posterior (with an emulator with 89 training simulations in total); there is no statistically significant change in the posterior in the final stages of building the emulator (from after thirty optimization simulations are added). The darker and lighter shaded areas respectively indicate the 68\% and 95\% credible regions. In each projection, crosses indicate the projected positions of emulator training simulations (expect for \(\tau_0 (z = 4.6)\) projections which are densely sampled by simulation post-processing). The crosses are colored according to the stage of the emulator by which the training simulations are added. For the dark matter parameters \([\mathrm{log}(m [\mathrm{eV}]), \mathrm{log}(\sigma [\mathrm{cm}^2])]\), the initial training simulations fully span the \([\alpha, \beta, \gamma]\) sub-volume and do not project onto those axes; however, the initial simulations contribute to the final emulator as we always emulate in \([\alpha, \beta, \gamma]\). We show, for clarity, the IGM parameters only at our central redshift bin \(z = 4.6\); there is similar convergence at the other redshifts we consider.}
\end{figure*}
%\(A_\mathrm{s}\) is in units of \(10^{-9}\); \(T_0 (z = 4.6)\) is in units of \(10^4\,\mathrm{K}\); \(u_0 (z = 4.6)\) is in units of \(\mathrm{eV}\,m_\mathrm{p}^{-1}\); \(m\) is in units of eV; and \(\sigma\) is in units of \(\mathrm{cm}^2\).

\begin{figure}
\includegraphics[width=\columnwidth]{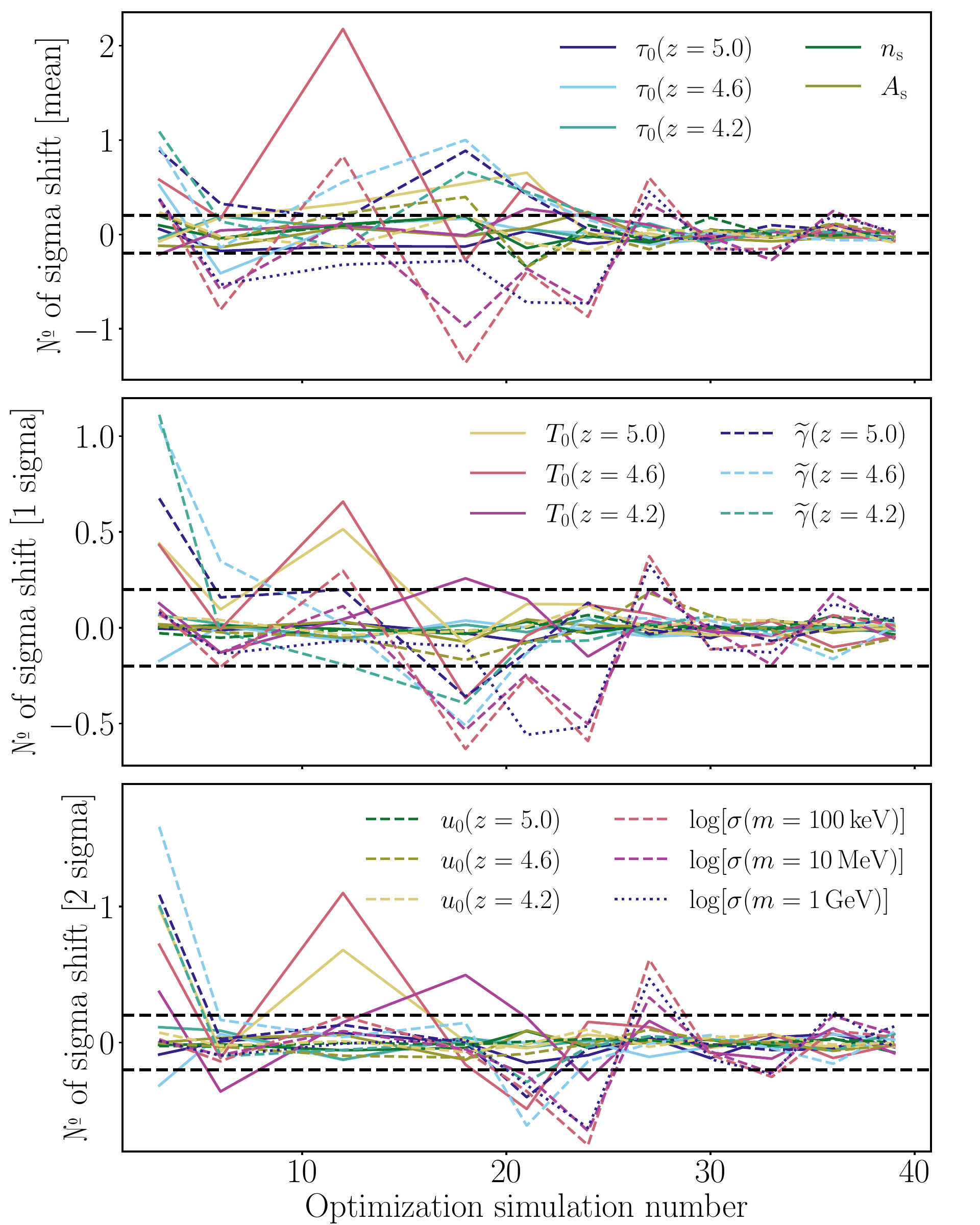}
\caption{\label{fig:convergence}The convergence of the posterior distribution (\textit{from top to bottom}, the 1D marginalized posterior means, \(1 \sigma\) and \(2 \sigma\) constraints) with respect to the construction of the flux power spectrum emulator. Each colored line shows the number of sigma shift (defined by the marginalized posteriors at a given optimization epoch) in the summary statistics from one iteration of the emulator from the previous. There is no statistically significant change in the posterior summaries in the final stages of building the emulator. The black dashed lines indicate shifts of \(0.2 \sigma\).}
\end{figure}
Figures \ref{fig:posterior} and \ref{fig:convergence} demonstrate the convergence of our bounds with respect to the construction of the flux power spectrum emulator by active learning. In both marginalized 1D and 2D posterior distributions (Fig.~\ref{fig:posterior}) and summary statistics of the posterior (Fig.~\ref{fig:convergence}), we see that the addition of further training simulations does not change the constraints. We also see convergence in the active learning as the exploration term of the acquisition function we use to select optimization simulations\footnote{For the selection of active learning simulations, we maximize an acquisition function which is a weighted sum of an exploration term \(\vec{\Sigma}_\mathrm{emu}^\mathrm{T} (\vec{\theta}) \Sigma_\mathrm{data}^{-1} \vec{\Sigma}_\mathrm{emu} (\vec{\theta})\) and an exploitation term \(\mathcal{P}(\vec{\theta}|\vec{d})\), where \(\vec{\Sigma}_\mathrm{emu}\) is the emulator error, \(\Sigma_\mathrm{data}\) is the data covariance and \(\mathcal{P}(\vec{\theta}|\vec{d})\) is the natural logarithm of the posterior probability given data \(\vec{d}\) \citep{2019JCAP...02..031R}. The exploration term tends to add simulations where the emulator model is least constrained, while the exploitation term tends to add simulations at the peak of the posterior.} tends towards zero. This indicates that the acquisition of training simulations is dominated by exploitation, \ie we tend to add further simulations only at the peak of the posterior and so the true posterior peak has been found \citep[see][]{2019JCAP...02..031R}. Since we anticipate no constraint on the dark matter mass \(m\) parameter, we restrict the addition of active learning simulations to a range of fixed values (as seen in Fig.~\ref{fig:posterior}) in order to guarantee that we span fully this axis. Our final results (the green contours in Fig.~\ref{fig:posterior}) use a flux power spectrum emulator with 89 training simulations in total: 50 from the initial set presented in Ref.~\cite{2020RogersPRD} and 39 optimization simulations added by active learning \citep{2019JCAP...02..031R}.

We compare our results to a previous bound on the mass of ultralight axion dark matter that we set \citep{2020RogersPRL} by comparing the linear matter power spectrum transfer function \(T(k)\) at the respective 95\% credible limits from the two analyses. These coincide at a wavenumber where the power spectrum in both models is suppressed by 25\% relative to the cold dark matter case. This indicates that the equivalent length-scale \(\lambda_{0.75}\) is, to first order, the scale driving our bounds both on the dark matter -- proton interaction and ultralight axion dark matter (see also Fig.~\ref{fig:scale}). This is consistent with previous results using both Lyman-alpha forest \citep{2017PhRvL.119c1302I} and Milky Way satellites \citep{2021PhRvL.126i1101N} data. However, in general, the exact scale driving dark matter limits will depend on the dataset and model under consideration. Therefore, approximate bounds derived using equivalent scales (\eg half-mode scale) to translate between different dark matter models are not as accurate as the forward modeling approach we use here. The IGM and cosmological model over which we marginalize is statistically consistent in our light dark matter and axion bounds. In Fig.~\ref{fig:posterior}, similar to the axion case and previous high-redshift (\(z \sim 5\)) Lyman-alpha forest analyses \citep[\eg][]{2017PhRvL.119c1302I}, we find no significant degeneracy between the dark matter and other parameters. As in our axion analysis, in order to map to the dark matter emulator we present in Ref.~\cite{2020RogersPRD}, we model the light dark matter transfer function using a parametric model fit to calculations from a modified Boltzmann code \citep{2014PhRvD..89b3519D}: \(T(k) \equiv \left[P_\mathrm{pDM}(k)/P_\mathrm{CDM}(k)\right]^\frac{1}{2} = [1 + (\alpha (m, \sigma) k)^{\beta (m)}]^{\gamma}\). Here, \(\log(\alpha [h^{-1}\,\mathrm{Mpc}]) = -7.5 \times 10^{-3} \mathcal{M}^3 - 2.3 \times 10^{-3} \mathcal{M}^2 + 0.013 \mathcal{M} - 0.018 \mathcal{S}^3 - 0.90 \mathcal{S}^2 - 4.2 \mathcal{S} - 24.1\); \(\beta = -2.3 \times 10^{-4} \mathcal{M}^2 - 0.082 \mathcal{M} + 2.4\); and \(\gamma = -4.46\), where \(\mathcal{M} = \log(m [\mathrm{eV}])\) and \(\mathcal{S} = \log(\sigma [\mathrm{cm}^2])\).

\begin{figure}
\includegraphics[width=\columnwidth]{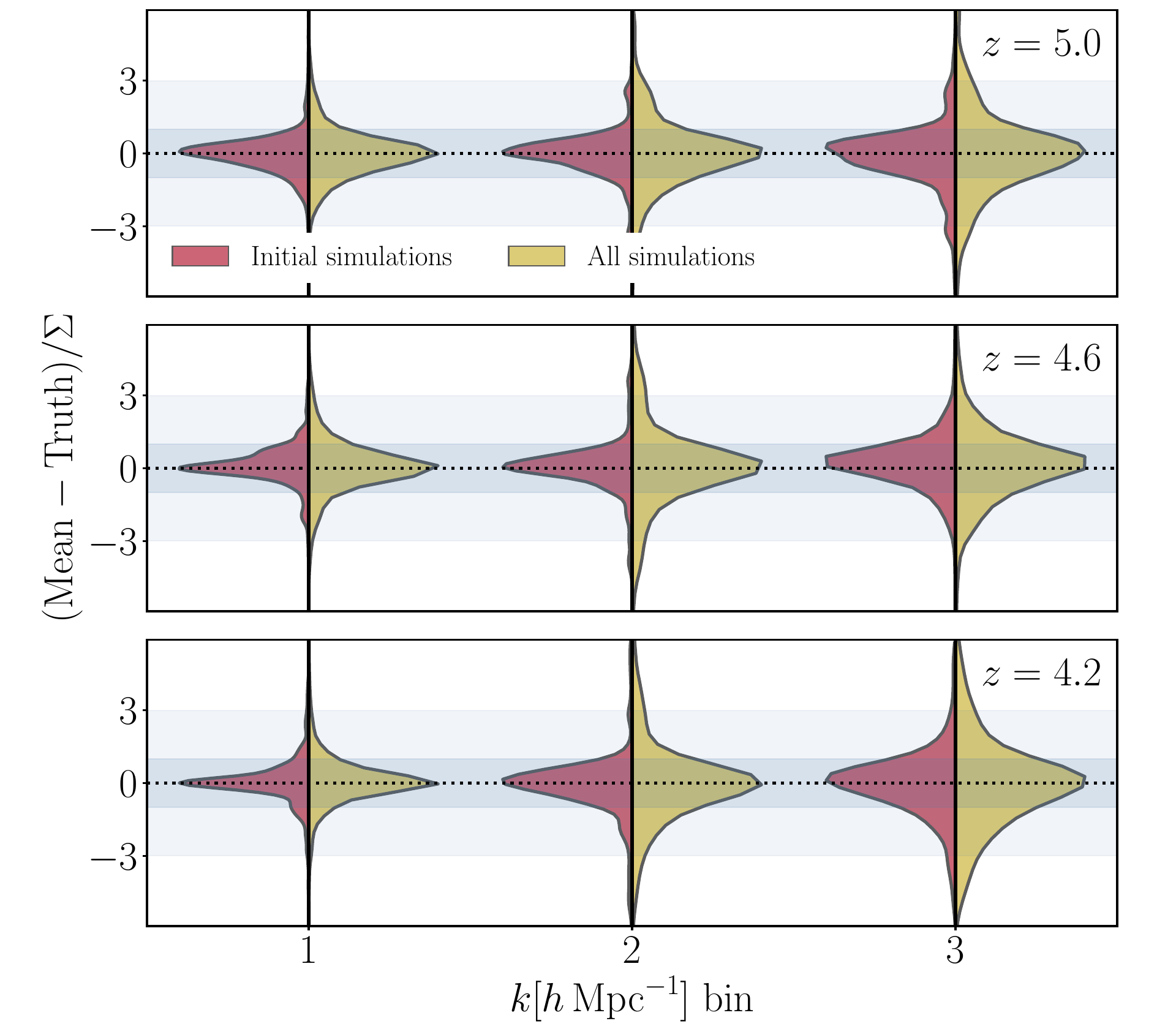}
\caption{\label{fig:validation}Leave-one-out cross-validation of the flux power spectrum emulator. Each violin plot shows the distribution of the ratio of empirical (emulator mean - truth) to predicted (\(\Sigma\)) emulator error for the leave-one-out cross-validation simulations. \textit{From left to right}, the cross-validation in different wavenumber \(k\) bins and \textit{from top to bottom}, for different redshifts \(z\). The left sides of the violins consider only the initial training simulations, while the right sides consider all the simulations.}
\end{figure}
Figure \ref{fig:validation} shows our leave-one-out cross-validation test for the flux power spectrum emulator. This leaves out in turn the training samples for each of the 89 simulations in our training set. We find that the distribution of the ratio of empirical error (difference between true flux power spectra and emulator prediction) to modeled error (as predicted by the emulator) is close to a unit Gaussian. This indicates a good fit for the emulator model. The deviation from a unit Gaussian is strongest in the lowest redshift bin, with the smallest wavenumber bin slightly over-fit (the predicted error overestimating the empirical error) and the largest wavenumber bin slightly under-fit (the predicted error underestimating the empirical error). We show the cross-validation test for the full training set, as well as the initial set of simulations only, in order to highlight that even the initial base emulator is well fit.

We note that the emulator fit (with the initial and final training sets) is improved with respect to the version used in our axion analysis \citep{2020RogersPRD}. This arises from some modifications made to the emulator model. We use a more complex and flexible Gaussian process kernel to model the covariance between training points \(K(\vec{\theta}, \vec{\theta'})\) \citep{rasmussen2003gaussian}, which is better suited to the multi-dimensional parameter space we consider. This kernel is the product of a rational quadratic kernel \(\sigma_\mathrm{RQ}^2 \left(1 + \sum_i \frac{(\theta_i - \theta'_i)^2}{2 l_i^2}\right)^{-a}\) and a linear kernel \(\sum_i \sigma_{\mathrm{linear},i}^2 \theta_i \theta'_i\), then added to a constant noise kernel \(\sigma_\mathrm{noise}^2\). Here, \(i\) indexes the physical parameters at each redshift bin (since we emulate the three redshift bins independently) and the kernel hyperparameters \([\sigma_\mathrm{RQ}, \sigma_{\mathrm{linear},i}, \sigma_\mathrm{noise}, a, l_i]\) are optimized by maximizing the marginal likelihood of the training data. The rational quadratic kernel allows for a spectrum of length-scales in the modeled covariance. For this more complex kernel, we use a different hyperparameter optimization method that runs a truncated Newton algorithm 35 times with different initial conditions. This explores more thoroughly the highly non-convex hyperparameter likelihood surface. We anticipate that these emulator model improvements will assist in future uses of our dark matter emulator to test other dark matter models.

%\nocite{*}
\bibliography{bDM}% Produces the bibliography via BibTeX.

\begin{thebibliography}{79}%
\makeatletter
\providecommand \@ifxundefined [1]{%
 \@ifx{#1\undefined}
}%
\providecommand \@ifnum [1]{%
 \ifnum #1\expandafter \@firstoftwo
 \else \expandafter \@secondoftwo
 \fi
}%
\providecommand \@ifx [1]{%
 \ifx #1\expandafter \@firstoftwo
 \else \expandafter \@secondoftwo
 \fi
}%
\providecommand \natexlab [1]{#1}%
\providecommand \enquote  [1]{``#1''}%
\providecommand \bibnamefont  [1]{#1}%
\providecommand \bibfnamefont [1]{#1}%
\providecommand \citenamefont [1]{#1}%
\providecommand \href@noop [0]{\@secondoftwo}%
\providecommand \href [0]{\begingroup \@sanitize@url \@href}%
\providecommand \@href[1]{\@@startlink{#1}\@@href}%
\providecommand \@@href[1]{\endgroup#1\@@endlink}%
\providecommand \@sanitize@url [0]{\catcode `\\12\catcode `\$12\catcode
  `\&12\catcode `\#12\catcode `\^12\catcode `\_12\catcode `\%12\relax}%
\providecommand \@@startlink[1]{}%
\providecommand \@@endlink[0]{}%
\providecommand \url  [0]{\begingroup\@sanitize@url \@url }%
\providecommand \@url [1]{\endgroup\@href {#1}{\urlprefix }}%
\providecommand \urlprefix  [0]{URL }%
\providecommand \Eprint [0]{\href }%
\providecommand \doibase [0]{https://doi.org/}%
\providecommand \selectlanguage [0]{\@gobble}%
\providecommand \bibinfo  [0]{\@secondoftwo}%
\providecommand \bibfield  [0]{\@secondoftwo}%
\providecommand \translation [1]{[#1]}%
\providecommand \BibitemOpen [0]{}%
\providecommand \bibitemStop [0]{}%
\providecommand \bibitemNoStop [0]{.\EOS\space}%
\providecommand \EOS [0]{\spacefactor3000\relax}%
\providecommand \BibitemShut  [1]{\csname bibitem#1\endcsname}%
\let\auto@bib@innerbib\@empty
%</preamble>
\bibitem [{\citenamefont {Emken}\ and\ \citenamefont
  {Kouvaris}(2018)}]{PhysRevD.97.115047}%
  \BibitemOpen
  \bibfield  {author} {\bibinfo {author} {\bibfnamefont {T.}~\bibnamefont
  {Emken}}\ and\ \bibinfo {author} {\bibfnamefont {C.}~\bibnamefont
  {Kouvaris}},\ }\bibfield  {title} {\bibinfo {title} {How blind are
  underground and surface detectors to strongly interacting dark matter?},\
  }\href {https://doi.org/10.1103/PhysRevD.97.115047} {\bibfield  {journal}
  {\bibinfo  {journal} {Phys. Rev. D}\ }\textbf {\bibinfo {volume} {97}},\
  \bibinfo {pages} {115047} (\bibinfo {year} {2018})}\BibitemShut {NoStop}%
\bibitem [{\citenamefont {{Feng}}\ and\ \citenamefont
  {{Kumar}}(2008)}]{2008PhRvL.101w1301F}%
  \BibitemOpen
  \bibfield  {author} {\bibinfo {author} {\bibfnamefont {J.~L.}\ \bibnamefont
  {{Feng}}}\ and\ \bibinfo {author} {\bibfnamefont {J.}~\bibnamefont
  {{Kumar}}},\ }\bibfield  {title} {\bibinfo {title} {{Dark-Matter Particles
  without Weak-Scale Masses or Weak Interactions}},\ }\href
  {https://doi.org/10.1103/PhysRevLett.101.231301} {\bibfield  {journal}
  {\bibinfo  {journal} {\prl}\ }\textbf {\bibinfo {volume} {101}},\ \bibinfo
  {eid} {231301} (\bibinfo {year} {2008})},\ \Eprint
  {https://arxiv.org/abs/0803.4196} {arXiv:0803.4196 [hep-ph]} \BibitemShut
  {NoStop}%
\bibitem [{\citenamefont {{Hochberg}}\ \emph {et~al.}(2014)\citenamefont
  {{Hochberg}}, \citenamefont {{Kuflik}}, \citenamefont {{Volansky}},\ and\
  \citenamefont {{Wacker}}}]{2014arXiv1402.5143H}%
  \BibitemOpen
  \bibfield  {author} {\bibinfo {author} {\bibfnamefont {Y.}~\bibnamefont
  {{Hochberg}}}, \bibinfo {author} {\bibfnamefont {E.}~\bibnamefont
  {{Kuflik}}}, \bibinfo {author} {\bibfnamefont {T.}~\bibnamefont
  {{Volansky}}},\ and\ \bibinfo {author} {\bibfnamefont {J.~G.}\ \bibnamefont
  {{Wacker}}},\ }\bibfield  {title} {\bibinfo {title} {{The SIMP Miracle}},\
  }\href@noop {} {\bibfield  {journal} {\bibinfo  {journal} {arXiv e-prints}\
  ,\ \bibinfo {eid} {arXiv:1402.5143}} (\bibinfo {year} {2014})},\ \Eprint
  {https://arxiv.org/abs/1402.5143} {arXiv:1402.5143 [hep-ph]} \BibitemShut
  {NoStop}%
\bibitem [{\citenamefont {{Hochberg}}\ \emph {et~al.}(2015)\citenamefont
  {{Hochberg}}, \citenamefont {{Kuflik}}, \citenamefont {{Murayama}},
  \citenamefont {{Volansky}},\ and\ \citenamefont
  {{Wacker}}}]{2015PhRvL.115b1301H}%
  \BibitemOpen
  \bibfield  {author} {\bibinfo {author} {\bibfnamefont {Y.}~\bibnamefont
  {{Hochberg}}}, \bibinfo {author} {\bibfnamefont {E.}~\bibnamefont
  {{Kuflik}}}, \bibinfo {author} {\bibfnamefont {H.}~\bibnamefont
  {{Murayama}}}, \bibinfo {author} {\bibfnamefont {T.}~\bibnamefont
  {{Volansky}}},\ and\ \bibinfo {author} {\bibfnamefont {J.~G.}\ \bibnamefont
  {{Wacker}}},\ }\bibfield  {title} {\bibinfo {title} {{Model for Thermal Relic
  Dark Matter of Strongly Interacting Massive Particles}},\ }\href
  {https://doi.org/10.1103/PhysRevLett.115.021301} {\bibfield  {journal}
  {\bibinfo  {journal} {\prl}\ }\textbf {\bibinfo {volume} {115}},\ \bibinfo
  {eid} {021301} (\bibinfo {year} {2015})},\ \Eprint
  {https://arxiv.org/abs/1411.3727} {arXiv:1411.3727 [hep-ph]} \BibitemShut
  {NoStop}%
\bibitem [{\citenamefont {{Kuflik}}\ \emph {et~al.}(2016)\citenamefont
  {{Kuflik}}, \citenamefont {{Perelstein}}, \citenamefont {{Lorier}},\ and\
  \citenamefont {{Tsai}}}]{2016PhRvL.116v1302K}%
  \BibitemOpen
  \bibfield  {author} {\bibinfo {author} {\bibfnamefont {E.}~\bibnamefont
  {{Kuflik}}}, \bibinfo {author} {\bibfnamefont {M.}~\bibnamefont
  {{Perelstein}}}, \bibinfo {author} {\bibfnamefont {N.~R.-L.}\ \bibnamefont
  {{Lorier}}},\ and\ \bibinfo {author} {\bibfnamefont {Y.-D.}\ \bibnamefont
  {{Tsai}}},\ }\bibfield  {title} {\bibinfo {title} {{Elastically Decoupling
  Dark Matter}},\ }\href {https://doi.org/10.1103/PhysRevLett.116.221302}
  {\bibfield  {journal} {\bibinfo  {journal} {\prl}\ }\textbf {\bibinfo
  {volume} {116}},\ \bibinfo {eid} {221302} (\bibinfo {year} {2016})},\ \Eprint
  {https://arxiv.org/abs/1512.04545} {arXiv:1512.04545 [hep-ph]} \BibitemShut
  {NoStop}%
\bibitem [{\citenamefont {{Knapen}}\ \emph {et~al.}(2017)\citenamefont
  {{Knapen}}, \citenamefont {{Lin}},\ and\ \citenamefont
  {{Zurek}}}]{2017PhRvD..96k5021K}%
  \BibitemOpen
  \bibfield  {author} {\bibinfo {author} {\bibfnamefont {S.}~\bibnamefont
  {{Knapen}}}, \bibinfo {author} {\bibfnamefont {T.}~\bibnamefont {{Lin}}},\
  and\ \bibinfo {author} {\bibfnamefont {K.~M.}\ \bibnamefont {{Zurek}}},\
  }\bibfield  {title} {\bibinfo {title} {{Light dark matter: Models and
  constraints}},\ }\href {https://doi.org/10.1103/PhysRevD.96.115021}
  {\bibfield  {journal} {\bibinfo  {journal} {\prd}\ }\textbf {\bibinfo
  {volume} {96}},\ \bibinfo {eid} {115021} (\bibinfo {year} {2017})},\ \Eprint
  {https://arxiv.org/abs/1709.07882} {arXiv:1709.07882 [hep-ph]} \BibitemShut
  {NoStop}%
\bibitem [{\citenamefont {Essig}\ \emph {et~al.}(2012)\citenamefont {Essig},
  \citenamefont {Mardon},\ and\ \citenamefont {Volansky}}]{PhysRevD.85.076007}%
  \BibitemOpen
  \bibfield  {author} {\bibinfo {author} {\bibfnamefont {R.}~\bibnamefont
  {Essig}}, \bibinfo {author} {\bibfnamefont {J.}~\bibnamefont {Mardon}},\ and\
  \bibinfo {author} {\bibfnamefont {T.}~\bibnamefont {Volansky}},\ }\bibfield
  {title} {\bibinfo {title} {Direct detection of sub-gev dark matter},\ }\href
  {https://doi.org/10.1103/PhysRevD.85.076007} {\bibfield  {journal} {\bibinfo
  {journal} {Phys. Rev. D}\ }\textbf {\bibinfo {volume} {85}},\ \bibinfo
  {pages} {076007} (\bibinfo {year} {2012})}\BibitemShut {NoStop}%
\bibitem [{\citenamefont {Derenzo}\ \emph {et~al.}(2017)\citenamefont
  {Derenzo}, \citenamefont {Essig}, \citenamefont {Massari}, \citenamefont
  {Soto},\ and\ \citenamefont {Yu}}]{PhysRevD.96.016026}%
  \BibitemOpen
  \bibfield  {author} {\bibinfo {author} {\bibfnamefont {S.}~\bibnamefont
  {Derenzo}}, \bibinfo {author} {\bibfnamefont {R.}~\bibnamefont {Essig}},
  \bibinfo {author} {\bibfnamefont {A.}~\bibnamefont {Massari}}, \bibinfo
  {author} {\bibfnamefont {A.}~\bibnamefont {Soto}},\ and\ \bibinfo {author}
  {\bibfnamefont {T.-T.}\ \bibnamefont {Yu}},\ }\bibfield  {title} {\bibinfo
  {title} {Direct detection of sub-gev dark matter with scintillating
  targets},\ }\href {https://doi.org/10.1103/PhysRevD.96.016026} {\bibfield
  {journal} {\bibinfo  {journal} {Phys. Rev. D}\ }\textbf {\bibinfo {volume}
  {96}},\ \bibinfo {pages} {016026} (\bibinfo {year} {2017})}\BibitemShut
  {NoStop}%
\bibitem [{\citenamefont {Graham}\ \emph {et~al.}(2012)\citenamefont {Graham},
  \citenamefont {Kaplan}, \citenamefont {Rajendran},\ and\ \citenamefont
  {Walters}}]{GRAHAM201232}%
  \BibitemOpen
  \bibfield  {author} {\bibinfo {author} {\bibfnamefont {P.~W.}\ \bibnamefont
  {Graham}}, \bibinfo {author} {\bibfnamefont {D.~E.}\ \bibnamefont {Kaplan}},
  \bibinfo {author} {\bibfnamefont {S.}~\bibnamefont {Rajendran}},\ and\
  \bibinfo {author} {\bibfnamefont {M.~T.}\ \bibnamefont {Walters}},\
  }\bibfield  {title} {\bibinfo {title} {Semiconductor probes of light dark
  matter},\ }\href {https://doi.org/https://doi.org/10.1016/j.dark.2012.09.001}
  {\bibfield  {journal} {\bibinfo  {journal} {Physics of the Dark Universe}\
  }\textbf {\bibinfo {volume} {1}},\ \bibinfo {pages} {32} (\bibinfo {year}
  {2012})},\ \bibinfo {note} {next Decade in Dark Matter and Dark
  Energy}\BibitemShut {NoStop}%
\bibitem [{\citenamefont {{Hochberg}}\ \emph {et~al.}(2016)\citenamefont
  {{Hochberg}}, \citenamefont {{Pyle}}, \citenamefont {{Zhao}},\ and\
  \citenamefont {{Zurek}}}]{2016JHEP...08..057H}%
  \BibitemOpen
  \bibfield  {author} {\bibinfo {author} {\bibfnamefont {Y.}~\bibnamefont
  {{Hochberg}}}, \bibinfo {author} {\bibfnamefont {M.}~\bibnamefont {{Pyle}}},
  \bibinfo {author} {\bibfnamefont {Y.}~\bibnamefont {{Zhao}}},\ and\ \bibinfo
  {author} {\bibfnamefont {K.~M.}\ \bibnamefont {{Zurek}}},\ }\bibfield
  {title} {\bibinfo {title} {{Detecting superlight dark matter with
  Fermi-degenerate materials}},\ }\href
  {https://doi.org/10.1007/JHEP08(2016)057} {\bibfield  {journal} {\bibinfo
  {journal} {Journal of High Energy Physics}\ }\textbf {\bibinfo {volume}
  {2016}},\ \bibinfo {eid} {57} (\bibinfo {year} {2016})},\ \Eprint
  {https://arxiv.org/abs/1512.04533} {arXiv:1512.04533 [hep-ph]} \BibitemShut
  {NoStop}%
\bibitem [{\citenamefont {Hochberg}\ \emph {et~al.}(2017)\citenamefont
  {Hochberg}, \citenamefont {Kahn}, \citenamefont {Lisanti}, \citenamefont
  {Tully},\ and\ \citenamefont {Zurek}}]{HOCHBERG2017239}%
  \BibitemOpen
  \bibfield  {author} {\bibinfo {author} {\bibfnamefont {Y.}~\bibnamefont
  {Hochberg}}, \bibinfo {author} {\bibfnamefont {Y.}~\bibnamefont {Kahn}},
  \bibinfo {author} {\bibfnamefont {M.}~\bibnamefont {Lisanti}}, \bibinfo
  {author} {\bibfnamefont {C.~G.}\ \bibnamefont {Tully}},\ and\ \bibinfo
  {author} {\bibfnamefont {K.~M.}\ \bibnamefont {Zurek}},\ }\bibfield  {title}
  {\bibinfo {title} {Directional detection of dark matter with two-dimensional
  targets},\ }\href
  {https://doi.org/https://doi.org/10.1016/j.physletb.2017.06.051} {\bibfield
  {journal} {\bibinfo  {journal} {Physics Letters B}\ }\textbf {\bibinfo
  {volume} {772}},\ \bibinfo {pages} {239} (\bibinfo {year}
  {2017})}\BibitemShut {NoStop}%
\bibitem [{\citenamefont {{Battaglieri}}\ \emph {et~al.}(2017)\citenamefont
  {{Battaglieri}}, \citenamefont {{Belloni}}, \citenamefont {{Chou}},
  \citenamefont {{Cushman}}, \citenamefont {{Echenard}}, \citenamefont
  {{Essig}}, \citenamefont {{Estrada}}, \citenamefont {{Feng}}, \citenamefont
  {{Flaugher}}, \citenamefont {{Fox}}, \citenamefont {{Graham}}, \citenamefont
  {{Hall}}, \citenamefont {{Harnik}}, \citenamefont {{Hewett}}, \citenamefont
  {{Incandela}}, \citenamefont {{Izaguirre}}, \citenamefont {{McKinsey}},
  \citenamefont {{Pyle}}, \citenamefont {{Roe}}, \citenamefont {{Rybka}},
  \citenamefont {{Sikivie}}, \citenamefont {{Tait}}, \citenamefont {{Toro}},
  \citenamefont {{Van De Water}}, \citenamefont {{Weiner}}, \citenamefont
  {{Zurek}}, \citenamefont {{Adelberger}}, \citenamefont {{Afanasev}},
  \citenamefont {{Alexander}}, \citenamefont {{Alexander}}, \citenamefont
  {{Cristian Antochi}}, \citenamefont {{Asner}}, \citenamefont {{Baer}},
  \citenamefont {{Banerjee}}, \citenamefont {{Baracchini}}, \citenamefont
  {{Barbeau}}, \citenamefont {{Barrow}}, \citenamefont {{Bastidon}},
  \citenamefont {{Battat}}, \citenamefont {{Benson}}, \citenamefont {{Berlin}},
  \citenamefont {{Bird}}, \citenamefont {{Blinov}}, \citenamefont {{Boddy}},
  \citenamefont {{Bondi}}, \citenamefont {{Bonivento}}, \citenamefont
  {{Boulay}}, \citenamefont {{Boyce}}, \citenamefont {{Brodeur}}, \citenamefont
  {{Broussard}}, \citenamefont {{Budnik}}, \citenamefont {{Bunting}},
  \citenamefont {{Caffee}}, \citenamefont {{Caiazza}}, \citenamefont
  {{Campbell}}, \citenamefont {{Cao}}, \citenamefont {{Carosi}}, \citenamefont
  {{Carpinelli}}, \citenamefont {{Cavoto}}, \citenamefont {{Celentano}},
  \citenamefont {{Hyeok Chang}}, \citenamefont {{Chattopadhyay}}, \citenamefont
  {{Chavarria}}, \citenamefont {{Chen}}, \citenamefont {{Clark}}, \citenamefont
  {{Clarke}}, \citenamefont {{Colegrove}}, \citenamefont {{Coleman}},
  \citenamefont {{Cooke}}, \citenamefont {{Cooper}}, \citenamefont {{Crisler}},
  \citenamefont {{Crivelli}}, \citenamefont {{D'Eramo}}, \citenamefont
  {{D'Urso}}, \citenamefont {{Dahl}}, \citenamefont {{Dawson}}, \citenamefont
  {{De Napoli}}, \citenamefont {{De Vita}}, \citenamefont {{DeNiverville}},
  \citenamefont {{Derenzo}}, \citenamefont {{Di Crescenzo}}, \citenamefont {{Di
  Marco}}, \citenamefont {{Dienes}}, \citenamefont {{Diwan}}, \citenamefont
  {{Handiipondola Dongwi}}, \citenamefont {{Drlica-Wagner}}, \citenamefont
  {{Ellis}}, \citenamefont {{Chigbo Ezeribe}}, \citenamefont {{Farrar}},
  \citenamefont {{Ferrer}}, \citenamefont {{Figueroa-Feliciano}}, \citenamefont
  {{Filippi}}, \citenamefont {{Fiorillo}}, \citenamefont {{Fornal}},
  \citenamefont {{Freyberger}}, \citenamefont {{Frugiuele}}, \citenamefont
  {{Galbiati}}, \citenamefont {{Galon}}, \citenamefont {{Gardner}},
  \citenamefont {{Geraci}}, \citenamefont {{Gerbier}}, \citenamefont
  {{Graham}}, \citenamefont {{Gschwendtner}}, \citenamefont {{Hearty}},
  \citenamefont {{Heise}}, \citenamefont {{Henning}}, \citenamefont {{Hill}},
  \citenamefont {{Hitlin}}, \citenamefont {{Hochberg}}, \citenamefont
  {{Hogan}}, \citenamefont {{Holtrop}}, \citenamefont {{Hong}}, \citenamefont
  {{Hossbach}}, \citenamefont {{Humensky}}, \citenamefont {{Ilten}},
  \citenamefont {{Irwin}}, \citenamefont {{Jaros}}, \citenamefont {{Johnson}},
  \citenamefont {{Jones}}, \citenamefont {{Kahn}}, \citenamefont
  {{Kalantarians}}, \citenamefont {{Kaplinghat}}, \citenamefont {{Khatiwada}},
  \citenamefont {{Knapen}}, \citenamefont {{Kohl}}, \citenamefont {{Kouvaris}},
  \citenamefont {{Kozaczuk}}, \citenamefont {{Krnjaic}}, \citenamefont
  {{Kubarovsky}}, \citenamefont {{Kuflik}}, \citenamefont {{Kusenko}},
  \citenamefont {{Lang}}, \citenamefont {{Leach}}, \citenamefont {{Lin}},
  \citenamefont {{Lisanti}}, \citenamefont {{Liu}}, \citenamefont {{Liu}},
  \citenamefont {{Liu}}, \citenamefont {{Loomba}}, \citenamefont {{Lykken}},
  \citenamefont {{Mack}}, \citenamefont {{Mans}}, \citenamefont {{Maris}},
  \citenamefont {{Markiewicz}}, \citenamefont {{Marsicano}}, \citenamefont
  {{Martoff}}, \citenamefont {{Mazzitelli}}, \citenamefont {{McCabe}},
  \citenamefont {{McDermott}}, \citenamefont {{McDonald}}, \citenamefont
  {{McKinnon}}, \citenamefont {{Mei}}, \citenamefont {{Melia}}, \citenamefont
  {{Miller}}, \citenamefont {{Miuchi}}, \citenamefont {{Nazeer}}, \citenamefont
  {{Moreno}}, \citenamefont {{Morozov}}, \citenamefont {{Mouton}},
  \citenamefont {{Mueller}}, \citenamefont {{Murphy}}, \citenamefont
  {{Neilson}}, \citenamefont {{Nelson}}, \citenamefont {{Neu}}, \citenamefont
  {{Nosochkov}}, \citenamefont {{O'Hare}}, \citenamefont {{Oblath}},
  \citenamefont {{Orrell}}, \citenamefont {{Ouellet}}, \citenamefont
  {{Pastore}}, \citenamefont {{Paul}}, \citenamefont {{Perelstein}},
  \citenamefont {{Peter}}, \citenamefont {{Phan}}, \citenamefont {{Phinney}},
  \citenamefont {{Pivovaroff}}, \citenamefont {{Pocar}}, \citenamefont
  {{Pospelov}}, \citenamefont {{Pradler}}, \citenamefont {{Privitera}},
  \citenamefont {{Profumo}}, \citenamefont {{Raggi}}, \citenamefont
  {{Rajendran}}, \citenamefont {{Randazzo}}, \citenamefont {{Raubenheimer}},
  \citenamefont {{Regenfus}}, \citenamefont {{Renshaw}}, \citenamefont
  {{Ritz}}, \citenamefont {{Rizzo}}, \citenamefont {{Rosenberg}}, \citenamefont
  {{Rubbia}}, \citenamefont {{Rybolt}}, \citenamefont {{Saab}}, \citenamefont
  {{Safdi}}, \citenamefont {{Santopinto}}, \citenamefont {{Scarff}},
  \citenamefont {{Schneider}}, \citenamefont {{Schuster}}, \citenamefont
  {{Seidel}}, \citenamefont {{Sekiya}}, \citenamefont {{Seong}}, \citenamefont
  {{Simi}}, \citenamefont {{Sipala}}, \citenamefont {{Slatyer}}, \citenamefont
  {{Slone}}, \citenamefont {{Smith}}, \citenamefont {{Smolinsky}},
  \citenamefont {{Snowden-Ifft}}, \citenamefont {{Solt}}, \citenamefont
  {{Sonnenschein}}, \citenamefont {{Sorensen}}, \citenamefont {{Spooner}},
  \citenamefont {{Srivastava}}, \citenamefont {{Stancu}}, \citenamefont
  {{Strigari}}, \citenamefont {{Strube}}, \citenamefont {{Sushkov}},
  \citenamefont {{Szydagis}}, \citenamefont {{Tanedo}}, \citenamefont
  {{Tanner}}, \citenamefont {{Tayloe}}, \citenamefont {{Terrano}},
  \citenamefont {{Thaler}}, \citenamefont {{Thomas}}, \citenamefont {{Thorpe}},
  \citenamefont {{Thorpe}}, \citenamefont {{Tiffenberg}}, \citenamefont
  {{Tran}}, \citenamefont {{Trovato}}, \citenamefont {{Tully}}, \citenamefont
  {{Tyson}}, \citenamefont {{Vachaspati}}, \citenamefont {{Vahsen}},
  \citenamefont {{van Bibber}}, \citenamefont {{Vandenbroucke}}, \citenamefont
  {{Villano}}, \citenamefont {{Volansky}}, \citenamefont {{Wang}},
  \citenamefont {{Ward}}, \citenamefont {{Wester}}, \citenamefont {{Whitbeck}},
  \citenamefont {{Williams}}, \citenamefont {{Wing}}, \citenamefont
  {{Winslow}}, \citenamefont {{Wojtsekhowski}}, \citenamefont {{Yu}},
  \citenamefont {{Yu}}, \citenamefont {{Yu}}, \citenamefont {{Zhang}},
  \citenamefont {{Zhao}},\ and\ \citenamefont {{Zhong}}}]{2017arXiv170704591B}%
  \BibitemOpen
  \bibfield  {author} {\bibinfo {author} {\bibfnamefont {M.}~\bibnamefont
  {{Battaglieri}}}, \bibinfo {author} {\bibfnamefont {A.}~\bibnamefont
  {{Belloni}}}, \bibinfo {author} {\bibfnamefont {A.}~\bibnamefont {{Chou}}},
  \bibinfo {author} {\bibfnamefont {P.}~\bibnamefont {{Cushman}}}, \bibinfo
  {author} {\bibfnamefont {B.}~\bibnamefont {{Echenard}}}, \bibinfo {author}
  {\bibfnamefont {R.}~\bibnamefont {{Essig}}}, \bibinfo {author} {\bibfnamefont
  {J.}~\bibnamefont {{Estrada}}}, \bibinfo {author} {\bibfnamefont {J.~L.}\
  \bibnamefont {{Feng}}}, \bibinfo {author} {\bibfnamefont {B.}~\bibnamefont
  {{Flaugher}}}, \bibinfo {author} {\bibfnamefont {P.~J.}\ \bibnamefont
  {{Fox}}}, \bibinfo {author} {\bibfnamefont {P.}~\bibnamefont {{Graham}}},
  \bibinfo {author} {\bibfnamefont {C.}~\bibnamefont {{Hall}}}, \bibinfo
  {author} {\bibfnamefont {R.}~\bibnamefont {{Harnik}}}, \bibinfo {author}
  {\bibfnamefont {J.}~\bibnamefont {{Hewett}}}, \bibinfo {author}
  {\bibfnamefont {J.}~\bibnamefont {{Incandela}}}, \bibinfo {author}
  {\bibfnamefont {E.}~\bibnamefont {{Izaguirre}}}, \bibinfo {author}
  {\bibfnamefont {D.}~\bibnamefont {{McKinsey}}}, \bibinfo {author}
  {\bibfnamefont {M.}~\bibnamefont {{Pyle}}}, \bibinfo {author} {\bibfnamefont
  {N.}~\bibnamefont {{Roe}}}, \bibinfo {author} {\bibfnamefont
  {G.}~\bibnamefont {{Rybka}}}, \bibinfo {author} {\bibfnamefont
  {P.}~\bibnamefont {{Sikivie}}}, \bibinfo {author} {\bibfnamefont {T.~M.~P.}\
  \bibnamefont {{Tait}}}, \bibinfo {author} {\bibfnamefont {N.}~\bibnamefont
  {{Toro}}}, \bibinfo {author} {\bibfnamefont {R.}~\bibnamefont {{Van De
  Water}}}, \bibinfo {author} {\bibfnamefont {N.}~\bibnamefont {{Weiner}}},
  \bibinfo {author} {\bibfnamefont {K.}~\bibnamefont {{Zurek}}}, \bibinfo
  {author} {\bibfnamefont {E.}~\bibnamefont {{Adelberger}}}, \bibinfo {author}
  {\bibfnamefont {A.}~\bibnamefont {{Afanasev}}}, \bibinfo {author}
  {\bibfnamefont {D.}~\bibnamefont {{Alexander}}}, \bibinfo {author}
  {\bibfnamefont {J.}~\bibnamefont {{Alexander}}}, \bibinfo {author}
  {\bibfnamefont {V.}~\bibnamefont {{Cristian Antochi}}}, \bibinfo {author}
  {\bibfnamefont {D.~M.}\ \bibnamefont {{Asner}}}, \bibinfo {author}
  {\bibfnamefont {H.}~\bibnamefont {{Baer}}}, \bibinfo {author} {\bibfnamefont
  {D.}~\bibnamefont {{Banerjee}}}, \bibinfo {author} {\bibfnamefont
  {E.}~\bibnamefont {{Baracchini}}}, \bibinfo {author} {\bibfnamefont
  {P.}~\bibnamefont {{Barbeau}}}, \bibinfo {author} {\bibfnamefont
  {J.}~\bibnamefont {{Barrow}}}, \bibinfo {author} {\bibfnamefont
  {N.}~\bibnamefont {{Bastidon}}}, \bibinfo {author} {\bibfnamefont
  {J.}~\bibnamefont {{Battat}}}, \bibinfo {author} {\bibfnamefont
  {S.}~\bibnamefont {{Benson}}}, \bibinfo {author} {\bibfnamefont
  {A.}~\bibnamefont {{Berlin}}}, \bibinfo {author} {\bibfnamefont
  {M.}~\bibnamefont {{Bird}}}, \bibinfo {author} {\bibfnamefont
  {N.}~\bibnamefont {{Blinov}}}, \bibinfo {author} {\bibfnamefont {K.~K.}\
  \bibnamefont {{Boddy}}}, \bibinfo {author} {\bibfnamefont {M.}~\bibnamefont
  {{Bondi}}}, \bibinfo {author} {\bibfnamefont {W.~M.}\ \bibnamefont
  {{Bonivento}}}, \bibinfo {author} {\bibfnamefont {M.}~\bibnamefont
  {{Boulay}}}, \bibinfo {author} {\bibfnamefont {J.}~\bibnamefont {{Boyce}}},
  \bibinfo {author} {\bibfnamefont {M.}~\bibnamefont {{Brodeur}}}, \bibinfo
  {author} {\bibfnamefont {L.}~\bibnamefont {{Broussard}}}, \bibinfo {author}
  {\bibfnamefont {R.}~\bibnamefont {{Budnik}}}, \bibinfo {author}
  {\bibfnamefont {P.}~\bibnamefont {{Bunting}}}, \bibinfo {author}
  {\bibfnamefont {M.}~\bibnamefont {{Caffee}}}, \bibinfo {author}
  {\bibfnamefont {S.~S.}\ \bibnamefont {{Caiazza}}}, \bibinfo {author}
  {\bibfnamefont {S.}~\bibnamefont {{Campbell}}}, \bibinfo {author}
  {\bibfnamefont {T.}~\bibnamefont {{Cao}}}, \bibinfo {author} {\bibfnamefont
  {G.}~\bibnamefont {{Carosi}}}, \bibinfo {author} {\bibfnamefont
  {M.}~\bibnamefont {{Carpinelli}}}, \bibinfo {author} {\bibfnamefont
  {G.}~\bibnamefont {{Cavoto}}}, \bibinfo {author} {\bibfnamefont
  {A.}~\bibnamefont {{Celentano}}}, \bibinfo {author} {\bibfnamefont
  {J.}~\bibnamefont {{Hyeok Chang}}}, \bibinfo {author} {\bibfnamefont
  {S.}~\bibnamefont {{Chattopadhyay}}}, \bibinfo {author} {\bibfnamefont
  {A.}~\bibnamefont {{Chavarria}}}, \bibinfo {author} {\bibfnamefont {C.-Y.}\
  \bibnamefont {{Chen}}}, \bibinfo {author} {\bibfnamefont {K.}~\bibnamefont
  {{Clark}}}, \bibinfo {author} {\bibfnamefont {J.}~\bibnamefont {{Clarke}}},
  \bibinfo {author} {\bibfnamefont {O.}~\bibnamefont {{Colegrove}}}, \bibinfo
  {author} {\bibfnamefont {J.}~\bibnamefont {{Coleman}}}, \bibinfo {author}
  {\bibfnamefont {D.}~\bibnamefont {{Cooke}}}, \bibinfo {author} {\bibfnamefont
  {R.}~\bibnamefont {{Cooper}}}, \bibinfo {author} {\bibfnamefont
  {M.}~\bibnamefont {{Crisler}}}, \bibinfo {author} {\bibfnamefont
  {P.}~\bibnamefont {{Crivelli}}}, \bibinfo {author} {\bibfnamefont
  {F.}~\bibnamefont {{D'Eramo}}}, \bibinfo {author} {\bibfnamefont
  {D.}~\bibnamefont {{D'Urso}}}, \bibinfo {author} {\bibfnamefont
  {E.}~\bibnamefont {{Dahl}}}, \bibinfo {author} {\bibfnamefont
  {W.}~\bibnamefont {{Dawson}}}, \bibinfo {author} {\bibfnamefont
  {M.}~\bibnamefont {{De Napoli}}}, \bibinfo {author} {\bibfnamefont
  {R.}~\bibnamefont {{De Vita}}}, \bibinfo {author} {\bibfnamefont
  {P.}~\bibnamefont {{DeNiverville}}}, \bibinfo {author} {\bibfnamefont
  {S.}~\bibnamefont {{Derenzo}}}, \bibinfo {author} {\bibfnamefont
  {A.}~\bibnamefont {{Di Crescenzo}}}, \bibinfo {author} {\bibfnamefont
  {E.}~\bibnamefont {{Di Marco}}}, \bibinfo {author} {\bibfnamefont {K.~R.}\
  \bibnamefont {{Dienes}}}, \bibinfo {author} {\bibfnamefont {M.}~\bibnamefont
  {{Diwan}}}, \bibinfo {author} {\bibfnamefont {D.}~\bibnamefont
  {{Handiipondola Dongwi}}}, \bibinfo {author} {\bibfnamefont {A.}~\bibnamefont
  {{Drlica-Wagner}}}, \bibinfo {author} {\bibfnamefont {S.}~\bibnamefont
  {{Ellis}}}, \bibinfo {author} {\bibfnamefont {A.}~\bibnamefont {{Chigbo
  Ezeribe}}}, \bibinfo {author} {\bibfnamefont {G.}~\bibnamefont {{Farrar}}},
  \bibinfo {author} {\bibfnamefont {F.}~\bibnamefont {{Ferrer}}}, \bibinfo
  {author} {\bibfnamefont {E.}~\bibnamefont {{Figueroa-Feliciano}}}, \bibinfo
  {author} {\bibfnamefont {A.}~\bibnamefont {{Filippi}}}, \bibinfo {author}
  {\bibfnamefont {G.}~\bibnamefont {{Fiorillo}}}, \bibinfo {author}
  {\bibfnamefont {B.}~\bibnamefont {{Fornal}}}, \bibinfo {author}
  {\bibfnamefont {A.}~\bibnamefont {{Freyberger}}}, \bibinfo {author}
  {\bibfnamefont {C.}~\bibnamefont {{Frugiuele}}}, \bibinfo {author}
  {\bibfnamefont {C.}~\bibnamefont {{Galbiati}}}, \bibinfo {author}
  {\bibfnamefont {I.}~\bibnamefont {{Galon}}}, \bibinfo {author} {\bibfnamefont
  {S.}~\bibnamefont {{Gardner}}}, \bibinfo {author} {\bibfnamefont
  {A.}~\bibnamefont {{Geraci}}}, \bibinfo {author} {\bibfnamefont
  {G.}~\bibnamefont {{Gerbier}}}, \bibinfo {author} {\bibfnamefont
  {M.}~\bibnamefont {{Graham}}}, \bibinfo {author} {\bibfnamefont
  {E.}~\bibnamefont {{Gschwendtner}}}, \bibinfo {author} {\bibfnamefont
  {C.}~\bibnamefont {{Hearty}}}, \bibinfo {author} {\bibfnamefont
  {J.}~\bibnamefont {{Heise}}}, \bibinfo {author} {\bibfnamefont
  {R.}~\bibnamefont {{Henning}}}, \bibinfo {author} {\bibfnamefont {R.~J.}\
  \bibnamefont {{Hill}}}, \bibinfo {author} {\bibfnamefont {D.}~\bibnamefont
  {{Hitlin}}}, \bibinfo {author} {\bibfnamefont {Y.}~\bibnamefont
  {{Hochberg}}}, \bibinfo {author} {\bibfnamefont {J.}~\bibnamefont {{Hogan}}},
  \bibinfo {author} {\bibfnamefont {M.}~\bibnamefont {{Holtrop}}}, \bibinfo
  {author} {\bibfnamefont {Z.}~\bibnamefont {{Hong}}}, \bibinfo {author}
  {\bibfnamefont {T.}~\bibnamefont {{Hossbach}}}, \bibinfo {author}
  {\bibfnamefont {T.~B.}\ \bibnamefont {{Humensky}}}, \bibinfo {author}
  {\bibfnamefont {P.}~\bibnamefont {{Ilten}}}, \bibinfo {author} {\bibfnamefont
  {K.}~\bibnamefont {{Irwin}}}, \bibinfo {author} {\bibfnamefont
  {J.}~\bibnamefont {{Jaros}}}, \bibinfo {author} {\bibfnamefont
  {R.}~\bibnamefont {{Johnson}}}, \bibinfo {author} {\bibfnamefont
  {M.}~\bibnamefont {{Jones}}}, \bibinfo {author} {\bibfnamefont
  {Y.}~\bibnamefont {{Kahn}}}, \bibinfo {author} {\bibfnamefont
  {N.}~\bibnamefont {{Kalantarians}}}, \bibinfo {author} {\bibfnamefont
  {M.}~\bibnamefont {{Kaplinghat}}}, \bibinfo {author} {\bibfnamefont
  {R.}~\bibnamefont {{Khatiwada}}}, \bibinfo {author} {\bibfnamefont
  {S.}~\bibnamefont {{Knapen}}}, \bibinfo {author} {\bibfnamefont
  {M.}~\bibnamefont {{Kohl}}}, \bibinfo {author} {\bibfnamefont
  {C.}~\bibnamefont {{Kouvaris}}}, \bibinfo {author} {\bibfnamefont
  {J.}~\bibnamefont {{Kozaczuk}}}, \bibinfo {author} {\bibfnamefont
  {G.}~\bibnamefont {{Krnjaic}}}, \bibinfo {author} {\bibfnamefont
  {V.}~\bibnamefont {{Kubarovsky}}}, \bibinfo {author} {\bibfnamefont
  {E.}~\bibnamefont {{Kuflik}}}, \bibinfo {author} {\bibfnamefont
  {A.}~\bibnamefont {{Kusenko}}}, \bibinfo {author} {\bibfnamefont
  {R.}~\bibnamefont {{Lang}}}, \bibinfo {author} {\bibfnamefont
  {K.}~\bibnamefont {{Leach}}}, \bibinfo {author} {\bibfnamefont
  {T.}~\bibnamefont {{Lin}}}, \bibinfo {author} {\bibfnamefont
  {M.}~\bibnamefont {{Lisanti}}}, \bibinfo {author} {\bibfnamefont
  {J.}~\bibnamefont {{Liu}}}, \bibinfo {author} {\bibfnamefont
  {K.}~\bibnamefont {{Liu}}}, \bibinfo {author} {\bibfnamefont
  {M.}~\bibnamefont {{Liu}}}, \bibinfo {author} {\bibfnamefont
  {D.}~\bibnamefont {{Loomba}}}, \bibinfo {author} {\bibfnamefont
  {J.}~\bibnamefont {{Lykken}}}, \bibinfo {author} {\bibfnamefont
  {K.}~\bibnamefont {{Mack}}}, \bibinfo {author} {\bibfnamefont
  {J.}~\bibnamefont {{Mans}}}, \bibinfo {author} {\bibfnamefont
  {H.}~\bibnamefont {{Maris}}}, \bibinfo {author} {\bibfnamefont
  {T.}~\bibnamefont {{Markiewicz}}}, \bibinfo {author} {\bibfnamefont
  {L.}~\bibnamefont {{Marsicano}}}, \bibinfo {author} {\bibfnamefont {C.~J.}\
  \bibnamefont {{Martoff}}}, \bibinfo {author} {\bibfnamefont {G.}~\bibnamefont
  {{Mazzitelli}}}, \bibinfo {author} {\bibfnamefont {C.}~\bibnamefont
  {{McCabe}}}, \bibinfo {author} {\bibfnamefont {S.~D.}\ \bibnamefont
  {{McDermott}}}, \bibinfo {author} {\bibfnamefont {A.}~\bibnamefont
  {{McDonald}}}, \bibinfo {author} {\bibfnamefont {B.}~\bibnamefont
  {{McKinnon}}}, \bibinfo {author} {\bibfnamefont {D.}~\bibnamefont {{Mei}}},
  \bibinfo {author} {\bibfnamefont {T.}~\bibnamefont {{Melia}}}, \bibinfo
  {author} {\bibfnamefont {G.~A.}\ \bibnamefont {{Miller}}}, \bibinfo {author}
  {\bibfnamefont {K.}~\bibnamefont {{Miuchi}}}, \bibinfo {author}
  {\bibfnamefont {S.~M.~P.}\ \bibnamefont {{Nazeer}}}, \bibinfo {author}
  {\bibfnamefont {O.}~\bibnamefont {{Moreno}}}, \bibinfo {author}
  {\bibfnamefont {V.}~\bibnamefont {{Morozov}}}, \bibinfo {author}
  {\bibfnamefont {F.}~\bibnamefont {{Mouton}}}, \bibinfo {author}
  {\bibfnamefont {H.}~\bibnamefont {{Mueller}}}, \bibinfo {author}
  {\bibfnamefont {A.}~\bibnamefont {{Murphy}}}, \bibinfo {author}
  {\bibfnamefont {R.}~\bibnamefont {{Neilson}}}, \bibinfo {author}
  {\bibfnamefont {T.}~\bibnamefont {{Nelson}}}, \bibinfo {author}
  {\bibfnamefont {C.}~\bibnamefont {{Neu}}}, \bibinfo {author} {\bibfnamefont
  {Y.}~\bibnamefont {{Nosochkov}}}, \bibinfo {author} {\bibfnamefont
  {C.}~\bibnamefont {{O'Hare}}}, \bibinfo {author} {\bibfnamefont
  {N.}~\bibnamefont {{Oblath}}}, \bibinfo {author} {\bibfnamefont
  {J.}~\bibnamefont {{Orrell}}}, \bibinfo {author} {\bibfnamefont
  {J.}~\bibnamefont {{Ouellet}}}, \bibinfo {author} {\bibfnamefont
  {S.}~\bibnamefont {{Pastore}}}, \bibinfo {author} {\bibfnamefont
  {S.}~\bibnamefont {{Paul}}}, \bibinfo {author} {\bibfnamefont
  {M.}~\bibnamefont {{Perelstein}}}, \bibinfo {author} {\bibfnamefont
  {A.}~\bibnamefont {{Peter}}}, \bibinfo {author} {\bibfnamefont
  {N.}~\bibnamefont {{Phan}}}, \bibinfo {author} {\bibfnamefont
  {N.}~\bibnamefont {{Phinney}}}, \bibinfo {author} {\bibfnamefont
  {M.}~\bibnamefont {{Pivovaroff}}}, \bibinfo {author} {\bibfnamefont
  {A.}~\bibnamefont {{Pocar}}}, \bibinfo {author} {\bibfnamefont
  {M.}~\bibnamefont {{Pospelov}}}, \bibinfo {author} {\bibfnamefont
  {J.}~\bibnamefont {{Pradler}}}, \bibinfo {author} {\bibfnamefont
  {P.}~\bibnamefont {{Privitera}}}, \bibinfo {author} {\bibfnamefont
  {S.}~\bibnamefont {{Profumo}}}, \bibinfo {author} {\bibfnamefont
  {M.}~\bibnamefont {{Raggi}}}, \bibinfo {author} {\bibfnamefont
  {S.}~\bibnamefont {{Rajendran}}}, \bibinfo {author} {\bibfnamefont
  {N.}~\bibnamefont {{Randazzo}}}, \bibinfo {author} {\bibfnamefont
  {T.}~\bibnamefont {{Raubenheimer}}}, \bibinfo {author} {\bibfnamefont
  {C.}~\bibnamefont {{Regenfus}}}, \bibinfo {author} {\bibfnamefont
  {A.}~\bibnamefont {{Renshaw}}}, \bibinfo {author} {\bibfnamefont
  {A.}~\bibnamefont {{Ritz}}}, \bibinfo {author} {\bibfnamefont
  {T.}~\bibnamefont {{Rizzo}}}, \bibinfo {author} {\bibfnamefont
  {L.}~\bibnamefont {{Rosenberg}}}, \bibinfo {author} {\bibfnamefont
  {A.}~\bibnamefont {{Rubbia}}}, \bibinfo {author} {\bibfnamefont
  {B.}~\bibnamefont {{Rybolt}}}, \bibinfo {author} {\bibfnamefont
  {T.}~\bibnamefont {{Saab}}}, \bibinfo {author} {\bibfnamefont {B.~R.}\
  \bibnamefont {{Safdi}}}, \bibinfo {author} {\bibfnamefont {E.}~\bibnamefont
  {{Santopinto}}}, \bibinfo {author} {\bibfnamefont {A.}~\bibnamefont
  {{Scarff}}}, \bibinfo {author} {\bibfnamefont {M.}~\bibnamefont
  {{Schneider}}}, \bibinfo {author} {\bibfnamefont {P.}~\bibnamefont
  {{Schuster}}}, \bibinfo {author} {\bibfnamefont {G.}~\bibnamefont
  {{Seidel}}}, \bibinfo {author} {\bibfnamefont {H.}~\bibnamefont {{Sekiya}}},
  \bibinfo {author} {\bibfnamefont {I.}~\bibnamefont {{Seong}}}, \bibinfo
  {author} {\bibfnamefont {G.}~\bibnamefont {{Simi}}}, \bibinfo {author}
  {\bibfnamefont {V.}~\bibnamefont {{Sipala}}}, \bibinfo {author}
  {\bibfnamefont {T.}~\bibnamefont {{Slatyer}}}, \bibinfo {author}
  {\bibfnamefont {O.}~\bibnamefont {{Slone}}}, \bibinfo {author} {\bibfnamefont
  {P.~F.}\ \bibnamefont {{Smith}}}, \bibinfo {author} {\bibfnamefont
  {J.}~\bibnamefont {{Smolinsky}}}, \bibinfo {author} {\bibfnamefont
  {D.}~\bibnamefont {{Snowden-Ifft}}}, \bibinfo {author} {\bibfnamefont
  {M.}~\bibnamefont {{Solt}}}, \bibinfo {author} {\bibfnamefont
  {A.}~\bibnamefont {{Sonnenschein}}}, \bibinfo {author} {\bibfnamefont
  {P.}~\bibnamefont {{Sorensen}}}, \bibinfo {author} {\bibfnamefont
  {N.}~\bibnamefont {{Spooner}}}, \bibinfo {author} {\bibfnamefont
  {B.}~\bibnamefont {{Srivastava}}}, \bibinfo {author} {\bibfnamefont
  {I.}~\bibnamefont {{Stancu}}}, \bibinfo {author} {\bibfnamefont
  {L.}~\bibnamefont {{Strigari}}}, \bibinfo {author} {\bibfnamefont
  {J.}~\bibnamefont {{Strube}}}, \bibinfo {author} {\bibfnamefont {A.~O.}\
  \bibnamefont {{Sushkov}}}, \bibinfo {author} {\bibfnamefont {M.}~\bibnamefont
  {{Szydagis}}}, \bibinfo {author} {\bibfnamefont {P.}~\bibnamefont
  {{Tanedo}}}, \bibinfo {author} {\bibfnamefont {D.}~\bibnamefont {{Tanner}}},
  \bibinfo {author} {\bibfnamefont {R.}~\bibnamefont {{Tayloe}}}, \bibinfo
  {author} {\bibfnamefont {W.}~\bibnamefont {{Terrano}}}, \bibinfo {author}
  {\bibfnamefont {J.}~\bibnamefont {{Thaler}}}, \bibinfo {author}
  {\bibfnamefont {B.}~\bibnamefont {{Thomas}}}, \bibinfo {author}
  {\bibfnamefont {B.}~\bibnamefont {{Thorpe}}}, \bibinfo {author}
  {\bibfnamefont {T.}~\bibnamefont {{Thorpe}}}, \bibinfo {author}
  {\bibfnamefont {J.}~\bibnamefont {{Tiffenberg}}}, \bibinfo {author}
  {\bibfnamefont {N.}~\bibnamefont {{Tran}}}, \bibinfo {author} {\bibfnamefont
  {M.}~\bibnamefont {{Trovato}}}, \bibinfo {author} {\bibfnamefont
  {C.}~\bibnamefont {{Tully}}}, \bibinfo {author} {\bibfnamefont
  {T.}~\bibnamefont {{Tyson}}}, \bibinfo {author} {\bibfnamefont
  {T.}~\bibnamefont {{Vachaspati}}}, \bibinfo {author} {\bibfnamefont
  {S.}~\bibnamefont {{Vahsen}}}, \bibinfo {author} {\bibfnamefont
  {K.}~\bibnamefont {{van Bibber}}}, \bibinfo {author} {\bibfnamefont
  {J.}~\bibnamefont {{Vandenbroucke}}}, \bibinfo {author} {\bibfnamefont
  {A.}~\bibnamefont {{Villano}}}, \bibinfo {author} {\bibfnamefont
  {T.}~\bibnamefont {{Volansky}}}, \bibinfo {author} {\bibfnamefont
  {G.}~\bibnamefont {{Wang}}}, \bibinfo {author} {\bibfnamefont
  {T.}~\bibnamefont {{Ward}}}, \bibinfo {author} {\bibfnamefont
  {W.}~\bibnamefont {{Wester}}}, \bibinfo {author} {\bibfnamefont
  {A.}~\bibnamefont {{Whitbeck}}}, \bibinfo {author} {\bibfnamefont {D.~A.}\
  \bibnamefont {{Williams}}}, \bibinfo {author} {\bibfnamefont
  {M.}~\bibnamefont {{Wing}}}, \bibinfo {author} {\bibfnamefont
  {L.}~\bibnamefont {{Winslow}}}, \bibinfo {author} {\bibfnamefont
  {B.}~\bibnamefont {{Wojtsekhowski}}}, \bibinfo {author} {\bibfnamefont
  {H.-B.}\ \bibnamefont {{Yu}}}, \bibinfo {author} {\bibfnamefont {S.-S.}\
  \bibnamefont {{Yu}}}, \bibinfo {author} {\bibfnamefont {T.-T.}\ \bibnamefont
  {{Yu}}}, \bibinfo {author} {\bibfnamefont {X.}~\bibnamefont {{Zhang}}},
  \bibinfo {author} {\bibfnamefont {Y.}~\bibnamefont {{Zhao}}},\ and\ \bibinfo
  {author} {\bibfnamefont {Y.-M.}\ \bibnamefont {{Zhong}}},\ }\bibfield
  {title} {\bibinfo {title} {{US Cosmic Visions: New Ideas in Dark Matter 2017:
  Community Report}},\ }\href@noop {} {\bibfield  {journal} {\bibinfo
  {journal} {arXiv e-prints}\ ,\ \bibinfo {eid} {arXiv:1707.04591}} (\bibinfo
  {year} {2017})},\ \Eprint {https://arxiv.org/abs/1707.04591}
  {arXiv:1707.04591 [hep-ph]} \BibitemShut {NoStop}%
\bibitem [{\citenamefont {{Chen}}\ \emph {et~al.}(2002)\citenamefont {{Chen}},
  \citenamefont {{Hannestad}},\ and\ \citenamefont
  {{Scherrer}}}]{2002astro.ph..2496C}%
  \BibitemOpen
  \bibfield  {author} {\bibinfo {author} {\bibfnamefont {X.}~\bibnamefont
  {{Chen}}}, \bibinfo {author} {\bibfnamefont {S.}~\bibnamefont
  {{Hannestad}}},\ and\ \bibinfo {author} {\bibfnamefont {R.~J.}\ \bibnamefont
  {{Scherrer}}},\ }\bibfield  {title} {\bibinfo {title} {{Cosmic microwave
  background and large scale structure limits on the interaction between dark
  matter and baryons}},\ }\href@noop {} {\bibfield  {journal} {\bibinfo
  {journal} {arXiv e-prints}\ ,\ \bibinfo {eid} {astro-ph/0202496}} (\bibinfo
  {year} {2002})},\ \Eprint {https://arxiv.org/abs/astro-ph/0202496}
  {arXiv:astro-ph/0202496 [astro-ph]} \BibitemShut {NoStop}%
\bibitem [{\citenamefont {{Dvorkin}}\ \emph {et~al.}(2014)\citenamefont
  {{Dvorkin}}, \citenamefont {{Blum}},\ and\ \citenamefont
  {{Kamionkowski}}}]{2014PhRvD..89b3519D}%
  \BibitemOpen
  \bibfield  {author} {\bibinfo {author} {\bibfnamefont {C.}~\bibnamefont
  {{Dvorkin}}}, \bibinfo {author} {\bibfnamefont {K.}~\bibnamefont {{Blum}}},\
  and\ \bibinfo {author} {\bibfnamefont {M.}~\bibnamefont {{Kamionkowski}}},\
  }\bibfield  {title} {\bibinfo {title} {{Constraining dark matter-baryon
  scattering with linear cosmology}},\ }\href
  {https://doi.org/10.1103/PhysRevD.89.023519} {\bibfield  {journal} {\bibinfo
  {journal} {\prd}\ }\textbf {\bibinfo {volume} {89}},\ \bibinfo {eid} {023519}
  (\bibinfo {year} {2014})},\ \Eprint {https://arxiv.org/abs/1311.2937}
  {arXiv:1311.2937 [astro-ph.CO]} \BibitemShut {NoStop}%
\bibitem [{\citenamefont {{Xu}}\ \emph {et~al.}(2018)\citenamefont {{Xu}},
  \citenamefont {{Dvorkin}},\ and\ \citenamefont
  {{Chael}}}]{2018PhRvD..97j3530X}%
  \BibitemOpen
  \bibfield  {author} {\bibinfo {author} {\bibfnamefont {W.~L.}\ \bibnamefont
  {{Xu}}}, \bibinfo {author} {\bibfnamefont {C.}~\bibnamefont {{Dvorkin}}},\
  and\ \bibinfo {author} {\bibfnamefont {A.}~\bibnamefont {{Chael}}},\
  }\bibfield  {title} {\bibinfo {title} {{Probing sub-GeV dark matter-baryon
  scattering with cosmological observables}},\ }\href
  {https://doi.org/10.1103/PhysRevD.97.103530} {\bibfield  {journal} {\bibinfo
  {journal} {\prd}\ }\textbf {\bibinfo {volume} {97}},\ \bibinfo {eid} {103530}
  (\bibinfo {year} {2018})},\ \Eprint {https://arxiv.org/abs/1802.06788}
  {arXiv:1802.06788 [astro-ph.CO]} \BibitemShut {NoStop}%
\bibitem [{\citenamefont {{Boddy}}\ and\ \citenamefont
  {{Gluscevic}}(2018)}]{2018PhRvD..98h3510B}%
  \BibitemOpen
  \bibfield  {author} {\bibinfo {author} {\bibfnamefont {K.~K.}\ \bibnamefont
  {{Boddy}}}\ and\ \bibinfo {author} {\bibfnamefont {V.}~\bibnamefont
  {{Gluscevic}}},\ }\bibfield  {title} {\bibinfo {title} {{First cosmological
  constraint on the effective theory of dark matter-proton interactions}},\
  }\href {https://doi.org/10.1103/PhysRevD.98.083510} {\bibfield  {journal}
  {\bibinfo  {journal} {\prd}\ }\textbf {\bibinfo {volume} {98}},\ \bibinfo
  {eid} {083510} (\bibinfo {year} {2018})},\ \Eprint
  {https://arxiv.org/abs/1801.08609} {arXiv:1801.08609 [astro-ph.CO]}
  \BibitemShut {NoStop}%
\bibitem [{\citenamefont {{Slatyer}}\ and\ \citenamefont
  {{Wu}}(2018)}]{2018PhRvD..98b3013S}%
  \BibitemOpen
  \bibfield  {author} {\bibinfo {author} {\bibfnamefont {T.~R.}\ \bibnamefont
  {{Slatyer}}}\ and\ \bibinfo {author} {\bibfnamefont {C.-L.}\ \bibnamefont
  {{Wu}}},\ }\bibfield  {title} {\bibinfo {title} {{Early-Universe constraints
  on dark matter-baryon scattering and their implications for a global 21 cm
  signal}},\ }\href {https://doi.org/10.1103/PhysRevD.98.023013} {\bibfield
  {journal} {\bibinfo  {journal} {\prd}\ }\textbf {\bibinfo {volume} {98}},\
  \bibinfo {eid} {023013} (\bibinfo {year} {2018})},\ \Eprint
  {https://arxiv.org/abs/1803.09734} {arXiv:1803.09734 [astro-ph.CO]}
  \BibitemShut {NoStop}%
\bibitem [{\citenamefont {{Nadler}}\ \emph
  {et~al.}(2021{\natexlab{a}})\citenamefont {{Nadler}}, \citenamefont
  {{Drlica-Wagner}}, \citenamefont {{Bechtol}}, \citenamefont {{Mau}},
  \citenamefont {{Wechsler}}, \citenamefont {{Gluscevic}}, \citenamefont
  {{Boddy}}, \citenamefont {{Pace}}, \citenamefont {{Li}}, \citenamefont
  {{McNanna}}, \citenamefont {{Riley}}, \citenamefont {{Garc{\'\i}a-Bellido}},
  \citenamefont {{Mao}}, \citenamefont {{Green}}, \citenamefont {{Burke}},
  \citenamefont {{Peter}}, \citenamefont {{Jain}}, \citenamefont {{Abbott}},
  \citenamefont {{Aguena}}, \citenamefont {{Allam}}, \citenamefont {{Annis}},
  \citenamefont {{Avila}}, \citenamefont {{Brooks}}, \citenamefont {{Carrasco
  Kind}}, \citenamefont {{Carretero}}, \citenamefont {{Costanzi}},
  \citenamefont {{da Costa}}, \citenamefont {{De Vicente}}, \citenamefont
  {{Desai}}, \citenamefont {{Diehl}}, \citenamefont {{Doel}}, \citenamefont
  {{Everett}}, \citenamefont {{Evrard}}, \citenamefont {{Flaugher}},
  \citenamefont {{Frieman}}, \citenamefont {{Gerdes}}, \citenamefont {{Gruen}},
  \citenamefont {{Gruendl}}, \citenamefont {{Gschwend}}, \citenamefont
  {{Gutierrez}}, \citenamefont {{Hinton}}, \citenamefont {{Honscheid}},
  \citenamefont {{Huterer}}, \citenamefont {{James}}, \citenamefont {{Krause}},
  \citenamefont {{Kuehn}}, \citenamefont {{Kuropatkin}}, \citenamefont
  {{Lahav}}, \citenamefont {{Maia}}, \citenamefont {{Marshall}}, \citenamefont
  {{Menanteau}}, \citenamefont {{Miquel}}, \citenamefont {{Palmese}},
  \citenamefont {{Paz-Chinch{\'o}n}}, \citenamefont {{Plazas}}, \citenamefont
  {{Romer}}, \citenamefont {{Sanchez}}, \citenamefont {{Scarpine}},
  \citenamefont {{Serrano}}, \citenamefont {{Sevilla-Noarbe}}, \citenamefont
  {{Smith}}, \citenamefont {{Soares-Santos}}, \citenamefont {{Suchyta}},
  \citenamefont {{Swanson}}, \citenamefont {{Tarle}}, \citenamefont {{Tucker}},
  \citenamefont {{Walker}}, \citenamefont {{Wester}},\ and\ \citenamefont {{DES
  Collaboration}}}]{2021PhRvL.126i1101N}%
  \BibitemOpen
  \bibfield  {author} {\bibinfo {author} {\bibfnamefont {E.~O.}\ \bibnamefont
  {{Nadler}}}, \bibinfo {author} {\bibfnamefont {A.}~\bibnamefont
  {{Drlica-Wagner}}}, \bibinfo {author} {\bibfnamefont {K.}~\bibnamefont
  {{Bechtol}}}, \bibinfo {author} {\bibfnamefont {S.}~\bibnamefont {{Mau}}},
  \bibinfo {author} {\bibfnamefont {R.~H.}\ \bibnamefont {{Wechsler}}},
  \bibinfo {author} {\bibfnamefont {V.}~\bibnamefont {{Gluscevic}}}, \bibinfo
  {author} {\bibfnamefont {K.}~\bibnamefont {{Boddy}}}, \bibinfo {author}
  {\bibfnamefont {A.~B.}\ \bibnamefont {{Pace}}}, \bibinfo {author}
  {\bibfnamefont {T.~S.}\ \bibnamefont {{Li}}}, \bibinfo {author}
  {\bibfnamefont {M.}~\bibnamefont {{McNanna}}}, \bibinfo {author}
  {\bibfnamefont {A.~H.}\ \bibnamefont {{Riley}}}, \bibinfo {author}
  {\bibfnamefont {J.}~\bibnamefont {{Garc{\'\i}a-Bellido}}}, \bibinfo {author}
  {\bibfnamefont {Y.~Y.}\ \bibnamefont {{Mao}}}, \bibinfo {author}
  {\bibfnamefont {G.}~\bibnamefont {{Green}}}, \bibinfo {author} {\bibfnamefont
  {D.~L.}\ \bibnamefont {{Burke}}}, \bibinfo {author} {\bibfnamefont
  {A.}~\bibnamefont {{Peter}}}, \bibinfo {author} {\bibfnamefont
  {B.}~\bibnamefont {{Jain}}}, \bibinfo {author} {\bibfnamefont {T.~M.~C.}\
  \bibnamefont {{Abbott}}}, \bibinfo {author} {\bibfnamefont {M.}~\bibnamefont
  {{Aguena}}}, \bibinfo {author} {\bibfnamefont {S.}~\bibnamefont {{Allam}}},
  \bibinfo {author} {\bibfnamefont {J.}~\bibnamefont {{Annis}}}, \bibinfo
  {author} {\bibfnamefont {S.}~\bibnamefont {{Avila}}}, \bibinfo {author}
  {\bibfnamefont {D.}~\bibnamefont {{Brooks}}}, \bibinfo {author}
  {\bibfnamefont {M.}~\bibnamefont {{Carrasco Kind}}}, \bibinfo {author}
  {\bibfnamefont {J.}~\bibnamefont {{Carretero}}}, \bibinfo {author}
  {\bibfnamefont {M.}~\bibnamefont {{Costanzi}}}, \bibinfo {author}
  {\bibfnamefont {L.~N.}\ \bibnamefont {{da Costa}}}, \bibinfo {author}
  {\bibfnamefont {J.}~\bibnamefont {{De Vicente}}}, \bibinfo {author}
  {\bibfnamefont {S.}~\bibnamefont {{Desai}}}, \bibinfo {author} {\bibfnamefont
  {H.~T.}\ \bibnamefont {{Diehl}}}, \bibinfo {author} {\bibfnamefont
  {P.}~\bibnamefont {{Doel}}}, \bibinfo {author} {\bibfnamefont
  {S.}~\bibnamefont {{Everett}}}, \bibinfo {author} {\bibfnamefont {A.~E.}\
  \bibnamefont {{Evrard}}}, \bibinfo {author} {\bibfnamefont {B.}~\bibnamefont
  {{Flaugher}}}, \bibinfo {author} {\bibfnamefont {J.}~\bibnamefont
  {{Frieman}}}, \bibinfo {author} {\bibfnamefont {D.~W.}\ \bibnamefont
  {{Gerdes}}}, \bibinfo {author} {\bibfnamefont {D.}~\bibnamefont {{Gruen}}},
  \bibinfo {author} {\bibfnamefont {R.~A.}\ \bibnamefont {{Gruendl}}}, \bibinfo
  {author} {\bibfnamefont {J.}~\bibnamefont {{Gschwend}}}, \bibinfo {author}
  {\bibfnamefont {G.}~\bibnamefont {{Gutierrez}}}, \bibinfo {author}
  {\bibfnamefont {S.~R.}\ \bibnamefont {{Hinton}}}, \bibinfo {author}
  {\bibfnamefont {K.}~\bibnamefont {{Honscheid}}}, \bibinfo {author}
  {\bibfnamefont {D.}~\bibnamefont {{Huterer}}}, \bibinfo {author}
  {\bibfnamefont {D.~J.}\ \bibnamefont {{James}}}, \bibinfo {author}
  {\bibfnamefont {E.}~\bibnamefont {{Krause}}}, \bibinfo {author}
  {\bibfnamefont {K.}~\bibnamefont {{Kuehn}}}, \bibinfo {author} {\bibfnamefont
  {N.}~\bibnamefont {{Kuropatkin}}}, \bibinfo {author} {\bibfnamefont
  {O.}~\bibnamefont {{Lahav}}}, \bibinfo {author} {\bibfnamefont {M.~A.~G.}\
  \bibnamefont {{Maia}}}, \bibinfo {author} {\bibfnamefont {J.~L.}\
  \bibnamefont {{Marshall}}}, \bibinfo {author} {\bibfnamefont
  {F.}~\bibnamefont {{Menanteau}}}, \bibinfo {author} {\bibfnamefont
  {R.}~\bibnamefont {{Miquel}}}, \bibinfo {author} {\bibfnamefont
  {A.}~\bibnamefont {{Palmese}}}, \bibinfo {author} {\bibfnamefont
  {F.}~\bibnamefont {{Paz-Chinch{\'o}n}}}, \bibinfo {author} {\bibfnamefont
  {A.~A.}\ \bibnamefont {{Plazas}}}, \bibinfo {author} {\bibfnamefont {A.~K.}\
  \bibnamefont {{Romer}}}, \bibinfo {author} {\bibfnamefont {E.}~\bibnamefont
  {{Sanchez}}}, \bibinfo {author} {\bibfnamefont {V.}~\bibnamefont
  {{Scarpine}}}, \bibinfo {author} {\bibfnamefont {S.}~\bibnamefont
  {{Serrano}}}, \bibinfo {author} {\bibfnamefont {I.}~\bibnamefont
  {{Sevilla-Noarbe}}}, \bibinfo {author} {\bibfnamefont {M.}~\bibnamefont
  {{Smith}}}, \bibinfo {author} {\bibfnamefont {M.}~\bibnamefont
  {{Soares-Santos}}}, \bibinfo {author} {\bibfnamefont {E.}~\bibnamefont
  {{Suchyta}}}, \bibinfo {author} {\bibfnamefont {M.~E.~C.}\ \bibnamefont
  {{Swanson}}}, \bibinfo {author} {\bibfnamefont {G.}~\bibnamefont {{Tarle}}},
  \bibinfo {author} {\bibfnamefont {D.~L.}\ \bibnamefont {{Tucker}}}, \bibinfo
  {author} {\bibfnamefont {A.~R.}\ \bibnamefont {{Walker}}}, \bibinfo {author}
  {\bibfnamefont {W.}~\bibnamefont {{Wester}}},\ and\ \bibinfo {author}
  {\bibnamefont {{DES Collaboration}}},\ }\bibfield  {title} {\bibinfo {title}
  {{Constraints on Dark Matter Properties from Observations of Milky Way
  Satellite Galaxies}},\ }\href
  {https://doi.org/10.1103/PhysRevLett.126.091101} {\bibfield  {journal}
  {\bibinfo  {journal} {\prl}\ }\textbf {\bibinfo {volume} {126}},\ \bibinfo
  {eid} {091101} (\bibinfo {year} {2021}{\natexlab{a}})},\ \Eprint
  {https://arxiv.org/abs/2008.00022} {arXiv:2008.00022 [astro-ph.CO]}
  \BibitemShut {NoStop}%
\bibitem [{\citenamefont {{Maamari}}\ \emph {et~al.}(2021)\citenamefont
  {{Maamari}}, \citenamefont {{Gluscevic}}, \citenamefont {{Boddy}},
  \citenamefont {{Nadler}},\ and\ \citenamefont
  {{Wechsler}}}]{2021ApJ...907L..46M}%
  \BibitemOpen
  \bibfield  {author} {\bibinfo {author} {\bibfnamefont {K.}~\bibnamefont
  {{Maamari}}}, \bibinfo {author} {\bibfnamefont {V.}~\bibnamefont
  {{Gluscevic}}}, \bibinfo {author} {\bibfnamefont {K.~K.}\ \bibnamefont
  {{Boddy}}}, \bibinfo {author} {\bibfnamefont {E.~O.}\ \bibnamefont
  {{Nadler}}},\ and\ \bibinfo {author} {\bibfnamefont {R.~H.}\ \bibnamefont
  {{Wechsler}}},\ }\bibfield  {title} {\bibinfo {title} {{Bounds on
  Velocity-dependent Dark Matter-Proton Scattering from Milky Way Satellite
  Abundance}},\ }\href {https://doi.org/10.3847/2041-8213/abd807} {\bibfield
  {journal} {\bibinfo  {journal} {\apjl}\ }\textbf {\bibinfo {volume} {907}},\
  \bibinfo {eid} {L46} (\bibinfo {year} {2021})},\ \Eprint
  {https://arxiv.org/abs/2010.02936} {arXiv:2010.02936 [astro-ph.CO]}
  \BibitemShut {NoStop}%
\bibitem [{\citenamefont {Dvorkin}\ \emph {et~al.}(2022)\citenamefont {Dvorkin}
  \emph {et~al.}}]{Dvorkin:2022bsc}%
  \BibitemOpen
  \bibfield  {author} {\bibinfo {author} {\bibfnamefont {C.}~\bibnamefont
  {Dvorkin}} \emph {et~al.},\ }\bibfield  {title} {\bibinfo {title} {{Dark
  Matter Physics from the CMB-S4 Experiment}},\ }in\ \href@noop {} {\emph
  {\bibinfo {booktitle} {{2022 Snowmass Summer Study}}}}\ (\bibinfo {year}
  {2022})\ \Eprint {https://arxiv.org/abs/2203.07064} {arXiv:2203.07064
  [hep-ph]} \BibitemShut {NoStop}%
\bibitem [{\citenamefont {{Sigurdson}}\ \emph {et~al.}(2004)\citenamefont
  {{Sigurdson}}, \citenamefont {{Doran}}, \citenamefont {{Kurylov}},
  \citenamefont {{Caldwell}},\ and\ \citenamefont
  {{Kamionkowski}}}]{2004PhRvD..70h3501S}%
  \BibitemOpen
  \bibfield  {author} {\bibinfo {author} {\bibfnamefont {K.}~\bibnamefont
  {{Sigurdson}}}, \bibinfo {author} {\bibfnamefont {M.}~\bibnamefont
  {{Doran}}}, \bibinfo {author} {\bibfnamefont {A.}~\bibnamefont {{Kurylov}}},
  \bibinfo {author} {\bibfnamefont {R.~R.}\ \bibnamefont {{Caldwell}}},\ and\
  \bibinfo {author} {\bibfnamefont {M.}~\bibnamefont {{Kamionkowski}}},\
  }\bibfield  {title} {\bibinfo {title} {{Dark-matter electric and magnetic
  dipole moments}},\ }\href {https://doi.org/10.1103/PhysRevD.70.083501}
  {\bibfield  {journal} {\bibinfo  {journal} {\prd}\ }\textbf {\bibinfo
  {volume} {70}},\ \bibinfo {eid} {083501} (\bibinfo {year} {2004})},\ \Eprint
  {https://arxiv.org/abs/astro-ph/0406355} {arXiv:astro-ph/0406355 [astro-ph]}
  \BibitemShut {NoStop}%
\bibitem [{\citenamefont {{Del Nobile}}\ \emph {et~al.}(2012)\citenamefont
  {{Del Nobile}}, \citenamefont {{Kouvaris}}, \citenamefont {{Panci}},
  \citenamefont {{Sannino}},\ and\ \citenamefont
  {{Virkaj{\"a}rvi}}}]{2012JCAP...08..010D}%
  \BibitemOpen
  \bibfield  {author} {\bibinfo {author} {\bibfnamefont {E.}~\bibnamefont {{Del
  Nobile}}}, \bibinfo {author} {\bibfnamefont {C.}~\bibnamefont {{Kouvaris}}},
  \bibinfo {author} {\bibfnamefont {P.}~\bibnamefont {{Panci}}}, \bibinfo
  {author} {\bibfnamefont {F.}~\bibnamefont {{Sannino}}},\ and\ \bibinfo
  {author} {\bibfnamefont {J.}~\bibnamefont {{Virkaj{\"a}rvi}}},\ }\bibfield
  {title} {\bibinfo {title} {{Light magnetic dark matter in direct detection
  searches}},\ }\href {https://doi.org/10.1088/1475-7516/2012/08/010}
  {\bibfield  {journal} {\bibinfo  {journal} {\jcap}\ }\textbf {\bibinfo
  {volume} {2012}},\ \bibinfo {eid} {010} (\bibinfo {year} {2012})},\ \Eprint
  {https://arxiv.org/abs/1203.6652} {arXiv:1203.6652 [hep-ph]} \BibitemShut
  {NoStop}%
\bibitem [{\citenamefont {{Gnedin}}\ and\ \citenamefont
  {{Hui}}(1998)}]{1998MNRAS.296...44G}%
  \BibitemOpen
  \bibfield  {author} {\bibinfo {author} {\bibfnamefont {N.~Y.}\ \bibnamefont
  {{Gnedin}}}\ and\ \bibinfo {author} {\bibfnamefont {L.}~\bibnamefont
  {{Hui}}},\ }\bibfield  {title} {\bibinfo {title} {{Probing the Universe with
  the Lyalpha forest - I. Hydrodynamics of the low-density intergalactic
  medium}},\ }\href {https://doi.org/10.1046/j.1365-8711.1998.01249.x}
  {\bibfield  {journal} {\bibinfo  {journal} {\mnras}\ }\textbf {\bibinfo
  {volume} {296}},\ \bibinfo {pages} {44} (\bibinfo {year} {1998})},\ \Eprint
  {https://arxiv.org/abs/astro-ph/9706219} {arXiv:astro-ph/9706219 [astro-ph]}
  \BibitemShut {NoStop}%
\bibitem [{\citenamefont {{Croft}}\ \emph {et~al.}(1998)\citenamefont
  {{Croft}}, \citenamefont {{Weinberg}}, \citenamefont {{Katz}},\ and\
  \citenamefont {{Hernquist}}}]{1998ApJ...495...44C}%
  \BibitemOpen
  \bibfield  {author} {\bibinfo {author} {\bibfnamefont {R.~A.~C.}\
  \bibnamefont {{Croft}}}, \bibinfo {author} {\bibfnamefont {D.~H.}\
  \bibnamefont {{Weinberg}}}, \bibinfo {author} {\bibfnamefont
  {N.}~\bibnamefont {{Katz}}},\ and\ \bibinfo {author} {\bibfnamefont
  {L.}~\bibnamefont {{Hernquist}}},\ }\bibfield  {title} {\bibinfo {title}
  {{Recovery of the Power Spectrum of Mass Fluctuations from Observations of
  the Ly{\ensuremath{\alpha}} Forest}},\ }\href
  {https://doi.org/10.1086/305289} {\bibfield  {journal} {\bibinfo  {journal}
  {\apj}\ }\textbf {\bibinfo {volume} {495}},\ \bibinfo {pages} {44} (\bibinfo
  {year} {1998})},\ \Eprint {https://arxiv.org/abs/astro-ph/9708018}
  {arXiv:astro-ph/9708018 [astro-ph]} \BibitemShut {NoStop}%
\bibitem [{\citenamefont {{Seljak}}\ \emph {et~al.}(2005)\citenamefont
  {{Seljak}}, \citenamefont {{Makarov}}, \citenamefont {{McDonald}},
  \citenamefont {{Anderson}}, \citenamefont {{Bahcall}}, \citenamefont
  {{Brinkmann}}, \citenamefont {{Burles}}, \citenamefont {{Cen}}, \citenamefont
  {{Doi}}, \citenamefont {{Gunn}},\ and\ \citenamefont
  {{others}}}]{2005PhRvD..71j3515S}%
  \BibitemOpen
  \bibfield  {author} {\bibinfo {author} {\bibfnamefont {U.}~\bibnamefont
  {{Seljak}}}, \bibinfo {author} {\bibfnamefont {A.}~\bibnamefont {{Makarov}}},
  \bibinfo {author} {\bibfnamefont {P.}~\bibnamefont {{McDonald}}}, \bibinfo
  {author} {\bibfnamefont {S.~F.}\ \bibnamefont {{Anderson}}}, \bibinfo
  {author} {\bibfnamefont {N.~A.}\ \bibnamefont {{Bahcall}}}, \bibinfo {author}
  {\bibfnamefont {J.}~\bibnamefont {{Brinkmann}}}, \bibinfo {author}
  {\bibfnamefont {S.}~\bibnamefont {{Burles}}}, \bibinfo {author}
  {\bibfnamefont {R.}~\bibnamefont {{Cen}}}, \bibinfo {author} {\bibfnamefont
  {M.}~\bibnamefont {{Doi}}}, \bibinfo {author} {\bibfnamefont {J.~E.}\
  \bibnamefont {{Gunn}}},\ and\ \bibinfo {author} {\bibnamefont {{others}}},\
  }\bibfield  {title} {\bibinfo {title} {{Cosmological parameter analysis
  including SDSS Ly{$\alpha$} forest and galaxy bias: Constraints on the
  primordial spectrum of fluctuations, neutrino mass, and dark energy}},\
  }\href {https://doi.org/10.1103/PhysRevD.71.103515} {\bibfield  {journal}
  {\bibinfo  {journal} {\prd}\ }\textbf {\bibinfo {volume} {71}},\ \bibinfo
  {eid} {103515} (\bibinfo {year} {2005})},\ \Eprint
  {https://arxiv.org/abs/astro-ph/0407372} {astro-ph/0407372} \BibitemShut
  {NoStop}%
\bibitem [{\citenamefont {{McDonald}}\ \emph {et~al.}(2005)\citenamefont
  {{McDonald}}, \citenamefont {{Seljak}}, \citenamefont {{Cen}}, \citenamefont
  {{Shih}}, \citenamefont {{Weinberg}}, \citenamefont {{Burles}}, \citenamefont
  {{Schneider}}, \citenamefont {{Schlegel}}, \citenamefont {{Bahcall}},
  \citenamefont {{Briggs}}, \citenamefont {{Brinkmann}}, \citenamefont
  {{Fukugita}}, \citenamefont {{Ivezi{\'c}}}, \citenamefont {{Kent}},\ and\
  \citenamefont {{Vanden Berk}}}]{2005ApJ...635..761M}%
  \BibitemOpen
  \bibfield  {author} {\bibinfo {author} {\bibfnamefont {P.}~\bibnamefont
  {{McDonald}}}, \bibinfo {author} {\bibfnamefont {U.}~\bibnamefont
  {{Seljak}}}, \bibinfo {author} {\bibfnamefont {R.}~\bibnamefont {{Cen}}},
  \bibinfo {author} {\bibfnamefont {D.}~\bibnamefont {{Shih}}}, \bibinfo
  {author} {\bibfnamefont {D.~H.}\ \bibnamefont {{Weinberg}}}, \bibinfo
  {author} {\bibfnamefont {S.}~\bibnamefont {{Burles}}}, \bibinfo {author}
  {\bibfnamefont {D.~P.}\ \bibnamefont {{Schneider}}}, \bibinfo {author}
  {\bibfnamefont {D.~J.}\ \bibnamefont {{Schlegel}}}, \bibinfo {author}
  {\bibfnamefont {N.~A.}\ \bibnamefont {{Bahcall}}}, \bibinfo {author}
  {\bibfnamefont {J.~W.}\ \bibnamefont {{Briggs}}}, \bibinfo {author}
  {\bibfnamefont {J.}~\bibnamefont {{Brinkmann}}}, \bibinfo {author}
  {\bibfnamefont {M.}~\bibnamefont {{Fukugita}}}, \bibinfo {author}
  {\bibfnamefont {{\v{Z}}.}~\bibnamefont {{Ivezi{\'c}}}}, \bibinfo {author}
  {\bibfnamefont {S.}~\bibnamefont {{Kent}}},\ and\ \bibinfo {author}
  {\bibfnamefont {D.~E.}\ \bibnamefont {{Vanden Berk}}},\ }\bibfield  {title}
  {\bibinfo {title} {{The Linear Theory Power Spectrum from the
  Ly{\ensuremath{\alpha}} Forest in the Sloan Digital Sky Survey}},\ }\href
  {https://doi.org/10.1086/497563} {\bibfield  {journal} {\bibinfo  {journal}
  {\apj}\ }\textbf {\bibinfo {volume} {635}},\ \bibinfo {pages} {761} (\bibinfo
  {year} {2005})},\ \Eprint {https://arxiv.org/abs/astro-ph/0407377}
  {arXiv:astro-ph/0407377 [astro-ph]} \BibitemShut {NoStop}%
\bibitem [{\citenamefont {{McDonald}}\ \emph {et~al.}(2006)\citenamefont
  {{McDonald}}, \citenamefont {{Seljak}}, \citenamefont {{Burles}},
  \citenamefont {{Schlegel}}, \citenamefont {{Weinberg}}, \citenamefont
  {{Cen}}, \citenamefont {{Shih}}, \citenamefont {{Schaye}}, \citenamefont
  {{Schneider}}, \citenamefont {{Bahcall}}, \citenamefont {{Briggs}},
  \citenamefont {{Brinkmann}}, \citenamefont {{Brunner}}, \citenamefont
  {{Fukugita}}, \citenamefont {{Gunn}}, \citenamefont {{Ivezi{\'c}}},
  \citenamefont {{Kent}}, \citenamefont {{Lupton}},\ and\ \citenamefont
  {{Vanden Berk}}}]{2006ApJS..163...80M}%
  \BibitemOpen
  \bibfield  {author} {\bibinfo {author} {\bibfnamefont {P.}~\bibnamefont
  {{McDonald}}}, \bibinfo {author} {\bibfnamefont {U.}~\bibnamefont
  {{Seljak}}}, \bibinfo {author} {\bibfnamefont {S.}~\bibnamefont {{Burles}}},
  \bibinfo {author} {\bibfnamefont {D.~J.}\ \bibnamefont {{Schlegel}}},
  \bibinfo {author} {\bibfnamefont {D.~H.}\ \bibnamefont {{Weinberg}}},
  \bibinfo {author} {\bibfnamefont {R.}~\bibnamefont {{Cen}}}, \bibinfo
  {author} {\bibfnamefont {D.}~\bibnamefont {{Shih}}}, \bibinfo {author}
  {\bibfnamefont {J.}~\bibnamefont {{Schaye}}}, \bibinfo {author}
  {\bibfnamefont {D.~P.}\ \bibnamefont {{Schneider}}}, \bibinfo {author}
  {\bibfnamefont {N.~A.}\ \bibnamefont {{Bahcall}}}, \bibinfo {author}
  {\bibfnamefont {J.~W.}\ \bibnamefont {{Briggs}}}, \bibinfo {author}
  {\bibfnamefont {J.}~\bibnamefont {{Brinkmann}}}, \bibinfo {author}
  {\bibfnamefont {R.~J.}\ \bibnamefont {{Brunner}}}, \bibinfo {author}
  {\bibfnamefont {M.}~\bibnamefont {{Fukugita}}}, \bibinfo {author}
  {\bibfnamefont {J.~E.}\ \bibnamefont {{Gunn}}}, \bibinfo {author}
  {\bibfnamefont {{\v{Z}}.}~\bibnamefont {{Ivezi{\'c}}}}, \bibinfo {author}
  {\bibfnamefont {S.}~\bibnamefont {{Kent}}}, \bibinfo {author} {\bibfnamefont
  {R.~H.}\ \bibnamefont {{Lupton}}},\ and\ \bibinfo {author} {\bibfnamefont
  {D.~E.}\ \bibnamefont {{Vanden Berk}}},\ }\bibfield  {title} {\bibinfo
  {title} {{The Ly{\ensuremath{\alpha}} Forest Power Spectrum from the Sloan
  Digital Sky Survey}},\ }\href {https://doi.org/10.1086/444361} {\bibfield
  {journal} {\bibinfo  {journal} {\apjs}\ }\textbf {\bibinfo {volume} {163}},\
  \bibinfo {pages} {80} (\bibinfo {year} {2006})},\ \Eprint
  {https://arxiv.org/abs/astro-ph/0405013} {arXiv:astro-ph/0405013 [astro-ph]}
  \BibitemShut {NoStop}%
\bibitem [{\citenamefont {{Boera}}\ \emph {et~al.}(2019)\citenamefont
  {{Boera}}, \citenamefont {{Becker}}, \citenamefont {{Bolton}},\ and\
  \citenamefont {{Nasir}}}]{2019ApJ...872..101B}%
  \BibitemOpen
  \bibfield  {author} {\bibinfo {author} {\bibfnamefont {E.}~\bibnamefont
  {{Boera}}}, \bibinfo {author} {\bibfnamefont {G.~D.}\ \bibnamefont
  {{Becker}}}, \bibinfo {author} {\bibfnamefont {J.~S.}\ \bibnamefont
  {{Bolton}}},\ and\ \bibinfo {author} {\bibfnamefont {F.}~\bibnamefont
  {{Nasir}}},\ }\bibfield  {title} {\bibinfo {title} {{Revealing Reionization
  with the Thermal History of the Intergalactic Medium: New Constraints from
  the Ly{\ensuremath{\alpha}} Flux Power Spectrum}},\ }\href
  {https://doi.org/10.3847/1538-4357/aafee4} {\bibfield  {journal} {\bibinfo
  {journal} {\apj}\ }\textbf {\bibinfo {volume} {872}},\ \bibinfo {eid} {101}
  (\bibinfo {year} {2019})},\ \Eprint {https://arxiv.org/abs/1809.06980}
  {arXiv:1809.06980 [astro-ph.CO]} \BibitemShut {NoStop}%
\bibitem [{\citenamefont {Feng}\ \emph {et~al.}(2018)\citenamefont {Feng},
  \citenamefont {Bird}, \citenamefont {Anderson}, \citenamefont {Font-Ribera},\
  and\ \citenamefont {Pedersen}}]{yu_feng_2018_1451799}%
  \BibitemOpen
  \bibfield  {author} {\bibinfo {author} {\bibfnamefont {Y.}~\bibnamefont
  {Feng}}, \bibinfo {author} {\bibfnamefont {S.}~\bibnamefont {Bird}}, \bibinfo
  {author} {\bibfnamefont {L.}~\bibnamefont {Anderson}}, \bibinfo {author}
  {\bibfnamefont {A.}~\bibnamefont {Font-Ribera}},\ and\ \bibinfo {author}
  {\bibfnamefont {C.}~\bibnamefont {Pedersen}},\ }\href
  {https://doi.org/10.5281/zenodo.1451799} {\bibinfo {title}
  {Mp-gadget/mp-gadget: A tag for getting a doi}} (\bibinfo {year}
  {2018})\BibitemShut {NoStop}%
\bibitem [{\citenamefont {{Springel}}\ \emph {et~al.}(2001)\citenamefont
  {{Springel}}, \citenamefont {{Yoshida}},\ and\ \citenamefont
  {{White}}}]{2001NewA....6...79S}%
  \BibitemOpen
  \bibfield  {author} {\bibinfo {author} {\bibfnamefont {V.}~\bibnamefont
  {{Springel}}}, \bibinfo {author} {\bibfnamefont {N.}~\bibnamefont
  {{Yoshida}}},\ and\ \bibinfo {author} {\bibfnamefont {S.~D.~M.}\ \bibnamefont
  {{White}}},\ }\bibfield  {title} {\bibinfo {title} {{GADGET: a code for
  collisionless and gasdynamical cosmological simulations}},\ }\href
  {https://doi.org/10.1016/S1384-1076(01)00042-2} {\bibfield  {journal}
  {\bibinfo  {journal} {\na}\ }\textbf {\bibinfo {volume} {6}},\ \bibinfo
  {pages} {79} (\bibinfo {year} {2001})},\ \Eprint
  {https://arxiv.org/abs/astro-ph/0003162} {arXiv:astro-ph/0003162 [astro-ph]}
  \BibitemShut {NoStop}%
\bibitem [{\citenamefont {{Springel}}(2005)}]{2005MNRAS.364.1105S}%
  \BibitemOpen
  \bibfield  {author} {\bibinfo {author} {\bibfnamefont {V.}~\bibnamefont
  {{Springel}}},\ }\bibfield  {title} {\bibinfo {title} {{The cosmological
  simulation code GADGET-2}},\ }\href
  {https://doi.org/10.1111/j.1365-2966.2005.09655.x} {\bibfield  {journal}
  {\bibinfo  {journal} {\mnras}\ }\textbf {\bibinfo {volume} {364}},\ \bibinfo
  {pages} {1105} (\bibinfo {year} {2005})},\ \Eprint
  {https://arxiv.org/abs/astro-ph/0505010} {arXiv:astro-ph/0505010 [astro-ph]}
  \BibitemShut {NoStop}%
\bibitem [{\citenamefont {Anderson}\ \emph {et~al.}(2019)\citenamefont
  {Anderson}, \citenamefont {Pontzen}, \citenamefont {Font-Ribera},
  \citenamefont {Villaescusa-Navarro}, \citenamefont {Rogers},\ and\
  \citenamefont {Genel}}]{Anderson:2018zkm}%
  \BibitemOpen
  \bibfield  {author} {\bibinfo {author} {\bibfnamefont {L.}~\bibnamefont
  {Anderson}}, \bibinfo {author} {\bibfnamefont {A.}~\bibnamefont {Pontzen}},
  \bibinfo {author} {\bibfnamefont {A.}~\bibnamefont {Font-Ribera}}, \bibinfo
  {author} {\bibfnamefont {F.}~\bibnamefont {Villaescusa-Navarro}}, \bibinfo
  {author} {\bibfnamefont {K.~K.}\ \bibnamefont {Rogers}},\ and\ \bibinfo
  {author} {\bibfnamefont {S.}~\bibnamefont {Genel}},\ }\bibfield  {title}
  {\bibinfo {title} {{Cosmological Hydrodynamic Simulations with Suppressed
  Variance in the Ly$\alpha$ Forest Power Spectrum}},\ }\href
  {https://doi.org/10.3847/1538-4357/aaf576} {\bibfield  {journal} {\bibinfo
  {journal} {Astrophys. J.}\ }\textbf {\bibinfo {volume} {871}},\ \bibinfo
  {pages} {144} (\bibinfo {year} {2019})},\ \Eprint
  {https://arxiv.org/abs/1811.00043} {arXiv:1811.00043 [astro-ph.CO]}
  \BibitemShut {NoStop}%
\bibitem [{\citenamefont {{Rogers}}\ and\ \citenamefont
  {{Peiris}}(2021{\natexlab{a}})}]{2020RogersPRD}%
  \BibitemOpen
  \bibfield  {author} {\bibinfo {author} {\bibfnamefont {K.~K.}\ \bibnamefont
  {{Rogers}}}\ and\ \bibinfo {author} {\bibfnamefont {H.~V.}\ \bibnamefont
  {{Peiris}}},\ }\bibfield  {title} {\bibinfo {title} {{General framework for
  cosmological dark matter bounds using N -body simulations}},\ }\href
  {https://doi.org/10.1103/PhysRevD.103.043526} {\bibfield  {journal} {\bibinfo
   {journal} {\prd}\ }\textbf {\bibinfo {volume} {103}},\ \bibinfo {eid}
  {043526} (\bibinfo {year} {2021}{\natexlab{a}})},\ \Eprint
  {https://arxiv.org/abs/2007.13751} {arXiv:2007.13751 [astro-ph.CO]}
  \BibitemShut {NoStop}%
\bibitem [{\citenamefont {{Rogers}}\ \emph {et~al.}(2019)\citenamefont
  {{Rogers}}, \citenamefont {{Peiris}}, \citenamefont {{Pontzen}},
  \citenamefont {{Bird}}, \citenamefont {{Verde}},\ and\ \citenamefont
  {{Font-Ribera}}}]{2019JCAP...02..031R}%
  \BibitemOpen
  \bibfield  {author} {\bibinfo {author} {\bibfnamefont {K.~K.}\ \bibnamefont
  {{Rogers}}}, \bibinfo {author} {\bibfnamefont {H.~V.}\ \bibnamefont
  {{Peiris}}}, \bibinfo {author} {\bibfnamefont {A.}~\bibnamefont {{Pontzen}}},
  \bibinfo {author} {\bibfnamefont {S.}~\bibnamefont {{Bird}}}, \bibinfo
  {author} {\bibfnamefont {L.}~\bibnamefont {{Verde}}},\ and\ \bibinfo {author}
  {\bibfnamefont {A.}~\bibnamefont {{Font-Ribera}}},\ }\bibfield  {title}
  {\bibinfo {title} {{Bayesian emulator optimisation for cosmology: application
  to the Lyman-alpha forest}},\ }\href
  {https://doi.org/10.1088/1475-7516/2019/02/031} {\bibfield  {journal}
  {\bibinfo  {journal} {\jcap}\ }\textbf {\bibinfo {volume} {2019}},\ \bibinfo
  {eid} {031} (\bibinfo {year} {2019})},\ \Eprint
  {https://arxiv.org/abs/1812.04631} {arXiv:1812.04631 [astro-ph.CO]}
  \BibitemShut {NoStop}%
\bibitem [{\citenamefont {{Bird}}\ \emph {et~al.}(2019)\citenamefont {{Bird}},
  \citenamefont {{Rogers}}, \citenamefont {{Peiris}}, \citenamefont {{Verde}},
  \citenamefont {{Font-Ribera}},\ and\ \citenamefont
  {{Pontzen}}}]{2019JCAP...02..050B}%
  \BibitemOpen
  \bibfield  {author} {\bibinfo {author} {\bibfnamefont {S.}~\bibnamefont
  {{Bird}}}, \bibinfo {author} {\bibfnamefont {K.~K.}\ \bibnamefont
  {{Rogers}}}, \bibinfo {author} {\bibfnamefont {H.~V.}\ \bibnamefont
  {{Peiris}}}, \bibinfo {author} {\bibfnamefont {L.}~\bibnamefont {{Verde}}},
  \bibinfo {author} {\bibfnamefont {A.}~\bibnamefont {{Font-Ribera}}},\ and\
  \bibinfo {author} {\bibfnamefont {A.}~\bibnamefont {{Pontzen}}},\ }\bibfield
  {title} {\bibinfo {title} {{An emulator for the Lyman-{\ensuremath{\alpha}}
  forest}},\ }\href {https://doi.org/10.1088/1475-7516/2019/02/050} {\bibfield
  {journal} {\bibinfo  {journal} {\jcap}\ }\textbf {\bibinfo {volume} {2019}},\
  \bibinfo {eid} {050} (\bibinfo {year} {2019})},\ \Eprint
  {https://arxiv.org/abs/1812.04654} {arXiv:1812.04654 [astro-ph.CO]}
  \BibitemShut {NoStop}%
\bibitem [{\citenamefont {{Pedersen}}\ \emph {et~al.}(2021)\citenamefont
  {{Pedersen}}, \citenamefont {{Font-Ribera}}, \citenamefont {{Rogers}},
  \citenamefont {{McDonald}}, \citenamefont {{Peiris}}, \citenamefont
  {{Pontzen}},\ and\ \citenamefont {{Slosar}}}]{2021JCAP...05..033P}%
  \BibitemOpen
  \bibfield  {author} {\bibinfo {author} {\bibfnamefont {C.}~\bibnamefont
  {{Pedersen}}}, \bibinfo {author} {\bibfnamefont {A.}~\bibnamefont
  {{Font-Ribera}}}, \bibinfo {author} {\bibfnamefont {K.~K.}\ \bibnamefont
  {{Rogers}}}, \bibinfo {author} {\bibfnamefont {P.}~\bibnamefont
  {{McDonald}}}, \bibinfo {author} {\bibfnamefont {H.~V.}\ \bibnamefont
  {{Peiris}}}, \bibinfo {author} {\bibfnamefont {A.}~\bibnamefont
  {{Pontzen}}},\ and\ \bibinfo {author} {\bibfnamefont {A.}~\bibnamefont
  {{Slosar}}},\ }\bibfield  {title} {\bibinfo {title} {{An emulator for the
  Lyman-{\ensuremath{\alpha}} forest in beyond-{\ensuremath{\Lambda}}CDM
  cosmologies}},\ }\href {https://doi.org/10.1088/1475-7516/2021/05/033}
  {\bibfield  {journal} {\bibinfo  {journal} {\jcap}\ }\textbf {\bibinfo
  {volume} {2021}},\ \bibinfo {eid} {033} (\bibinfo {year} {2021})},\ \Eprint
  {https://arxiv.org/abs/2011.15127} {arXiv:2011.15127 [astro-ph.CO]}
  \BibitemShut {NoStop}%
\bibitem [{\citenamefont {{Murgia}}\ \emph {et~al.}(2017)\citenamefont
  {{Murgia}}, \citenamefont {{Merle}}, \citenamefont {{Viel}}, \citenamefont
  {{Totzauer}},\ and\ \citenamefont {{Schneider}}}]{2017JCAP...11..046M}%
  \BibitemOpen
  \bibfield  {author} {\bibinfo {author} {\bibfnamefont {R.}~\bibnamefont
  {{Murgia}}}, \bibinfo {author} {\bibfnamefont {A.}~\bibnamefont {{Merle}}},
  \bibinfo {author} {\bibfnamefont {M.}~\bibnamefont {{Viel}}}, \bibinfo
  {author} {\bibfnamefont {M.}~\bibnamefont {{Totzauer}}},\ and\ \bibinfo
  {author} {\bibfnamefont {A.}~\bibnamefont {{Schneider}}},\ }\bibfield
  {title} {\bibinfo {title} {{``Non-cold'' dark matter at small scales: a
  general approach}},\ }\href {https://doi.org/10.1088/1475-7516/2017/11/046}
  {\bibfield  {journal} {\bibinfo  {journal} {\jcap}\ }\textbf {\bibinfo
  {volume} {2017}},\ \bibinfo {eid} {046} (\bibinfo {year} {2017})},\ \Eprint
  {https://arxiv.org/abs/1704.07838} {arXiv:1704.07838 [astro-ph.CO]}
  \BibitemShut {NoStop}%
\bibitem [{\citenamefont {{Rogers}}\ and\ \citenamefont
  {{Peiris}}(2021{\natexlab{b}})}]{2020RogersPRL}%
  \BibitemOpen
  \bibfield  {author} {\bibinfo {author} {\bibfnamefont {K.~K.}\ \bibnamefont
  {{Rogers}}}\ and\ \bibinfo {author} {\bibfnamefont {H.~V.}\ \bibnamefont
  {{Peiris}}},\ }\bibfield  {title} {\bibinfo {title} {{Strong Bound on
  Canonical Ultralight Axion Dark Matter from the Lyman-Alpha Forest}},\ }\href
  {https://doi.org/10.1103/PhysRevLett.126.071302} {\bibfield  {journal}
  {\bibinfo  {journal} {\prl}\ }\textbf {\bibinfo {volume} {126}},\ \bibinfo
  {eid} {071302} (\bibinfo {year} {2021}{\natexlab{b}})},\ \Eprint
  {https://arxiv.org/abs/2007.12705} {arXiv:2007.12705 [astro-ph.CO]}
  \BibitemShut {NoStop}%
\bibitem [{\citenamefont {{Ir{\v{s}}i{\v{c}}}}\ \emph
  {et~al.}(2017)\citenamefont {{Ir{\v{s}}i{\v{c}}}}, \citenamefont {{Viel}},
  \citenamefont {{Haehnelt}}, \citenamefont {{Bolton}},\ and\ \citenamefont
  {{Becker}}}]{2017PhRvL.119c1302I}%
  \BibitemOpen
  \bibfield  {author} {\bibinfo {author} {\bibfnamefont {V.}~\bibnamefont
  {{Ir{\v{s}}i{\v{c}}}}}, \bibinfo {author} {\bibfnamefont {M.}~\bibnamefont
  {{Viel}}}, \bibinfo {author} {\bibfnamefont {M.~G.}\ \bibnamefont
  {{Haehnelt}}}, \bibinfo {author} {\bibfnamefont {J.~S.}\ \bibnamefont
  {{Bolton}}},\ and\ \bibinfo {author} {\bibfnamefont {G.~D.}\ \bibnamefont
  {{Becker}}},\ }\bibfield  {title} {\bibinfo {title} {{First Constraints on
  Fuzzy Dark Matter from Lyman-{\ensuremath{\alpha}} Forest Data and
  Hydrodynamical Simulations}},\ }\href
  {https://doi.org/10.1103/PhysRevLett.119.031302} {\bibfield  {journal}
  {\bibinfo  {journal} {\prl}\ }\textbf {\bibinfo {volume} {119}},\ \bibinfo
  {eid} {031302} (\bibinfo {year} {2017})},\ \Eprint
  {https://arxiv.org/abs/1703.04683} {arXiv:1703.04683 [astro-ph.CO]}
  \BibitemShut {NoStop}%
\bibitem [{\citenamefont {Rasmussen}(2003)}]{rasmussen2003gaussian}%
  \BibitemOpen
  \bibfield  {author} {\bibinfo {author} {\bibfnamefont {C.~E.}\ \bibnamefont
  {Rasmussen}},\ }\bibfield  {title} {\bibinfo {title} {Gaussian processes in
  machine learning},\ }in\ \href@noop {} {\emph {\bibinfo {booktitle} {Summer
  School on Machine Learning}}}\ (\bibinfo {organization} {Springer},\ \bibinfo
  {year} {2003})\ pp.\ \bibinfo {pages} {63--71}\BibitemShut {NoStop}%
\bibitem [{\citenamefont {{Hui}}\ and\ \citenamefont
  {{Gnedin}}(1997)}]{1997MNRAS.292...27H}%
  \BibitemOpen
  \bibfield  {author} {\bibinfo {author} {\bibfnamefont {L.}~\bibnamefont
  {{Hui}}}\ and\ \bibinfo {author} {\bibfnamefont {N.~Y.}\ \bibnamefont
  {{Gnedin}}},\ }\bibfield  {title} {\bibinfo {title} {{Equation of state of
  the photoionized intergalactic medium}},\ }\href
  {https://doi.org/10.1093/mnras/292.1.27} {\bibfield  {journal} {\bibinfo
  {journal} {\mnras}\ }\textbf {\bibinfo {volume} {292}},\ \bibinfo {pages}
  {27} (\bibinfo {year} {1997})},\ \Eprint
  {https://arxiv.org/abs/astro-ph/9612232} {arXiv:astro-ph/9612232 [astro-ph]}
  \BibitemShut {NoStop}%
\bibitem [{\citenamefont {{Nasir}}\ \emph {et~al.}(2016)\citenamefont
  {{Nasir}}, \citenamefont {{Bolton}},\ and\ \citenamefont
  {{Becker}}}]{2016MNRAS.463.2335N}%
  \BibitemOpen
  \bibfield  {author} {\bibinfo {author} {\bibfnamefont {F.}~\bibnamefont
  {{Nasir}}}, \bibinfo {author} {\bibfnamefont {J.~S.}\ \bibnamefont
  {{Bolton}}},\ and\ \bibinfo {author} {\bibfnamefont {G.~D.}\ \bibnamefont
  {{Becker}}},\ }\bibfield  {title} {\bibinfo {title} {{Inferring the IGM
  thermal history during reionization with the Lyman {\ensuremath{\alpha}}
  forest power spectrum at redshift \(z \simeq 5\)}},\ }\href
  {https://doi.org/10.1093/mnras/stw2147} {\bibfield  {journal} {\bibinfo
  {journal} {\mnras}\ }\textbf {\bibinfo {volume} {463}},\ \bibinfo {pages}
  {2335} (\bibinfo {year} {2016})},\ \Eprint {https://arxiv.org/abs/1605.04155}
  {arXiv:1605.04155 [astro-ph.CO]} \BibitemShut {NoStop}%
\bibitem [{\citenamefont {{Kulkarni}}\ \emph {et~al.}(2015)\citenamefont
  {{Kulkarni}}, \citenamefont {{Hennawi}}, \citenamefont {{O{\~n}orbe}},
  \citenamefont {{Rorai}},\ and\ \citenamefont
  {{Springel}}}]{2015ApJ...812...30K}%
  \BibitemOpen
  \bibfield  {author} {\bibinfo {author} {\bibfnamefont {G.}~\bibnamefont
  {{Kulkarni}}}, \bibinfo {author} {\bibfnamefont {J.~F.}\ \bibnamefont
  {{Hennawi}}}, \bibinfo {author} {\bibfnamefont {J.}~\bibnamefont
  {{O{\~n}orbe}}}, \bibinfo {author} {\bibfnamefont {A.}~\bibnamefont
  {{Rorai}}},\ and\ \bibinfo {author} {\bibfnamefont {V.}~\bibnamefont
  {{Springel}}},\ }\bibfield  {title} {\bibinfo {title} {{Characterizing the
  Pressure Smoothing Scale of the Intergalactic Medium}},\ }\href
  {https://doi.org/10.1088/0004-637X/812/1/30} {\bibfield  {journal} {\bibinfo
  {journal} {\apj}\ }\textbf {\bibinfo {volume} {812}},\ \bibinfo {eid} {30}
  (\bibinfo {year} {2015})},\ \Eprint {https://arxiv.org/abs/1504.00366}
  {arXiv:1504.00366 [astro-ph.CO]} \BibitemShut {NoStop}%
\bibitem [{\citenamefont {{Planck Collaboration}}(2020)}]{refId0}%
  \BibitemOpen
  \bibfield  {author} {\bibinfo {author} {\bibnamefont {{Planck
  Collaboration}}},\ }\bibfield  {title} {\bibinfo {title} {Planck 2018
  results. vi. cosmological parameters},\ }\bibfield  {journal} {\bibinfo
  {journal} {A\&A}\ }\href {https://doi.org/10.1051/0004-6361/201833910}
  {10.1051/0004-6361/201833910} (\bibinfo {year} {2020})\BibitemShut {NoStop}%
\bibitem [{\citenamefont {{Bird}}(2017)}]{2017ascl.soft10012B}%
  \BibitemOpen
  \bibfield  {author} {\bibinfo {author} {\bibfnamefont {S.}~\bibnamefont
  {{Bird}}},\ }\href@noop {} {\bibinfo {title} {{FSFE: Fake Spectra Flux
  Extractor}}} (\bibinfo {year} {2017}),\ \Eprint
  {https://arxiv.org/abs/1710.012} {ascl:1710.012} \BibitemShut {NoStop}%
\bibitem [{\citenamefont {{Haardt}}\ and\ \citenamefont
  {{Madau}}(2012)}]{2012ApJ...746..125H}%
  \BibitemOpen
  \bibfield  {author} {\bibinfo {author} {\bibfnamefont {F.}~\bibnamefont
  {{Haardt}}}\ and\ \bibinfo {author} {\bibfnamefont {P.}~\bibnamefont
  {{Madau}}},\ }\bibfield  {title} {\bibinfo {title} {{Radiative Transfer in a
  Clumpy Universe. IV. New Synthesis Models of the Cosmic UV/X-Ray
  Background}},\ }\href {https://doi.org/10.1088/0004-637X/746/2/125}
  {\bibfield  {journal} {\bibinfo  {journal} {\apj}\ }\textbf {\bibinfo
  {volume} {746}},\ \bibinfo {eid} {125} (\bibinfo {year} {2012})},\ \Eprint
  {https://arxiv.org/abs/1105.2039} {arXiv:1105.2039 [astro-ph.CO]}
  \BibitemShut {NoStop}%
\bibitem [{\citenamefont {{Wu}}\ \emph {et~al.}(2019)\citenamefont {{Wu}},
  \citenamefont {{McQuinn}}, \citenamefont {{Kannan}}, \citenamefont
  {{D'Aloisio}}, \citenamefont {{Bird}}, \citenamefont {{Marinacci}},
  \citenamefont {{Dav{\'e}}},\ and\ \citenamefont
  {{Hernquist}}}]{2019MNRAS.490.3177W}%
  \BibitemOpen
  \bibfield  {author} {\bibinfo {author} {\bibfnamefont {X.}~\bibnamefont
  {{Wu}}}, \bibinfo {author} {\bibfnamefont {M.}~\bibnamefont {{McQuinn}}},
  \bibinfo {author} {\bibfnamefont {R.}~\bibnamefont {{Kannan}}}, \bibinfo
  {author} {\bibfnamefont {A.}~\bibnamefont {{D'Aloisio}}}, \bibinfo {author}
  {\bibfnamefont {S.}~\bibnamefont {{Bird}}}, \bibinfo {author} {\bibfnamefont
  {F.}~\bibnamefont {{Marinacci}}}, \bibinfo {author} {\bibfnamefont
  {R.}~\bibnamefont {{Dav{\'e}}}},\ and\ \bibinfo {author} {\bibfnamefont
  {L.}~\bibnamefont {{Hernquist}}},\ }\bibfield  {title} {\bibinfo {title}
  {{Imprints of temperature fluctuations on the z {\ensuremath{\sim}} 5
  Lyman-{\ensuremath{\alpha}} forest: a view from radiation-hydrodynamic
  simulations of reionization}},\ }\href
  {https://doi.org/10.1093/mnras/stz2807} {\bibfield  {journal} {\bibinfo
  {journal} {\mnras}\ }\textbf {\bibinfo {volume} {490}},\ \bibinfo {pages}
  {3177} (\bibinfo {year} {2019})},\ \Eprint {https://arxiv.org/abs/1907.04860}
  {arXiv:1907.04860 [astro-ph.CO]} \BibitemShut {NoStop}%
\bibitem [{\citenamefont {{Molaro}}\ \emph {et~al.}(2021)\citenamefont
  {{Molaro}}, \citenamefont {{Ir{\v{s}}i{\v{c}}}}, \citenamefont {{Bolton}},
  \citenamefont {{Keating}}, \citenamefont {{Puchwein}}, \citenamefont
  {{Gaikwad}}, \citenamefont {{Haehnelt}}, \citenamefont {{Kulkarni}},\ and\
  \citenamefont {{Viel}}}]{2021arXiv210906897M}%
  \BibitemOpen
  \bibfield  {author} {\bibinfo {author} {\bibfnamefont {M.}~\bibnamefont
  {{Molaro}}}, \bibinfo {author} {\bibfnamefont {V.}~\bibnamefont
  {{Ir{\v{s}}i{\v{c}}}}}, \bibinfo {author} {\bibfnamefont {J.~S.}\
  \bibnamefont {{Bolton}}}, \bibinfo {author} {\bibfnamefont {L.~C.}\
  \bibnamefont {{Keating}}}, \bibinfo {author} {\bibfnamefont {E.}~\bibnamefont
  {{Puchwein}}}, \bibinfo {author} {\bibfnamefont {P.}~\bibnamefont
  {{Gaikwad}}}, \bibinfo {author} {\bibfnamefont {M.~G.}\ \bibnamefont
  {{Haehnelt}}}, \bibinfo {author} {\bibfnamefont {G.}~\bibnamefont
  {{Kulkarni}}},\ and\ \bibinfo {author} {\bibfnamefont {M.}~\bibnamefont
  {{Viel}}},\ }\bibfield  {title} {\bibinfo {title} {{The effect of
  inhomogeneous reionisation on the Lyman-$\alpha$ forest power spectrum at
  redshift $z>4$: implications for thermal parameter recovery}},\ }\href@noop
  {} {\bibfield  {journal} {\bibinfo  {journal} {arXiv e-prints}\ ,\ \bibinfo
  {eid} {arXiv:2109.06897}} (\bibinfo {year} {2021})},\ \Eprint
  {https://arxiv.org/abs/2109.06897} {arXiv:2109.06897 [astro-ph.CO]}
  \BibitemShut {NoStop}%
\bibitem [{\citenamefont {{O{\~n}orbe}}\ \emph {et~al.}(2017)\citenamefont
  {{O{\~n}orbe}}, \citenamefont {{Hennawi}},\ and\ \citenamefont
  {{Luki{\'c}}}}]{2017ApJ...837..106O}%
  \BibitemOpen
  \bibfield  {author} {\bibinfo {author} {\bibfnamefont {J.}~\bibnamefont
  {{O{\~n}orbe}}}, \bibinfo {author} {\bibfnamefont {J.~F.}\ \bibnamefont
  {{Hennawi}}},\ and\ \bibinfo {author} {\bibfnamefont {Z.}~\bibnamefont
  {{Luki{\'c}}}},\ }\bibfield  {title} {\bibinfo {title} {{Self-consistent
  Modeling of Reionization in Cosmological Hydrodynamical Simulations}},\
  }\href {https://doi.org/10.3847/1538-4357/aa6031} {\bibfield  {journal}
  {\bibinfo  {journal} {\apj}\ }\textbf {\bibinfo {volume} {837}},\ \bibinfo
  {eid} {106} (\bibinfo {year} {2017})},\ \Eprint
  {https://arxiv.org/abs/1607.04218} {arXiv:1607.04218 [astro-ph.CO]}
  \BibitemShut {NoStop}%
\bibitem [{\citenamefont {{Vogt}}\ \emph {et~al.}(1994)\citenamefont {{Vogt}},
  \citenamefont {{Allen}}, \citenamefont {{Bigelow}}, \citenamefont {{Bresee}},
  \citenamefont {{Brown}}, \citenamefont {{Cantrall}}, \citenamefont
  {{Conrad}}, \citenamefont {{Couture}}, \citenamefont {{Delaney}},
  \citenamefont {{Epps}}, \citenamefont {{Hilyard}}, \citenamefont {{Hilyard}},
  \citenamefont {{Horn}}, \citenamefont {{Jern}}, \citenamefont {{Kanto}},
  \citenamefont {{Keane}}, \citenamefont {{Kibrick}}, \citenamefont {{Lewis}},
  \citenamefont {{Osborne}}, \citenamefont {{Pardeilhan}}, \citenamefont
  {{Pfister}}, \citenamefont {{Ricketts}}, \citenamefont {{Robinson}},
  \citenamefont {{Stover}}, \citenamefont {{Tucker}}, \citenamefont {{Ward}},\
  and\ \citenamefont {{Wei}}}]{1994SPIE.2198..362V}%
  \BibitemOpen
  \bibfield  {author} {\bibinfo {author} {\bibfnamefont {S.~S.}\ \bibnamefont
  {{Vogt}}}, \bibinfo {author} {\bibfnamefont {S.~L.}\ \bibnamefont {{Allen}}},
  \bibinfo {author} {\bibfnamefont {B.~C.}\ \bibnamefont {{Bigelow}}}, \bibinfo
  {author} {\bibfnamefont {L.}~\bibnamefont {{Bresee}}}, \bibinfo {author}
  {\bibfnamefont {B.}~\bibnamefont {{Brown}}}, \bibinfo {author} {\bibfnamefont
  {T.}~\bibnamefont {{Cantrall}}}, \bibinfo {author} {\bibfnamefont
  {A.}~\bibnamefont {{Conrad}}}, \bibinfo {author} {\bibfnamefont
  {M.}~\bibnamefont {{Couture}}}, \bibinfo {author} {\bibfnamefont
  {C.}~\bibnamefont {{Delaney}}}, \bibinfo {author} {\bibfnamefont {H.~W.}\
  \bibnamefont {{Epps}}}, \bibinfo {author} {\bibfnamefont {D.}~\bibnamefont
  {{Hilyard}}}, \bibinfo {author} {\bibfnamefont {D.~F.}\ \bibnamefont
  {{Hilyard}}}, \bibinfo {author} {\bibfnamefont {E.}~\bibnamefont {{Horn}}},
  \bibinfo {author} {\bibfnamefont {N.}~\bibnamefont {{Jern}}}, \bibinfo
  {author} {\bibfnamefont {D.}~\bibnamefont {{Kanto}}}, \bibinfo {author}
  {\bibfnamefont {M.~J.}\ \bibnamefont {{Keane}}}, \bibinfo {author}
  {\bibfnamefont {R.~I.}\ \bibnamefont {{Kibrick}}}, \bibinfo {author}
  {\bibfnamefont {J.~W.}\ \bibnamefont {{Lewis}}}, \bibinfo {author}
  {\bibfnamefont {J.}~\bibnamefont {{Osborne}}}, \bibinfo {author}
  {\bibfnamefont {G.~H.}\ \bibnamefont {{Pardeilhan}}}, \bibinfo {author}
  {\bibfnamefont {T.}~\bibnamefont {{Pfister}}}, \bibinfo {author}
  {\bibfnamefont {T.}~\bibnamefont {{Ricketts}}}, \bibinfo {author}
  {\bibfnamefont {L.~B.}\ \bibnamefont {{Robinson}}}, \bibinfo {author}
  {\bibfnamefont {R.~J.}\ \bibnamefont {{Stover}}}, \bibinfo {author}
  {\bibfnamefont {D.}~\bibnamefont {{Tucker}}}, \bibinfo {author}
  {\bibfnamefont {J.}~\bibnamefont {{Ward}}},\ and\ \bibinfo {author}
  {\bibfnamefont {M.~Z.}\ \bibnamefont {{Wei}}},\ }\bibfield  {title} {\bibinfo
  {title} {{HIRES: the high-resolution echelle spectrometer on the Keck 10-m
  Telescope}},\ }in\ \href {https://doi.org/10.1117/12.176725} {\emph {\bibinfo
  {booktitle} {\procspie}}},\ \bibinfo {series} {Society of Photo-Optical
  Instrumentation Engineers (SPIE) Conference Series}, Vol.\ \bibinfo {volume}
  {2198},\ \bibinfo {editor} {edited by\ \bibinfo {editor} {\bibfnamefont
  {D.~L.}\ \bibnamefont {{Crawford}}}\ and\ \bibinfo {editor} {\bibfnamefont
  {E.~R.}\ \bibnamefont {{Craine}}}}\ (\bibinfo {year} {1994})\ p.\ \bibinfo
  {pages} {362}\BibitemShut {NoStop}%
\bibitem [{\citenamefont {{Dekker}}\ \emph {et~al.}(2000)\citenamefont
  {{Dekker}}, \citenamefont {{D'Odorico}}, \citenamefont {{Kaufer}},
  \citenamefont {{Delabre}},\ and\ \citenamefont
  {{Kotzlowski}}}]{2000SPIE.4008..534D}%
  \BibitemOpen
  \bibfield  {author} {\bibinfo {author} {\bibfnamefont {H.}~\bibnamefont
  {{Dekker}}}, \bibinfo {author} {\bibfnamefont {S.}~\bibnamefont
  {{D'Odorico}}}, \bibinfo {author} {\bibfnamefont {A.}~\bibnamefont
  {{Kaufer}}}, \bibinfo {author} {\bibfnamefont {B.}~\bibnamefont
  {{Delabre}}},\ and\ \bibinfo {author} {\bibfnamefont {H.}~\bibnamefont
  {{Kotzlowski}}},\ }\bibfield  {title} {\bibinfo {title} {{Design,
  construction, and performance of UVES, the echelle spectrograph for the UT2
  Kueyen Telescope at the ESO Paranal Observatory}},\ }in\ \href
  {https://doi.org/10.1117/12.395512} {\emph {\bibinfo {booktitle}
  {\procspie}}},\ \bibinfo {series} {Society of Photo-Optical Instrumentation
  Engineers (SPIE) Conference Series}, Vol.\ \bibinfo {volume} {4008},\
  \bibinfo {editor} {edited by\ \bibinfo {editor} {\bibfnamefont
  {M.}~\bibnamefont {{Iye}}}\ and\ \bibinfo {editor} {\bibfnamefont {A.~F.}\
  \bibnamefont {{Moorwood}}}}\ (\bibinfo {year} {2000})\ pp.\ \bibinfo {pages}
  {534--545}\BibitemShut {NoStop}%
\bibitem [{\citenamefont {{Chabanier}}\ \emph {et~al.}(2019)\citenamefont
  {{Chabanier}}, \citenamefont {{Palanque-Delabrouille}}, \citenamefont
  {{Y{\`e}che}}, \citenamefont {{Le Goff}}, \citenamefont {{Armengaud}},
  \citenamefont {{Bautista}}, \citenamefont {{Blomqvist}}, \citenamefont
  {{Busca}}, \citenamefont {{Dawson}}, \citenamefont {{Etourneau}},\ and\
  \citenamefont {{others}}}]{2019JCAP...07..017C}%
  \BibitemOpen
  \bibfield  {author} {\bibinfo {author} {\bibfnamefont {S.}~\bibnamefont
  {{Chabanier}}}, \bibinfo {author} {\bibfnamefont {N.}~\bibnamefont
  {{Palanque-Delabrouille}}}, \bibinfo {author} {\bibfnamefont
  {C.}~\bibnamefont {{Y{\`e}che}}}, \bibinfo {author} {\bibfnamefont {J.-M.}\
  \bibnamefont {{Le Goff}}}, \bibinfo {author} {\bibfnamefont {E.}~\bibnamefont
  {{Armengaud}}}, \bibinfo {author} {\bibfnamefont {J.}~\bibnamefont
  {{Bautista}}}, \bibinfo {author} {\bibfnamefont {M.}~\bibnamefont
  {{Blomqvist}}}, \bibinfo {author} {\bibfnamefont {N.}~\bibnamefont
  {{Busca}}}, \bibinfo {author} {\bibfnamefont {K.}~\bibnamefont {{Dawson}}},
  \bibinfo {author} {\bibfnamefont {T.}~\bibnamefont {{Etourneau}}},\ and\
  \bibinfo {author} {\bibnamefont {{others}}},\ }\bibfield  {title} {\bibinfo
  {title} {{The one-dimensional power spectrum from the SDSS DR14
  Ly{\ensuremath{\alpha}} forests}},\ }\href
  {https://doi.org/10.1088/1475-7516/2019/07/017} {\bibfield  {journal}
  {\bibinfo  {journal} {\jcap}\ }\textbf {\bibinfo {volume} {2019}},\ \bibinfo
  {eid} {017} (\bibinfo {year} {2019})},\ \Eprint
  {https://arxiv.org/abs/1812.03554} {arXiv:1812.03554 [astro-ph.CO]}
  \BibitemShut {NoStop}%
\bibitem [{\citenamefont {{Pepe}}\ \emph {et~al.}(2014)\citenamefont {{Pepe}},
  \citenamefont {{Molaro}}, \citenamefont {{Cristiani}}, \citenamefont
  {{Rebolo}}, \citenamefont {{Santos}}, \citenamefont {{Dekker}}, \citenamefont
  {{M{\'e}gevand}}, \citenamefont {{Zerbi}}, \citenamefont {{Cabral}},
  \citenamefont {{Di Marcantonio}}, \citenamefont {{Abreu}}, \citenamefont
  {{Affolter}}, \citenamefont {{Aliverti}}, \citenamefont {{Allende Prieto}},
  \citenamefont {{Amate}}, \citenamefont {{Avila}}, \citenamefont {{Baldini}},
  \citenamefont {{Bristow}}, \citenamefont {{Broeg}}, \citenamefont {{Cirami}},
  \citenamefont {{Coelho}}, \citenamefont {{Conconi}}, \citenamefont
  {{Coretti}}, \citenamefont {{Cupani}}, \citenamefont {{D'Odorico}},
  \citenamefont {{De Caprio}}, \citenamefont {{Delabre}}, \citenamefont
  {{Dorn}}, \citenamefont {{Figueira}}, \citenamefont {{Fragoso}},
  \citenamefont {{Galeotta}}, \citenamefont {{Genolet}}, \citenamefont
  {{Gomes}}, \citenamefont {{Gonz{\'a}lez Hern{\'a}ndez}}, \citenamefont
  {{Hughes}}, \citenamefont {{Iwert}}, \citenamefont {{Kerber}}, \citenamefont
  {{Landoni}}, \citenamefont {{Lizon}}, \citenamefont {{Lovis}}, \citenamefont
  {{Maire}}, \citenamefont {{Mannetta}}, \citenamefont {{Martins}},
  \citenamefont {{Monteiro}}, \citenamefont {{Oliveira}}, \citenamefont
  {{Poretti}}, \citenamefont {{Rasilla}}, \citenamefont {{Riva}}, \citenamefont
  {{Santana Tschudi}}, \citenamefont {{Santos}}, \citenamefont {{Sosnowska}},
  \citenamefont {{Sousa}}, \citenamefont {{Span{\'o}}}, \citenamefont
  {{Tenegi}}, \citenamefont {{Toso}}, \citenamefont {{Vanzella}}, \citenamefont
  {{Viel}},\ and\ \citenamefont {{Zapatero Osorio}}}]{2014arXiv1401.5918P}%
  \BibitemOpen
  \bibfield  {author} {\bibinfo {author} {\bibfnamefont {F.}~\bibnamefont
  {{Pepe}}}, \bibinfo {author} {\bibfnamefont {P.}~\bibnamefont {{Molaro}}},
  \bibinfo {author} {\bibfnamefont {S.}~\bibnamefont {{Cristiani}}}, \bibinfo
  {author} {\bibfnamefont {R.}~\bibnamefont {{Rebolo}}}, \bibinfo {author}
  {\bibfnamefont {N.~C.}\ \bibnamefont {{Santos}}}, \bibinfo {author}
  {\bibfnamefont {H.}~\bibnamefont {{Dekker}}}, \bibinfo {author}
  {\bibfnamefont {D.}~\bibnamefont {{M{\'e}gevand}}}, \bibinfo {author}
  {\bibfnamefont {F.~M.}\ \bibnamefont {{Zerbi}}}, \bibinfo {author}
  {\bibfnamefont {A.}~\bibnamefont {{Cabral}}}, \bibinfo {author}
  {\bibfnamefont {P.}~\bibnamefont {{Di Marcantonio}}}, \bibinfo {author}
  {\bibfnamefont {M.}~\bibnamefont {{Abreu}}}, \bibinfo {author} {\bibfnamefont
  {M.}~\bibnamefont {{Affolter}}}, \bibinfo {author} {\bibfnamefont
  {M.}~\bibnamefont {{Aliverti}}}, \bibinfo {author} {\bibfnamefont
  {C.}~\bibnamefont {{Allende Prieto}}}, \bibinfo {author} {\bibfnamefont
  {M.}~\bibnamefont {{Amate}}}, \bibinfo {author} {\bibfnamefont
  {G.}~\bibnamefont {{Avila}}}, \bibinfo {author} {\bibfnamefont
  {V.}~\bibnamefont {{Baldini}}}, \bibinfo {author} {\bibfnamefont
  {P.}~\bibnamefont {{Bristow}}}, \bibinfo {author} {\bibfnamefont
  {C.}~\bibnamefont {{Broeg}}}, \bibinfo {author} {\bibfnamefont
  {R.}~\bibnamefont {{Cirami}}}, \bibinfo {author} {\bibfnamefont
  {J.}~\bibnamefont {{Coelho}}}, \bibinfo {author} {\bibfnamefont
  {P.}~\bibnamefont {{Conconi}}}, \bibinfo {author} {\bibfnamefont
  {I.}~\bibnamefont {{Coretti}}}, \bibinfo {author} {\bibfnamefont
  {G.}~\bibnamefont {{Cupani}}}, \bibinfo {author} {\bibfnamefont
  {V.}~\bibnamefont {{D'Odorico}}}, \bibinfo {author} {\bibfnamefont
  {V.}~\bibnamefont {{De Caprio}}}, \bibinfo {author} {\bibfnamefont
  {B.}~\bibnamefont {{Delabre}}}, \bibinfo {author} {\bibfnamefont
  {R.}~\bibnamefont {{Dorn}}}, \bibinfo {author} {\bibfnamefont
  {P.}~\bibnamefont {{Figueira}}}, \bibinfo {author} {\bibfnamefont
  {A.}~\bibnamefont {{Fragoso}}}, \bibinfo {author} {\bibfnamefont
  {S.}~\bibnamefont {{Galeotta}}}, \bibinfo {author} {\bibfnamefont
  {L.}~\bibnamefont {{Genolet}}}, \bibinfo {author} {\bibfnamefont
  {R.}~\bibnamefont {{Gomes}}}, \bibinfo {author} {\bibfnamefont {J.~I.}\
  \bibnamefont {{Gonz{\'a}lez Hern{\'a}ndez}}}, \bibinfo {author}
  {\bibfnamefont {I.}~\bibnamefont {{Hughes}}}, \bibinfo {author}
  {\bibfnamefont {O.}~\bibnamefont {{Iwert}}}, \bibinfo {author} {\bibfnamefont
  {F.}~\bibnamefont {{Kerber}}}, \bibinfo {author} {\bibfnamefont
  {M.}~\bibnamefont {{Landoni}}}, \bibinfo {author} {\bibfnamefont {J.~L.}\
  \bibnamefont {{Lizon}}}, \bibinfo {author} {\bibfnamefont {C.}~\bibnamefont
  {{Lovis}}}, \bibinfo {author} {\bibfnamefont {C.}~\bibnamefont {{Maire}}},
  \bibinfo {author} {\bibfnamefont {M.}~\bibnamefont {{Mannetta}}}, \bibinfo
  {author} {\bibfnamefont {C.}~\bibnamefont {{Martins}}}, \bibinfo {author}
  {\bibfnamefont {M.}~\bibnamefont {{Monteiro}}}, \bibinfo {author}
  {\bibfnamefont {A.}~\bibnamefont {{Oliveira}}}, \bibinfo {author}
  {\bibfnamefont {E.}~\bibnamefont {{Poretti}}}, \bibinfo {author}
  {\bibfnamefont {J.~L.}\ \bibnamefont {{Rasilla}}}, \bibinfo {author}
  {\bibfnamefont {M.}~\bibnamefont {{Riva}}}, \bibinfo {author} {\bibfnamefont
  {S.}~\bibnamefont {{Santana Tschudi}}}, \bibinfo {author} {\bibfnamefont
  {P.}~\bibnamefont {{Santos}}}, \bibinfo {author} {\bibfnamefont
  {D.}~\bibnamefont {{Sosnowska}}}, \bibinfo {author} {\bibfnamefont
  {S.}~\bibnamefont {{Sousa}}}, \bibinfo {author} {\bibfnamefont
  {P.}~\bibnamefont {{Span{\'o}}}}, \bibinfo {author} {\bibfnamefont
  {F.}~\bibnamefont {{Tenegi}}}, \bibinfo {author} {\bibfnamefont
  {G.}~\bibnamefont {{Toso}}}, \bibinfo {author} {\bibfnamefont
  {E.}~\bibnamefont {{Vanzella}}}, \bibinfo {author} {\bibfnamefont
  {M.}~\bibnamefont {{Viel}}},\ and\ \bibinfo {author} {\bibfnamefont {M.~R.}\
  \bibnamefont {{Zapatero Osorio}}},\ }\bibfield  {title} {\bibinfo {title}
  {{ESPRESSO: The next European exoplanet hunter}},\ }\href@noop {} {\bibfield
  {journal} {\bibinfo  {journal} {arXiv e-prints}\ ,\ \bibinfo {eid}
  {arXiv:1401.5918}} (\bibinfo {year} {2014})},\ \Eprint
  {https://arxiv.org/abs/1401.5918} {arXiv:1401.5918 [astro-ph.IM]}
  \BibitemShut {NoStop}%
\bibitem [{\citenamefont {{Pepe}}\ \emph {et~al.}(2021)\citenamefont {{Pepe}},
  \citenamefont {{Cristiani}}, \citenamefont {{Rebolo}}, \citenamefont
  {{Santos}}, \citenamefont {{Dekker}}, \citenamefont {{Cabral}}, \citenamefont
  {{Di Marcantonio}}, \citenamefont {{Figueira}}, \citenamefont {{Lo Curto}},
  \citenamefont {{Lovis}}, \citenamefont {{Mayor}}, \citenamefont
  {{M{\'e}gevand}}, \citenamefont {{Molaro}}, \citenamefont {{Riva}},
  \citenamefont {{Zapatero Osorio}}, \citenamefont {{Amate}}, \citenamefont
  {{Manescau}}, \citenamefont {{Pasquini}}, \citenamefont {{Zerbi}},
  \citenamefont {{Adibekyan}}, \citenamefont {{Abreu}}, \citenamefont
  {{Affolter}}, \citenamefont {{Alibert}}, \citenamefont {{Aliverti}},
  \citenamefont {{Allart}}, \citenamefont {{Allende Prieto}}, \citenamefont
  {{{\'A}lvarez}}, \citenamefont {{Alves}}, \citenamefont {{Avila}},
  \citenamefont {{Baldini}}, \citenamefont {{Bandy}}, \citenamefont {{Barros}},
  \citenamefont {{Benz}}, \citenamefont {{Bianco}}, \citenamefont {{Borsa}},
  \citenamefont {{Bourrier}}, \citenamefont {{Bouchy}}, \citenamefont
  {{Broeg}}, \citenamefont {{Calderone}}, \citenamefont {{Cirami}},
  \citenamefont {{Coelho}}, \citenamefont {{Conconi}}, \citenamefont
  {{Coretti}}, \citenamefont {{Cumani}}, \citenamefont {{Cupani}},
  \citenamefont {{D'Odorico}}, \citenamefont {{Damasso}}, \citenamefont
  {{Deiries}}, \citenamefont {{Delabre}}, \citenamefont {{Demangeon}},
  \citenamefont {{Dumusque}}, \citenamefont {{Ehrenreich}}, \citenamefont
  {{Faria}}, \citenamefont {{Fragoso}}, \citenamefont {{Genolet}},
  \citenamefont {{Genoni}}, \citenamefont {{G{\'e}nova Santos}}, \citenamefont
  {{Gonz{\'a}lez Hern{\'a}ndez}}, \citenamefont {{Hughes}}, \citenamefont
  {{Iwert}}, \citenamefont {{Kerber}}, \citenamefont {{Knudstrup}},
  \citenamefont {{Landoni}}, \citenamefont {{Lavie}}, \citenamefont
  {{Lillo-Box}}, \citenamefont {{Lizon}}, \citenamefont {{Maire}},
  \citenamefont {{Martins}}, \citenamefont {{Mehner}}, \citenamefont
  {{Micela}}, \citenamefont {{Modigliani}}, \citenamefont {{Monteiro}},
  \citenamefont {{Monteiro}}, \citenamefont {{Moschetti}}, \citenamefont
  {{Murphy}}, \citenamefont {{Nunes}}, \citenamefont {{Oggioni}}, \citenamefont
  {{Oliveira}}, \citenamefont {{Oshagh}}, \citenamefont {{Pall{\'e}}},
  \citenamefont {{Pariani}}, \citenamefont {{Poretti}}, \citenamefont
  {{Rasilla}}, \citenamefont {{Rebord{\~a}o}}, \citenamefont {{Redaelli}},
  \citenamefont {{Santana Tschudi}}, \citenamefont {{Santin}}, \citenamefont
  {{Santos}}, \citenamefont {{S{\'e}gransan}}, \citenamefont {{Schmidt}},
  \citenamefont {{Segovia}}, \citenamefont {{Sosnowska}}, \citenamefont
  {{Sozzetti}}, \citenamefont {{Sousa}}, \citenamefont {{Span{\`o}}},
  \citenamefont {{Su{\'a}rez Mascare{\~n}o}}, \citenamefont {{Tabernero}},
  \citenamefont {{Tenegi}}, \citenamefont {{Udry}},\ and\ \citenamefont
  {{Zanutta}}}]{2021A&A...645A..96P}%
  \BibitemOpen
  \bibfield  {author} {\bibinfo {author} {\bibfnamefont {F.}~\bibnamefont
  {{Pepe}}}, \bibinfo {author} {\bibfnamefont {S.}~\bibnamefont {{Cristiani}}},
  \bibinfo {author} {\bibfnamefont {R.}~\bibnamefont {{Rebolo}}}, \bibinfo
  {author} {\bibfnamefont {N.~C.}\ \bibnamefont {{Santos}}}, \bibinfo {author}
  {\bibfnamefont {H.}~\bibnamefont {{Dekker}}}, \bibinfo {author}
  {\bibfnamefont {A.}~\bibnamefont {{Cabral}}}, \bibinfo {author}
  {\bibfnamefont {P.}~\bibnamefont {{Di Marcantonio}}}, \bibinfo {author}
  {\bibfnamefont {P.}~\bibnamefont {{Figueira}}}, \bibinfo {author}
  {\bibfnamefont {G.}~\bibnamefont {{Lo Curto}}}, \bibinfo {author}
  {\bibfnamefont {C.}~\bibnamefont {{Lovis}}}, \bibinfo {author} {\bibfnamefont
  {M.}~\bibnamefont {{Mayor}}}, \bibinfo {author} {\bibfnamefont
  {D.}~\bibnamefont {{M{\'e}gevand}}}, \bibinfo {author} {\bibfnamefont
  {P.}~\bibnamefont {{Molaro}}}, \bibinfo {author} {\bibfnamefont
  {M.}~\bibnamefont {{Riva}}}, \bibinfo {author} {\bibfnamefont {M.~R.}\
  \bibnamefont {{Zapatero Osorio}}}, \bibinfo {author} {\bibfnamefont
  {M.}~\bibnamefont {{Amate}}}, \bibinfo {author} {\bibfnamefont
  {A.}~\bibnamefont {{Manescau}}}, \bibinfo {author} {\bibfnamefont
  {L.}~\bibnamefont {{Pasquini}}}, \bibinfo {author} {\bibfnamefont {F.~M.}\
  \bibnamefont {{Zerbi}}}, \bibinfo {author} {\bibfnamefont {V.}~\bibnamefont
  {{Adibekyan}}}, \bibinfo {author} {\bibfnamefont {M.}~\bibnamefont
  {{Abreu}}}, \bibinfo {author} {\bibfnamefont {M.}~\bibnamefont {{Affolter}}},
  \bibinfo {author} {\bibfnamefont {Y.}~\bibnamefont {{Alibert}}}, \bibinfo
  {author} {\bibfnamefont {M.}~\bibnamefont {{Aliverti}}}, \bibinfo {author}
  {\bibfnamefont {R.}~\bibnamefont {{Allart}}}, \bibinfo {author}
  {\bibfnamefont {C.}~\bibnamefont {{Allende Prieto}}}, \bibinfo {author}
  {\bibfnamefont {D.}~\bibnamefont {{{\'A}lvarez}}}, \bibinfo {author}
  {\bibfnamefont {D.}~\bibnamefont {{Alves}}}, \bibinfo {author} {\bibfnamefont
  {G.}~\bibnamefont {{Avila}}}, \bibinfo {author} {\bibfnamefont
  {V.}~\bibnamefont {{Baldini}}}, \bibinfo {author} {\bibfnamefont
  {T.}~\bibnamefont {{Bandy}}}, \bibinfo {author} {\bibfnamefont {S.~C.~C.}\
  \bibnamefont {{Barros}}}, \bibinfo {author} {\bibfnamefont {W.}~\bibnamefont
  {{Benz}}}, \bibinfo {author} {\bibfnamefont {A.}~\bibnamefont {{Bianco}}},
  \bibinfo {author} {\bibfnamefont {F.}~\bibnamefont {{Borsa}}}, \bibinfo
  {author} {\bibfnamefont {V.}~\bibnamefont {{Bourrier}}}, \bibinfo {author}
  {\bibfnamefont {F.}~\bibnamefont {{Bouchy}}}, \bibinfo {author}
  {\bibfnamefont {C.}~\bibnamefont {{Broeg}}}, \bibinfo {author} {\bibfnamefont
  {G.}~\bibnamefont {{Calderone}}}, \bibinfo {author} {\bibfnamefont
  {R.}~\bibnamefont {{Cirami}}}, \bibinfo {author} {\bibfnamefont
  {J.}~\bibnamefont {{Coelho}}}, \bibinfo {author} {\bibfnamefont
  {P.}~\bibnamefont {{Conconi}}}, \bibinfo {author} {\bibfnamefont
  {I.}~\bibnamefont {{Coretti}}}, \bibinfo {author} {\bibfnamefont
  {C.}~\bibnamefont {{Cumani}}}, \bibinfo {author} {\bibfnamefont
  {G.}~\bibnamefont {{Cupani}}}, \bibinfo {author} {\bibfnamefont
  {V.}~\bibnamefont {{D'Odorico}}}, \bibinfo {author} {\bibfnamefont
  {M.}~\bibnamefont {{Damasso}}}, \bibinfo {author} {\bibfnamefont
  {S.}~\bibnamefont {{Deiries}}}, \bibinfo {author} {\bibfnamefont
  {B.}~\bibnamefont {{Delabre}}}, \bibinfo {author} {\bibfnamefont {O.~D.~S.}\
  \bibnamefont {{Demangeon}}}, \bibinfo {author} {\bibfnamefont
  {X.}~\bibnamefont {{Dumusque}}}, \bibinfo {author} {\bibfnamefont
  {D.}~\bibnamefont {{Ehrenreich}}}, \bibinfo {author} {\bibfnamefont {J.~P.}\
  \bibnamefont {{Faria}}}, \bibinfo {author} {\bibfnamefont {A.}~\bibnamefont
  {{Fragoso}}}, \bibinfo {author} {\bibfnamefont {L.}~\bibnamefont
  {{Genolet}}}, \bibinfo {author} {\bibfnamefont {M.}~\bibnamefont {{Genoni}}},
  \bibinfo {author} {\bibfnamefont {R.}~\bibnamefont {{G{\'e}nova Santos}}},
  \bibinfo {author} {\bibfnamefont {J.~I.}\ \bibnamefont {{Gonz{\'a}lez
  Hern{\'a}ndez}}}, \bibinfo {author} {\bibfnamefont {I.}~\bibnamefont
  {{Hughes}}}, \bibinfo {author} {\bibfnamefont {O.}~\bibnamefont {{Iwert}}},
  \bibinfo {author} {\bibfnamefont {F.}~\bibnamefont {{Kerber}}}, \bibinfo
  {author} {\bibfnamefont {J.}~\bibnamefont {{Knudstrup}}}, \bibinfo {author}
  {\bibfnamefont {M.}~\bibnamefont {{Landoni}}}, \bibinfo {author}
  {\bibfnamefont {B.}~\bibnamefont {{Lavie}}}, \bibinfo {author} {\bibfnamefont
  {J.}~\bibnamefont {{Lillo-Box}}}, \bibinfo {author} {\bibfnamefont {J.~L.}\
  \bibnamefont {{Lizon}}}, \bibinfo {author} {\bibfnamefont {C.}~\bibnamefont
  {{Maire}}}, \bibinfo {author} {\bibfnamefont {C.~J.~A.~P.}\ \bibnamefont
  {{Martins}}}, \bibinfo {author} {\bibfnamefont {A.}~\bibnamefont {{Mehner}}},
  \bibinfo {author} {\bibfnamefont {G.}~\bibnamefont {{Micela}}}, \bibinfo
  {author} {\bibfnamefont {A.}~\bibnamefont {{Modigliani}}}, \bibinfo {author}
  {\bibfnamefont {M.~A.}\ \bibnamefont {{Monteiro}}}, \bibinfo {author}
  {\bibfnamefont {M.~J.~P.~F.~G.}\ \bibnamefont {{Monteiro}}}, \bibinfo
  {author} {\bibfnamefont {M.}~\bibnamefont {{Moschetti}}}, \bibinfo {author}
  {\bibfnamefont {M.~T.}\ \bibnamefont {{Murphy}}}, \bibinfo {author}
  {\bibfnamefont {N.}~\bibnamefont {{Nunes}}}, \bibinfo {author} {\bibfnamefont
  {L.}~\bibnamefont {{Oggioni}}}, \bibinfo {author} {\bibfnamefont
  {A.}~\bibnamefont {{Oliveira}}}, \bibinfo {author} {\bibfnamefont
  {M.}~\bibnamefont {{Oshagh}}}, \bibinfo {author} {\bibfnamefont
  {E.}~\bibnamefont {{Pall{\'e}}}}, \bibinfo {author} {\bibfnamefont
  {G.}~\bibnamefont {{Pariani}}}, \bibinfo {author} {\bibfnamefont
  {E.}~\bibnamefont {{Poretti}}}, \bibinfo {author} {\bibfnamefont {J.~L.}\
  \bibnamefont {{Rasilla}}}, \bibinfo {author} {\bibfnamefont {J.}~\bibnamefont
  {{Rebord{\~a}o}}}, \bibinfo {author} {\bibfnamefont {E.~M.}\ \bibnamefont
  {{Redaelli}}}, \bibinfo {author} {\bibfnamefont {S.}~\bibnamefont {{Santana
  Tschudi}}}, \bibinfo {author} {\bibfnamefont {P.}~\bibnamefont {{Santin}}},
  \bibinfo {author} {\bibfnamefont {P.}~\bibnamefont {{Santos}}}, \bibinfo
  {author} {\bibfnamefont {D.}~\bibnamefont {{S{\'e}gransan}}}, \bibinfo
  {author} {\bibfnamefont {T.~M.}\ \bibnamefont {{Schmidt}}}, \bibinfo {author}
  {\bibfnamefont {A.}~\bibnamefont {{Segovia}}}, \bibinfo {author}
  {\bibfnamefont {D.}~\bibnamefont {{Sosnowska}}}, \bibinfo {author}
  {\bibfnamefont {A.}~\bibnamefont {{Sozzetti}}}, \bibinfo {author}
  {\bibfnamefont {S.~G.}\ \bibnamefont {{Sousa}}}, \bibinfo {author}
  {\bibfnamefont {P.}~\bibnamefont {{Span{\`o}}}}, \bibinfo {author}
  {\bibfnamefont {A.}~\bibnamefont {{Su{\'a}rez Mascare{\~n}o}}}, \bibinfo
  {author} {\bibfnamefont {H.}~\bibnamefont {{Tabernero}}}, \bibinfo {author}
  {\bibfnamefont {F.}~\bibnamefont {{Tenegi}}}, \bibinfo {author}
  {\bibfnamefont {S.}~\bibnamefont {{Udry}}},\ and\ \bibinfo {author}
  {\bibfnamefont {A.}~\bibnamefont {{Zanutta}}},\ }\bibfield  {title} {\bibinfo
  {title} {{ESPRESSO at VLT. On-sky performance and first results}},\ }\href
  {https://doi.org/10.1051/0004-6361/202038306} {\bibfield  {journal} {\bibinfo
   {journal} {\aap}\ }\textbf {\bibinfo {volume} {645}},\ \bibinfo {eid} {A96}
  (\bibinfo {year} {2021})},\ \Eprint {https://arxiv.org/abs/2010.00316}
  {arXiv:2010.00316 [astro-ph.IM]} \BibitemShut {NoStop}%
\bibitem [{\citenamefont {{Foreman-Mackey}}\ \emph {et~al.}(2013)\citenamefont
  {{Foreman-Mackey}}, \citenamefont {{Hogg}}, \citenamefont {{Lang}},\ and\
  \citenamefont {{Goodman}}}]{2013PASP..125..306F}%
  \BibitemOpen
  \bibfield  {author} {\bibinfo {author} {\bibfnamefont {D.}~\bibnamefont
  {{Foreman-Mackey}}}, \bibinfo {author} {\bibfnamefont {D.~W.}\ \bibnamefont
  {{Hogg}}}, \bibinfo {author} {\bibfnamefont {D.}~\bibnamefont {{Lang}}},\
  and\ \bibinfo {author} {\bibfnamefont {J.}~\bibnamefont {{Goodman}}},\
  }\bibfield  {title} {\bibinfo {title} {{emcee: The MCMC Hammer}},\ }\href
  {https://doi.org/10.1086/670067} {\bibfield  {journal} {\bibinfo  {journal}
  {\pasp}\ }\textbf {\bibinfo {volume} {125}},\ \bibinfo {pages} {306}
  (\bibinfo {year} {2013})},\ \Eprint {https://arxiv.org/abs/1202.3665}
  {arXiv:1202.3665 [astro-ph.IM]} \BibitemShut {NoStop}%
\bibitem [{\citenamefont {{Planck Collaboration}}\ \emph
  {et~al.}(2016)\citenamefont {{Planck Collaboration}}, \citenamefont
  {{Aghanim}}, \citenamefont {{Arnaud}}, \citenamefont {{Ashdown}},
  \citenamefont {{Aumont}}, \citenamefont {{Baccigalupi}}, \citenamefont
  {{Banday}}, \citenamefont {{Barreiro}}, \citenamefont {{Bartlett}},
  \citenamefont {{Bartolo}}, \citenamefont {{Battaner}}, \citenamefont
  {{Benabed}}, \citenamefont {{Beno{\^\i}t}}, \citenamefont
  {{Benoit-L{\'e}vy}}, \citenamefont {{Bernard}}, \citenamefont {{Bersanelli}},
  \citenamefont {{Bielewicz}}, \citenamefont {{Bock}}, \citenamefont
  {{Bonaldi}}, \citenamefont {{Bonavera}}, \citenamefont {{Bond}},
  \citenamefont {{Borrill}}, \citenamefont {{Bouchet}}, \citenamefont
  {{Boulanger}}, \citenamefont {{Bucher}}, \citenamefont {{Burigana}},
  \citenamefont {{Butler}}, \citenamefont {{Calabrese}}, \citenamefont
  {{Cardoso}}, \citenamefont {{Catalano}}, \citenamefont {{Challinor}},
  \citenamefont {{Chiang}}, \citenamefont {{Christensen}}, \citenamefont
  {{Clements}}, \citenamefont {{Colombo}}, \citenamefont {{Combet}},
  \citenamefont {{Coulais}}, \citenamefont {{Crill}}, \citenamefont {{Curto}},
  \citenamefont {{Cuttaia}}, \citenamefont {{Danese}}, \citenamefont
  {{Davies}}, \citenamefont {{Davis}}, \citenamefont {{de Bernardis}},
  \citenamefont {{de Rosa}}, \citenamefont {{de Zotti}}, \citenamefont
  {{Delabrouille}}, \citenamefont {{D{\'e}sert}}, \citenamefont {{Di
  Valentino}}, \citenamefont {{Dickinson}}, \citenamefont {{Diego}},
  \citenamefont {{Dolag}}, \citenamefont {{Dole}}, \citenamefont {{Donzelli}},
  \citenamefont {{Dor{\'e}}}, \citenamefont {{Douspis}}, \citenamefont
  {{Ducout}}, \citenamefont {{Dunkley}}, \citenamefont {{Dupac}}, \citenamefont
  {{Efstathiou}}, \citenamefont {{Elsner}}, \citenamefont {{En{\ss}lin}},
  \citenamefont {{Eriksen}}, \citenamefont {{Fergusson}}, \citenamefont
  {{Finelli}}, \citenamefont {{Forni}}, \citenamefont {{Frailis}},
  \citenamefont {{Fraisse}}, \citenamefont {{Franceschi}}, \citenamefont
  {{Frejsel}}, \citenamefont {{Galeotta}}, \citenamefont {{Galli}},
  \citenamefont {{Ganga}}, \citenamefont {{Gauthier}}, \citenamefont
  {{Gerbino}}, \citenamefont {{Giard}}, \citenamefont {{Gjerl{\o}w}},
  \citenamefont {{Gonz{\'a}lez-Nuevo}}, \citenamefont {{G{\'o}rski}},
  \citenamefont {{Gratton}}, \citenamefont {{Gregorio}}, \citenamefont
  {{Gruppuso}}, \citenamefont {{Gudmundsson}}, \citenamefont {{Hamann}},
  \citenamefont {{Hansen}}, \citenamefont {{Harrison}}, \citenamefont
  {{Helou}}, \citenamefont {{Henrot-Versill{\'e}}}, \citenamefont
  {{Hern{\'a}ndez-Monteagudo}}, \citenamefont {{Herranz}}, \citenamefont
  {{Hildebrandt}}, \citenamefont {{Hivon}}, \citenamefont {{Holmes}},
  \citenamefont {{Hornstrup}}, \citenamefont {{Huffenberger}}, \citenamefont
  {{Hurier}}, \citenamefont {{Jaffe}}, \citenamefont {{Jones}}, \citenamefont
  {{Juvela}}, \citenamefont {{Keih{\"a}nen}}, \citenamefont {{Keskitalo}},
  \citenamefont {{Kiiveri}}, \citenamefont {{Knoche}}, \citenamefont {{Knox}},
  \citenamefont {{Kunz}}, \citenamefont {{Kurki-Suonio}}, \citenamefont
  {{Lagache}}, \citenamefont {{L{\"a}hteenm{\"a}ki}}, \citenamefont
  {{Lamarre}}, \citenamefont {{Lasenby}}, \citenamefont {{Lattanzi}},
  \citenamefont {{Lawrence}}, \citenamefont {{Le Jeune}}, \citenamefont
  {{Leonardi}}, \citenamefont {{Lesgourgues}}, \citenamefont {{Levrier}},
  \citenamefont {{Lewis}}, \citenamefont {{Liguori}}, \citenamefont {{Lilje}},
  \citenamefont {{Lilley}}, \citenamefont {{Linden-V{\o}rnle}}, \citenamefont
  {{Lindholm}}, \citenamefont {{L{\'o}pez-Caniego}}, \citenamefont
  {{Mac{\'\i}as-P{\'e}rez}}, \citenamefont {{Maffei}}, \citenamefont
  {{Maggio}}, \citenamefont {{Maino}}, \citenamefont {{Mandolesi}},
  \citenamefont {{Mangilli}}, \citenamefont {{Maris}}, \citenamefont
  {{Martin}}, \citenamefont {{Mart{\'\i}nez-Gonz{\'a}lez}}, \citenamefont
  {{Masi}}, \citenamefont {{Matarrese}}, \citenamefont {{Meinhold}},
  \citenamefont {{Melchiorri}}, \citenamefont {{Migliaccio}}, \citenamefont
  {{Millea}}, \citenamefont {{Mitra}}, \citenamefont {{Miville-Desch{\^e}nes}},
  \citenamefont {{Moneti}}, \citenamefont {{Montier}}, \citenamefont
  {{Morgante}}, \citenamefont {{Mortlock}}, \citenamefont {{Mottet}},
  \citenamefont {{Munshi}}, \citenamefont {{Murphy}}, \citenamefont
  {{Narimani}}, \citenamefont {{Naselsky}}, \citenamefont {{Nati}},
  \citenamefont {{Natoli}}, \citenamefont {{Noviello}}, \citenamefont
  {{Novikov}}, \citenamefont {{Novikov}}, \citenamefont {{Oxborrow}},
  \citenamefont {{Paci}}, \citenamefont {{Pagano}}, \citenamefont {{Pajot}},
  \citenamefont {{Paoletti}}, \citenamefont {{Partridge}}, \citenamefont
  {{Pasian}}, \citenamefont {{Patanchon}}, \citenamefont {{Pearson}},
  \citenamefont {{Perdereau}}, \citenamefont {{Perotto}}, \citenamefont
  {{Pettorino}}, \citenamefont {{Piacentini}}, \citenamefont {{Piat}},
  \citenamefont {{Pierpaoli}}, \citenamefont {{Pietrobon}}, \citenamefont
  {{Plaszczynski}}, \citenamefont {{Pointecouteau}}, \citenamefont {{Polenta}},
  \citenamefont {{Ponthieu}}, \citenamefont {{Pratt}}, \citenamefont
  {{Prunet}}, \citenamefont {{Puget}}, \citenamefont {{Rachen}}, \citenamefont
  {{Reinecke}}, \citenamefont {{Remazeilles}}, \citenamefont {{Renault}},
  \citenamefont {{Renzi}}, \citenamefont {{Ristorcelli}}, \citenamefont
  {{Rocha}}, \citenamefont {{Rossetti}}, \citenamefont {{Roudier}},
  \citenamefont {{Rouill{\'e} d'Orfeuil}}, \citenamefont
  {{Rubi{\~n}o-Mart{\'\i}n}}, \citenamefont {{Rusholme}}, \citenamefont
  {{Salvati}}, \citenamefont {{Sandri}}, \citenamefont {{Santos}},
  \citenamefont {{Savelainen}}, \citenamefont {{Savini}}, \citenamefont
  {{Scott}}, \citenamefont {{Serra}}, \citenamefont {{Spencer}}, \citenamefont
  {{Spinelli}}, \citenamefont {{Stolyarov}}, \citenamefont {{Stompor}},
  \citenamefont {{Sunyaev}}, \citenamefont {{Sutton}}, \citenamefont
  {{Suur-Uski}}, \citenamefont {{Sygnet}}, \citenamefont {{Tauber}},
  \citenamefont {{Terenzi}}, \citenamefont {{Toffolatti}}, \citenamefont
  {{Tomasi}}, \citenamefont {{Tristram}}, \citenamefont {{Trombetti}},
  \citenamefont {{Tucci}}, \citenamefont {{Tuovinen}}, \citenamefont {{Umana}},
  \citenamefont {{Valenziano}}, \citenamefont {{Valiviita}}, \citenamefont
  {{Van Tent}}, \citenamefont {{Vielva}}, \citenamefont {{Villa}},
  \citenamefont {{Wade}}, \citenamefont {{Wandelt}}, \citenamefont {{Wehus}},
  \citenamefont {{Yvon}}, \citenamefont {{Zacchei}},\ and\ \citenamefont
  {{Zonca}}}]{2016A&A...594A..11P}%
  \BibitemOpen
  \bibfield  {author} {\bibinfo {author} {\bibnamefont {{Planck
  Collaboration}}}, \bibinfo {author} {\bibfnamefont {N.}~\bibnamefont
  {{Aghanim}}}, \bibinfo {author} {\bibfnamefont {M.}~\bibnamefont {{Arnaud}}},
  \bibinfo {author} {\bibfnamefont {M.}~\bibnamefont {{Ashdown}}}, \bibinfo
  {author} {\bibfnamefont {J.}~\bibnamefont {{Aumont}}}, \bibinfo {author}
  {\bibfnamefont {C.}~\bibnamefont {{Baccigalupi}}}, \bibinfo {author}
  {\bibfnamefont {A.~J.}\ \bibnamefont {{Banday}}}, \bibinfo {author}
  {\bibfnamefont {R.~B.}\ \bibnamefont {{Barreiro}}}, \bibinfo {author}
  {\bibfnamefont {J.~G.}\ \bibnamefont {{Bartlett}}}, \bibinfo {author}
  {\bibfnamefont {N.}~\bibnamefont {{Bartolo}}}, \bibinfo {author}
  {\bibfnamefont {E.}~\bibnamefont {{Battaner}}}, \bibinfo {author}
  {\bibfnamefont {K.}~\bibnamefont {{Benabed}}}, \bibinfo {author}
  {\bibfnamefont {A.}~\bibnamefont {{Beno{\^\i}t}}}, \bibinfo {author}
  {\bibfnamefont {A.}~\bibnamefont {{Benoit-L{\'e}vy}}}, \bibinfo {author}
  {\bibfnamefont {J.~P.}\ \bibnamefont {{Bernard}}}, \bibinfo {author}
  {\bibfnamefont {M.}~\bibnamefont {{Bersanelli}}}, \bibinfo {author}
  {\bibfnamefont {P.}~\bibnamefont {{Bielewicz}}}, \bibinfo {author}
  {\bibfnamefont {J.~J.}\ \bibnamefont {{Bock}}}, \bibinfo {author}
  {\bibfnamefont {A.}~\bibnamefont {{Bonaldi}}}, \bibinfo {author}
  {\bibfnamefont {L.}~\bibnamefont {{Bonavera}}}, \bibinfo {author}
  {\bibfnamefont {J.~R.}\ \bibnamefont {{Bond}}}, \bibinfo {author}
  {\bibfnamefont {J.}~\bibnamefont {{Borrill}}}, \bibinfo {author}
  {\bibfnamefont {F.~R.}\ \bibnamefont {{Bouchet}}}, \bibinfo {author}
  {\bibfnamefont {F.}~\bibnamefont {{Boulanger}}}, \bibinfo {author}
  {\bibfnamefont {M.}~\bibnamefont {{Bucher}}}, \bibinfo {author}
  {\bibfnamefont {C.}~\bibnamefont {{Burigana}}}, \bibinfo {author}
  {\bibfnamefont {R.~C.}\ \bibnamefont {{Butler}}}, \bibinfo {author}
  {\bibfnamefont {E.}~\bibnamefont {{Calabrese}}}, \bibinfo {author}
  {\bibfnamefont {J.~F.}\ \bibnamefont {{Cardoso}}}, \bibinfo {author}
  {\bibfnamefont {A.}~\bibnamefont {{Catalano}}}, \bibinfo {author}
  {\bibfnamefont {A.}~\bibnamefont {{Challinor}}}, \bibinfo {author}
  {\bibfnamefont {H.~C.}\ \bibnamefont {{Chiang}}}, \bibinfo {author}
  {\bibfnamefont {P.~R.}\ \bibnamefont {{Christensen}}}, \bibinfo {author}
  {\bibfnamefont {D.~L.}\ \bibnamefont {{Clements}}}, \bibinfo {author}
  {\bibfnamefont {L.~P.~L.}\ \bibnamefont {{Colombo}}}, \bibinfo {author}
  {\bibfnamefont {C.}~\bibnamefont {{Combet}}}, \bibinfo {author}
  {\bibfnamefont {A.}~\bibnamefont {{Coulais}}}, \bibinfo {author}
  {\bibfnamefont {B.~P.}\ \bibnamefont {{Crill}}}, \bibinfo {author}
  {\bibfnamefont {A.}~\bibnamefont {{Curto}}}, \bibinfo {author} {\bibfnamefont
  {F.}~\bibnamefont {{Cuttaia}}}, \bibinfo {author} {\bibfnamefont
  {L.}~\bibnamefont {{Danese}}}, \bibinfo {author} {\bibfnamefont {R.~D.}\
  \bibnamefont {{Davies}}}, \bibinfo {author} {\bibfnamefont {R.~J.}\
  \bibnamefont {{Davis}}}, \bibinfo {author} {\bibfnamefont {P.}~\bibnamefont
  {{de Bernardis}}}, \bibinfo {author} {\bibfnamefont {A.}~\bibnamefont {{de
  Rosa}}}, \bibinfo {author} {\bibfnamefont {G.}~\bibnamefont {{de Zotti}}},
  \bibinfo {author} {\bibfnamefont {J.}~\bibnamefont {{Delabrouille}}},
  \bibinfo {author} {\bibfnamefont {F.~X.}\ \bibnamefont {{D{\'e}sert}}},
  \bibinfo {author} {\bibfnamefont {E.}~\bibnamefont {{Di Valentino}}},
  \bibinfo {author} {\bibfnamefont {C.}~\bibnamefont {{Dickinson}}}, \bibinfo
  {author} {\bibfnamefont {J.~M.}\ \bibnamefont {{Diego}}}, \bibinfo {author}
  {\bibfnamefont {K.}~\bibnamefont {{Dolag}}}, \bibinfo {author} {\bibfnamefont
  {H.}~\bibnamefont {{Dole}}}, \bibinfo {author} {\bibfnamefont
  {S.}~\bibnamefont {{Donzelli}}}, \bibinfo {author} {\bibfnamefont
  {O.}~\bibnamefont {{Dor{\'e}}}}, \bibinfo {author} {\bibfnamefont
  {M.}~\bibnamefont {{Douspis}}}, \bibinfo {author} {\bibfnamefont
  {A.}~\bibnamefont {{Ducout}}}, \bibinfo {author} {\bibfnamefont
  {J.}~\bibnamefont {{Dunkley}}}, \bibinfo {author} {\bibfnamefont
  {X.}~\bibnamefont {{Dupac}}}, \bibinfo {author} {\bibfnamefont
  {G.}~\bibnamefont {{Efstathiou}}}, \bibinfo {author} {\bibfnamefont
  {F.}~\bibnamefont {{Elsner}}}, \bibinfo {author} {\bibfnamefont {T.~A.}\
  \bibnamefont {{En{\ss}lin}}}, \bibinfo {author} {\bibfnamefont {H.~K.}\
  \bibnamefont {{Eriksen}}}, \bibinfo {author} {\bibfnamefont {J.}~\bibnamefont
  {{Fergusson}}}, \bibinfo {author} {\bibfnamefont {F.}~\bibnamefont
  {{Finelli}}}, \bibinfo {author} {\bibfnamefont {O.}~\bibnamefont {{Forni}}},
  \bibinfo {author} {\bibfnamefont {M.}~\bibnamefont {{Frailis}}}, \bibinfo
  {author} {\bibfnamefont {A.~A.}\ \bibnamefont {{Fraisse}}}, \bibinfo {author}
  {\bibfnamefont {E.}~\bibnamefont {{Franceschi}}}, \bibinfo {author}
  {\bibfnamefont {A.}~\bibnamefont {{Frejsel}}}, \bibinfo {author}
  {\bibfnamefont {S.}~\bibnamefont {{Galeotta}}}, \bibinfo {author}
  {\bibfnamefont {S.}~\bibnamefont {{Galli}}}, \bibinfo {author} {\bibfnamefont
  {K.}~\bibnamefont {{Ganga}}}, \bibinfo {author} {\bibfnamefont
  {C.}~\bibnamefont {{Gauthier}}}, \bibinfo {author} {\bibfnamefont
  {M.}~\bibnamefont {{Gerbino}}}, \bibinfo {author} {\bibfnamefont
  {M.}~\bibnamefont {{Giard}}}, \bibinfo {author} {\bibfnamefont
  {E.}~\bibnamefont {{Gjerl{\o}w}}}, \bibinfo {author} {\bibfnamefont
  {J.}~\bibnamefont {{Gonz{\'a}lez-Nuevo}}}, \bibinfo {author} {\bibfnamefont
  {K.~M.}\ \bibnamefont {{G{\'o}rski}}}, \bibinfo {author} {\bibfnamefont
  {S.}~\bibnamefont {{Gratton}}}, \bibinfo {author} {\bibfnamefont
  {A.}~\bibnamefont {{Gregorio}}}, \bibinfo {author} {\bibfnamefont
  {A.}~\bibnamefont {{Gruppuso}}}, \bibinfo {author} {\bibfnamefont {J.~E.}\
  \bibnamefont {{Gudmundsson}}}, \bibinfo {author} {\bibfnamefont
  {J.}~\bibnamefont {{Hamann}}}, \bibinfo {author} {\bibfnamefont {F.~K.}\
  \bibnamefont {{Hansen}}}, \bibinfo {author} {\bibfnamefont {D.~L.}\
  \bibnamefont {{Harrison}}}, \bibinfo {author} {\bibfnamefont
  {G.}~\bibnamefont {{Helou}}}, \bibinfo {author} {\bibfnamefont
  {S.}~\bibnamefont {{Henrot-Versill{\'e}}}}, \bibinfo {author} {\bibfnamefont
  {C.}~\bibnamefont {{Hern{\'a}ndez-Monteagudo}}}, \bibinfo {author}
  {\bibfnamefont {D.}~\bibnamefont {{Herranz}}}, \bibinfo {author}
  {\bibfnamefont {S.~R.}\ \bibnamefont {{Hildebrandt}}}, \bibinfo {author}
  {\bibfnamefont {E.}~\bibnamefont {{Hivon}}}, \bibinfo {author} {\bibfnamefont
  {W.~A.}\ \bibnamefont {{Holmes}}}, \bibinfo {author} {\bibfnamefont
  {A.}~\bibnamefont {{Hornstrup}}}, \bibinfo {author} {\bibfnamefont {K.~M.}\
  \bibnamefont {{Huffenberger}}}, \bibinfo {author} {\bibfnamefont
  {G.}~\bibnamefont {{Hurier}}}, \bibinfo {author} {\bibfnamefont {A.~H.}\
  \bibnamefont {{Jaffe}}}, \bibinfo {author} {\bibfnamefont {W.~C.}\
  \bibnamefont {{Jones}}}, \bibinfo {author} {\bibfnamefont {M.}~\bibnamefont
  {{Juvela}}}, \bibinfo {author} {\bibfnamefont {E.}~\bibnamefont
  {{Keih{\"a}nen}}}, \bibinfo {author} {\bibfnamefont {R.}~\bibnamefont
  {{Keskitalo}}}, \bibinfo {author} {\bibfnamefont {K.}~\bibnamefont
  {{Kiiveri}}}, \bibinfo {author} {\bibfnamefont {J.}~\bibnamefont {{Knoche}}},
  \bibinfo {author} {\bibfnamefont {L.}~\bibnamefont {{Knox}}}, \bibinfo
  {author} {\bibfnamefont {M.}~\bibnamefont {{Kunz}}}, \bibinfo {author}
  {\bibfnamefont {H.}~\bibnamefont {{Kurki-Suonio}}}, \bibinfo {author}
  {\bibfnamefont {G.}~\bibnamefont {{Lagache}}}, \bibinfo {author}
  {\bibfnamefont {A.}~\bibnamefont {{L{\"a}hteenm{\"a}ki}}}, \bibinfo {author}
  {\bibfnamefont {J.~M.}\ \bibnamefont {{Lamarre}}}, \bibinfo {author}
  {\bibfnamefont {A.}~\bibnamefont {{Lasenby}}}, \bibinfo {author}
  {\bibfnamefont {M.}~\bibnamefont {{Lattanzi}}}, \bibinfo {author}
  {\bibfnamefont {C.~R.}\ \bibnamefont {{Lawrence}}}, \bibinfo {author}
  {\bibfnamefont {M.}~\bibnamefont {{Le Jeune}}}, \bibinfo {author}
  {\bibfnamefont {R.}~\bibnamefont {{Leonardi}}}, \bibinfo {author}
  {\bibfnamefont {J.}~\bibnamefont {{Lesgourgues}}}, \bibinfo {author}
  {\bibfnamefont {F.}~\bibnamefont {{Levrier}}}, \bibinfo {author}
  {\bibfnamefont {A.}~\bibnamefont {{Lewis}}}, \bibinfo {author} {\bibfnamefont
  {M.}~\bibnamefont {{Liguori}}}, \bibinfo {author} {\bibfnamefont {P.~B.}\
  \bibnamefont {{Lilje}}}, \bibinfo {author} {\bibfnamefont {M.}~\bibnamefont
  {{Lilley}}}, \bibinfo {author} {\bibfnamefont {M.}~\bibnamefont
  {{Linden-V{\o}rnle}}}, \bibinfo {author} {\bibfnamefont {V.}~\bibnamefont
  {{Lindholm}}}, \bibinfo {author} {\bibfnamefont {M.}~\bibnamefont
  {{L{\'o}pez-Caniego}}}, \bibinfo {author} {\bibfnamefont {J.~F.}\
  \bibnamefont {{Mac{\'\i}as-P{\'e}rez}}}, \bibinfo {author} {\bibfnamefont
  {B.}~\bibnamefont {{Maffei}}}, \bibinfo {author} {\bibfnamefont
  {G.}~\bibnamefont {{Maggio}}}, \bibinfo {author} {\bibfnamefont
  {D.}~\bibnamefont {{Maino}}}, \bibinfo {author} {\bibfnamefont
  {N.}~\bibnamefont {{Mandolesi}}}, \bibinfo {author} {\bibfnamefont
  {A.}~\bibnamefont {{Mangilli}}}, \bibinfo {author} {\bibfnamefont
  {M.}~\bibnamefont {{Maris}}}, \bibinfo {author} {\bibfnamefont {P.~G.}\
  \bibnamefont {{Martin}}}, \bibinfo {author} {\bibfnamefont {E.}~\bibnamefont
  {{Mart{\'\i}nez-Gonz{\'a}lez}}}, \bibinfo {author} {\bibfnamefont
  {S.}~\bibnamefont {{Masi}}}, \bibinfo {author} {\bibfnamefont
  {S.}~\bibnamefont {{Matarrese}}}, \bibinfo {author} {\bibfnamefont {P.~R.}\
  \bibnamefont {{Meinhold}}}, \bibinfo {author} {\bibfnamefont
  {A.}~\bibnamefont {{Melchiorri}}}, \bibinfo {author} {\bibfnamefont
  {M.}~\bibnamefont {{Migliaccio}}}, \bibinfo {author} {\bibfnamefont
  {M.}~\bibnamefont {{Millea}}}, \bibinfo {author} {\bibfnamefont
  {S.}~\bibnamefont {{Mitra}}}, \bibinfo {author} {\bibfnamefont {M.~A.}\
  \bibnamefont {{Miville-Desch{\^e}nes}}}, \bibinfo {author} {\bibfnamefont
  {A.}~\bibnamefont {{Moneti}}}, \bibinfo {author} {\bibfnamefont
  {L.}~\bibnamefont {{Montier}}}, \bibinfo {author} {\bibfnamefont
  {G.}~\bibnamefont {{Morgante}}}, \bibinfo {author} {\bibfnamefont
  {D.}~\bibnamefont {{Mortlock}}}, \bibinfo {author} {\bibfnamefont
  {S.}~\bibnamefont {{Mottet}}}, \bibinfo {author} {\bibfnamefont
  {D.}~\bibnamefont {{Munshi}}}, \bibinfo {author} {\bibfnamefont {J.~A.}\
  \bibnamefont {{Murphy}}}, \bibinfo {author} {\bibfnamefont {A.}~\bibnamefont
  {{Narimani}}}, \bibinfo {author} {\bibfnamefont {P.}~\bibnamefont
  {{Naselsky}}}, \bibinfo {author} {\bibfnamefont {F.}~\bibnamefont {{Nati}}},
  \bibinfo {author} {\bibfnamefont {P.}~\bibnamefont {{Natoli}}}, \bibinfo
  {author} {\bibfnamefont {F.}~\bibnamefont {{Noviello}}}, \bibinfo {author}
  {\bibfnamefont {D.}~\bibnamefont {{Novikov}}}, \bibinfo {author}
  {\bibfnamefont {I.}~\bibnamefont {{Novikov}}}, \bibinfo {author}
  {\bibfnamefont {C.~A.}\ \bibnamefont {{Oxborrow}}}, \bibinfo {author}
  {\bibfnamefont {F.}~\bibnamefont {{Paci}}}, \bibinfo {author} {\bibfnamefont
  {L.}~\bibnamefont {{Pagano}}}, \bibinfo {author} {\bibfnamefont
  {F.}~\bibnamefont {{Pajot}}}, \bibinfo {author} {\bibfnamefont
  {D.}~\bibnamefont {{Paoletti}}}, \bibinfo {author} {\bibfnamefont
  {B.}~\bibnamefont {{Partridge}}}, \bibinfo {author} {\bibfnamefont
  {F.}~\bibnamefont {{Pasian}}}, \bibinfo {author} {\bibfnamefont
  {G.}~\bibnamefont {{Patanchon}}}, \bibinfo {author} {\bibfnamefont {T.~J.}\
  \bibnamefont {{Pearson}}}, \bibinfo {author} {\bibfnamefont {O.}~\bibnamefont
  {{Perdereau}}}, \bibinfo {author} {\bibfnamefont {L.}~\bibnamefont
  {{Perotto}}}, \bibinfo {author} {\bibfnamefont {V.}~\bibnamefont
  {{Pettorino}}}, \bibinfo {author} {\bibfnamefont {F.}~\bibnamefont
  {{Piacentini}}}, \bibinfo {author} {\bibfnamefont {M.}~\bibnamefont
  {{Piat}}}, \bibinfo {author} {\bibfnamefont {E.}~\bibnamefont {{Pierpaoli}}},
  \bibinfo {author} {\bibfnamefont {D.}~\bibnamefont {{Pietrobon}}}, \bibinfo
  {author} {\bibfnamefont {S.}~\bibnamefont {{Plaszczynski}}}, \bibinfo
  {author} {\bibfnamefont {E.}~\bibnamefont {{Pointecouteau}}}, \bibinfo
  {author} {\bibfnamefont {G.}~\bibnamefont {{Polenta}}}, \bibinfo {author}
  {\bibfnamefont {N.}~\bibnamefont {{Ponthieu}}}, \bibinfo {author}
  {\bibfnamefont {G.~W.}\ \bibnamefont {{Pratt}}}, \bibinfo {author}
  {\bibfnamefont {S.}~\bibnamefont {{Prunet}}}, \bibinfo {author}
  {\bibfnamefont {J.~L.}\ \bibnamefont {{Puget}}}, \bibinfo {author}
  {\bibfnamefont {J.~P.}\ \bibnamefont {{Rachen}}}, \bibinfo {author}
  {\bibfnamefont {M.}~\bibnamefont {{Reinecke}}}, \bibinfo {author}
  {\bibfnamefont {M.}~\bibnamefont {{Remazeilles}}}, \bibinfo {author}
  {\bibfnamefont {C.}~\bibnamefont {{Renault}}}, \bibinfo {author}
  {\bibfnamefont {A.}~\bibnamefont {{Renzi}}}, \bibinfo {author} {\bibfnamefont
  {I.}~\bibnamefont {{Ristorcelli}}}, \bibinfo {author} {\bibfnamefont
  {G.}~\bibnamefont {{Rocha}}}, \bibinfo {author} {\bibfnamefont
  {M.}~\bibnamefont {{Rossetti}}}, \bibinfo {author} {\bibfnamefont
  {G.}~\bibnamefont {{Roudier}}}, \bibinfo {author} {\bibfnamefont
  {B.}~\bibnamefont {{Rouill{\'e} d'Orfeuil}}}, \bibinfo {author}
  {\bibfnamefont {J.~A.}\ \bibnamefont {{Rubi{\~n}o-Mart{\'\i}n}}}, \bibinfo
  {author} {\bibfnamefont {B.}~\bibnamefont {{Rusholme}}}, \bibinfo {author}
  {\bibfnamefont {L.}~\bibnamefont {{Salvati}}}, \bibinfo {author}
  {\bibfnamefont {M.}~\bibnamefont {{Sandri}}}, \bibinfo {author}
  {\bibfnamefont {D.}~\bibnamefont {{Santos}}}, \bibinfo {author}
  {\bibfnamefont {M.}~\bibnamefont {{Savelainen}}}, \bibinfo {author}
  {\bibfnamefont {G.}~\bibnamefont {{Savini}}}, \bibinfo {author}
  {\bibfnamefont {D.}~\bibnamefont {{Scott}}}, \bibinfo {author} {\bibfnamefont
  {P.}~\bibnamefont {{Serra}}}, \bibinfo {author} {\bibfnamefont {L.~D.}\
  \bibnamefont {{Spencer}}}, \bibinfo {author} {\bibfnamefont {M.}~\bibnamefont
  {{Spinelli}}}, \bibinfo {author} {\bibfnamefont {V.}~\bibnamefont
  {{Stolyarov}}}, \bibinfo {author} {\bibfnamefont {R.}~\bibnamefont
  {{Stompor}}}, \bibinfo {author} {\bibfnamefont {R.}~\bibnamefont
  {{Sunyaev}}}, \bibinfo {author} {\bibfnamefont {D.}~\bibnamefont {{Sutton}}},
  \bibinfo {author} {\bibfnamefont {A.~S.}\ \bibnamefont {{Suur-Uski}}},
  \bibinfo {author} {\bibfnamefont {J.~F.}\ \bibnamefont {{Sygnet}}}, \bibinfo
  {author} {\bibfnamefont {J.~A.}\ \bibnamefont {{Tauber}}}, \bibinfo {author}
  {\bibfnamefont {L.}~\bibnamefont {{Terenzi}}}, \bibinfo {author}
  {\bibfnamefont {L.}~\bibnamefont {{Toffolatti}}}, \bibinfo {author}
  {\bibfnamefont {M.}~\bibnamefont {{Tomasi}}}, \bibinfo {author}
  {\bibfnamefont {M.}~\bibnamefont {{Tristram}}}, \bibinfo {author}
  {\bibfnamefont {T.}~\bibnamefont {{Trombetti}}}, \bibinfo {author}
  {\bibfnamefont {M.}~\bibnamefont {{Tucci}}}, \bibinfo {author} {\bibfnamefont
  {J.}~\bibnamefont {{Tuovinen}}}, \bibinfo {author} {\bibfnamefont
  {G.}~\bibnamefont {{Umana}}}, \bibinfo {author} {\bibfnamefont
  {L.}~\bibnamefont {{Valenziano}}}, \bibinfo {author} {\bibfnamefont
  {J.}~\bibnamefont {{Valiviita}}}, \bibinfo {author} {\bibfnamefont
  {F.}~\bibnamefont {{Van Tent}}}, \bibinfo {author} {\bibfnamefont
  {P.}~\bibnamefont {{Vielva}}}, \bibinfo {author} {\bibfnamefont
  {F.}~\bibnamefont {{Villa}}}, \bibinfo {author} {\bibfnamefont {L.~A.}\
  \bibnamefont {{Wade}}}, \bibinfo {author} {\bibfnamefont {B.~D.}\
  \bibnamefont {{Wandelt}}}, \bibinfo {author} {\bibfnamefont {I.~K.}\
  \bibnamefont {{Wehus}}}, \bibinfo {author} {\bibfnamefont {D.}~\bibnamefont
  {{Yvon}}}, \bibinfo {author} {\bibfnamefont {A.}~\bibnamefont {{Zacchei}}},\
  and\ \bibinfo {author} {\bibfnamefont {A.}~\bibnamefont {{Zonca}}},\
  }\bibfield  {title} {\bibinfo {title} {{Planck 2015 results. XI. CMB power
  spectra, likelihoods, and robustness of parameters}},\ }\href
  {https://doi.org/10.1051/0004-6361/201526926} {\bibfield  {journal} {\bibinfo
   {journal} {\aap}\ }\textbf {\bibinfo {volume} {594}},\ \bibinfo {eid} {A11}
  (\bibinfo {year} {2016})},\ \Eprint {https://arxiv.org/abs/1507.02704}
  {arXiv:1507.02704 [astro-ph.CO]} \BibitemShut {NoStop}%
\bibitem [{\citenamefont {{Drlica-Wagner}}\ \emph {et~al.}(2020)\citenamefont
  {{Drlica-Wagner}}, \citenamefont {{Bechtol}}, \citenamefont {{Mau}},
  \citenamefont {{McNanna}}, \citenamefont {{Nadler}}, \citenamefont {{Pace}},
  \citenamefont {{Li}}, \citenamefont {{Pieres}}, \citenamefont {{Rozo}},
  \citenamefont {{Simon}}, \citenamefont {{Walker}}, \citenamefont
  {{Wechsler}}, \citenamefont {{Abbott}}, \citenamefont {{Allam}},
  \citenamefont {{Annis}}, \citenamefont {{Bertin}}, \citenamefont {{Brooks}},
  \citenamefont {{Burke}}, \citenamefont {{Rosell}}, \citenamefont {{Carrasco
  Kind}}, \citenamefont {{Carretero}}, \citenamefont {{Costanzi}},
  \citenamefont {{da Costa}}, \citenamefont {{De Vicente}}, \citenamefont
  {{Desai}}, \citenamefont {{Diehl}}, \citenamefont {{Doel}}, \citenamefont
  {{Eifler}}, \citenamefont {{Everett}}, \citenamefont {{Flaugher}},
  \citenamefont {{Frieman}}, \citenamefont {{Garc{\'\i}a-Bellido}},
  \citenamefont {{Gaztanaga}}, \citenamefont {{Gruen}}, \citenamefont
  {{Gruendl}}, \citenamefont {{Gschwend}}, \citenamefont {{Gutierrez}},
  \citenamefont {{Honscheid}}, \citenamefont {{James}}, \citenamefont
  {{Krause}}, \citenamefont {{Kuehn}}, \citenamefont {{Kuropatkin}},
  \citenamefont {{Lahav}}, \citenamefont {{Maia}}, \citenamefont {{Marshall}},
  \citenamefont {{Melchior}}, \citenamefont {{Menanteau}}, \citenamefont
  {{Miquel}}, \citenamefont {{Palmese}}, \citenamefont {{Plazas}},
  \citenamefont {{Sanchez}}, \citenamefont {{Scarpine}}, \citenamefont
  {{Schubnell}}, \citenamefont {{Serrano}}, \citenamefont {{Sevilla-Noarbe}},
  \citenamefont {{Smith}}, \citenamefont {{Suchyta}}, \citenamefont {{Tarle}},\
  and\ \citenamefont {{DES Collaboration}}}]{2020ApJ...893...47D}%
  \BibitemOpen
  \bibfield  {author} {\bibinfo {author} {\bibfnamefont {A.}~\bibnamefont
  {{Drlica-Wagner}}}, \bibinfo {author} {\bibfnamefont {K.}~\bibnamefont
  {{Bechtol}}}, \bibinfo {author} {\bibfnamefont {S.}~\bibnamefont {{Mau}}},
  \bibinfo {author} {\bibfnamefont {M.}~\bibnamefont {{McNanna}}}, \bibinfo
  {author} {\bibfnamefont {E.~O.}\ \bibnamefont {{Nadler}}}, \bibinfo {author}
  {\bibfnamefont {A.~B.}\ \bibnamefont {{Pace}}}, \bibinfo {author}
  {\bibfnamefont {T.~S.}\ \bibnamefont {{Li}}}, \bibinfo {author}
  {\bibfnamefont {A.}~\bibnamefont {{Pieres}}}, \bibinfo {author}
  {\bibfnamefont {E.}~\bibnamefont {{Rozo}}}, \bibinfo {author} {\bibfnamefont
  {J.~D.}\ \bibnamefont {{Simon}}}, \bibinfo {author} {\bibfnamefont {A.~R.}\
  \bibnamefont {{Walker}}}, \bibinfo {author} {\bibfnamefont {R.~H.}\
  \bibnamefont {{Wechsler}}}, \bibinfo {author} {\bibfnamefont {T.~M.~C.}\
  \bibnamefont {{Abbott}}}, \bibinfo {author} {\bibfnamefont {S.}~\bibnamefont
  {{Allam}}}, \bibinfo {author} {\bibfnamefont {J.}~\bibnamefont {{Annis}}},
  \bibinfo {author} {\bibfnamefont {E.}~\bibnamefont {{Bertin}}}, \bibinfo
  {author} {\bibfnamefont {D.}~\bibnamefont {{Brooks}}}, \bibinfo {author}
  {\bibfnamefont {D.~L.}\ \bibnamefont {{Burke}}}, \bibinfo {author}
  {\bibfnamefont {A.~C.}\ \bibnamefont {{Rosell}}}, \bibinfo {author}
  {\bibfnamefont {M.}~\bibnamefont {{Carrasco Kind}}}, \bibinfo {author}
  {\bibfnamefont {J.}~\bibnamefont {{Carretero}}}, \bibinfo {author}
  {\bibfnamefont {M.}~\bibnamefont {{Costanzi}}}, \bibinfo {author}
  {\bibfnamefont {L.~N.}\ \bibnamefont {{da Costa}}}, \bibinfo {author}
  {\bibfnamefont {J.}~\bibnamefont {{De Vicente}}}, \bibinfo {author}
  {\bibfnamefont {S.}~\bibnamefont {{Desai}}}, \bibinfo {author} {\bibfnamefont
  {H.~T.}\ \bibnamefont {{Diehl}}}, \bibinfo {author} {\bibfnamefont
  {P.}~\bibnamefont {{Doel}}}, \bibinfo {author} {\bibfnamefont {T.~F.}\
  \bibnamefont {{Eifler}}}, \bibinfo {author} {\bibfnamefont {S.}~\bibnamefont
  {{Everett}}}, \bibinfo {author} {\bibfnamefont {B.}~\bibnamefont
  {{Flaugher}}}, \bibinfo {author} {\bibfnamefont {J.}~\bibnamefont
  {{Frieman}}}, \bibinfo {author} {\bibfnamefont {J.}~\bibnamefont
  {{Garc{\'\i}a-Bellido}}}, \bibinfo {author} {\bibfnamefont {E.}~\bibnamefont
  {{Gaztanaga}}}, \bibinfo {author} {\bibfnamefont {D.}~\bibnamefont
  {{Gruen}}}, \bibinfo {author} {\bibfnamefont {R.~A.}\ \bibnamefont
  {{Gruendl}}}, \bibinfo {author} {\bibfnamefont {J.}~\bibnamefont
  {{Gschwend}}}, \bibinfo {author} {\bibfnamefont {G.}~\bibnamefont
  {{Gutierrez}}}, \bibinfo {author} {\bibfnamefont {K.}~\bibnamefont
  {{Honscheid}}}, \bibinfo {author} {\bibfnamefont {D.~J.}\ \bibnamefont
  {{James}}}, \bibinfo {author} {\bibfnamefont {E.}~\bibnamefont {{Krause}}},
  \bibinfo {author} {\bibfnamefont {K.}~\bibnamefont {{Kuehn}}}, \bibinfo
  {author} {\bibfnamefont {N.}~\bibnamefont {{Kuropatkin}}}, \bibinfo {author}
  {\bibfnamefont {O.}~\bibnamefont {{Lahav}}}, \bibinfo {author} {\bibfnamefont
  {M.~A.~G.}\ \bibnamefont {{Maia}}}, \bibinfo {author} {\bibfnamefont {J.~L.}\
  \bibnamefont {{Marshall}}}, \bibinfo {author} {\bibfnamefont
  {P.}~\bibnamefont {{Melchior}}}, \bibinfo {author} {\bibfnamefont
  {F.}~\bibnamefont {{Menanteau}}}, \bibinfo {author} {\bibfnamefont
  {R.}~\bibnamefont {{Miquel}}}, \bibinfo {author} {\bibfnamefont
  {A.}~\bibnamefont {{Palmese}}}, \bibinfo {author} {\bibfnamefont {A.~A.}\
  \bibnamefont {{Plazas}}}, \bibinfo {author} {\bibfnamefont {E.}~\bibnamefont
  {{Sanchez}}}, \bibinfo {author} {\bibfnamefont {V.}~\bibnamefont
  {{Scarpine}}}, \bibinfo {author} {\bibfnamefont {M.}~\bibnamefont
  {{Schubnell}}}, \bibinfo {author} {\bibfnamefont {S.}~\bibnamefont
  {{Serrano}}}, \bibinfo {author} {\bibfnamefont {I.}~\bibnamefont
  {{Sevilla-Noarbe}}}, \bibinfo {author} {\bibfnamefont {M.}~\bibnamefont
  {{Smith}}}, \bibinfo {author} {\bibfnamefont {E.}~\bibnamefont {{Suchyta}}},
  \bibinfo {author} {\bibfnamefont {G.}~\bibnamefont {{Tarle}}},\ and\ \bibinfo
  {author} {\bibnamefont {{DES Collaboration}}},\ }\bibfield  {title} {\bibinfo
  {title} {{Milky Way Satellite Census. I. The Observational Selection Function
  for Milky Way Satellites in DES Y3 and Pan-STARRS DR1}},\ }\href
  {https://doi.org/10.3847/1538-4357/ab7eb9} {\bibfield  {journal} {\bibinfo
  {journal} {\apj}\ }\textbf {\bibinfo {volume} {893}},\ \bibinfo {eid} {47}
  (\bibinfo {year} {2020})},\ \Eprint {https://arxiv.org/abs/1912.03302}
  {arXiv:1912.03302 [astro-ph.GA]} \BibitemShut {NoStop}%
\bibitem [{\citenamefont {{Angloher}}\ \emph {et~al.}(2017)\citenamefont
  {{Angloher}}, \citenamefont {{Bauer}}, \citenamefont {{Bento}}, \citenamefont
  {{Bucci}}, \citenamefont {{Canonica}}, \citenamefont {{Defay}}, \citenamefont
  {{Erb}}, \citenamefont {{Feilitzsch}}, \citenamefont {{Iachellini}},
  \citenamefont {{Gorla}}, \citenamefont {{G{\"u}tlein}}, \citenamefont
  {{Hauff}}, \citenamefont {{Jochum}}, \citenamefont {{Kiefer}}, \citenamefont
  {{Kluck}}, \citenamefont {{Kraus}}, \citenamefont {{Lanfranchi}},
  \citenamefont {{Langenk{\"a}mper}}, \citenamefont {{Loebell}}, \citenamefont
  {{Mancuso}}, \citenamefont {{Mondragon}}, \citenamefont {{M{\"u}nster}},
  \citenamefont {{Oberauer}}, \citenamefont {{Pagliarone}}, \citenamefont
  {{Petricca}}, \citenamefont {{Potzel}}, \citenamefont {{Pr{\"o}bst}},
  \citenamefont {{Puig}}, \citenamefont {{Reindl}}, \citenamefont {{Rothe}},
  \citenamefont {{Sch{\"a}ffner}}, \citenamefont {{Schieck}}, \citenamefont
  {{Sch{\"o}nert}}, \citenamefont {{Seidel}}, \citenamefont {{Stahlberg}},
  \citenamefont {{Stodolsky}}, \citenamefont {{Strandhagen}}, \citenamefont
  {{Strauss}}, \citenamefont {{Tanzke}}, \citenamefont {{Thi}}, \citenamefont
  {{T{\"u}rko{\v{g}}lu}}, \citenamefont {{Uffinger}}, \citenamefont {{Ulrich}},
  \citenamefont {{Usherov}}, \citenamefont {{Wawoczny}}, \citenamefont
  {{Willers}}, \citenamefont {{W{\"u}strich}},\ and\ \citenamefont
  {{Z{\"o}ller}}}]{2017EPJC...77..637A}%
  \BibitemOpen
  \bibfield  {author} {\bibinfo {author} {\bibfnamefont {G.}~\bibnamefont
  {{Angloher}}}, \bibinfo {author} {\bibfnamefont {P.}~\bibnamefont {{Bauer}}},
  \bibinfo {author} {\bibfnamefont {A.}~\bibnamefont {{Bento}}}, \bibinfo
  {author} {\bibfnamefont {C.}~\bibnamefont {{Bucci}}}, \bibinfo {author}
  {\bibfnamefont {L.}~\bibnamefont {{Canonica}}}, \bibinfo {author}
  {\bibfnamefont {X.}~\bibnamefont {{Defay}}}, \bibinfo {author} {\bibfnamefont
  {A.}~\bibnamefont {{Erb}}}, \bibinfo {author} {\bibfnamefont {F.~v.}\
  \bibnamefont {{Feilitzsch}}}, \bibinfo {author} {\bibfnamefont {N.~F.}\
  \bibnamefont {{Iachellini}}}, \bibinfo {author} {\bibfnamefont
  {P.}~\bibnamefont {{Gorla}}}, \bibinfo {author} {\bibfnamefont
  {A.}~\bibnamefont {{G{\"u}tlein}}}, \bibinfo {author} {\bibfnamefont
  {D.}~\bibnamefont {{Hauff}}}, \bibinfo {author} {\bibfnamefont
  {J.}~\bibnamefont {{Jochum}}}, \bibinfo {author} {\bibfnamefont
  {M.}~\bibnamefont {{Kiefer}}}, \bibinfo {author} {\bibfnamefont
  {H.}~\bibnamefont {{Kluck}}}, \bibinfo {author} {\bibfnamefont
  {H.}~\bibnamefont {{Kraus}}}, \bibinfo {author} {\bibfnamefont {J.~C.}\
  \bibnamefont {{Lanfranchi}}}, \bibinfo {author} {\bibfnamefont
  {A.}~\bibnamefont {{Langenk{\"a}mper}}}, \bibinfo {author} {\bibfnamefont
  {J.}~\bibnamefont {{Loebell}}}, \bibinfo {author} {\bibfnamefont
  {M.}~\bibnamefont {{Mancuso}}}, \bibinfo {author} {\bibfnamefont
  {E.}~\bibnamefont {{Mondragon}}}, \bibinfo {author} {\bibfnamefont
  {A.}~\bibnamefont {{M{\"u}nster}}}, \bibinfo {author} {\bibfnamefont
  {L.}~\bibnamefont {{Oberauer}}}, \bibinfo {author} {\bibfnamefont
  {C.}~\bibnamefont {{Pagliarone}}}, \bibinfo {author} {\bibfnamefont
  {F.}~\bibnamefont {{Petricca}}}, \bibinfo {author} {\bibfnamefont
  {W.}~\bibnamefont {{Potzel}}}, \bibinfo {author} {\bibfnamefont
  {F.}~\bibnamefont {{Pr{\"o}bst}}}, \bibinfo {author} {\bibfnamefont
  {R.}~\bibnamefont {{Puig}}}, \bibinfo {author} {\bibfnamefont
  {F.}~\bibnamefont {{Reindl}}}, \bibinfo {author} {\bibfnamefont
  {J.}~\bibnamefont {{Rothe}}}, \bibinfo {author} {\bibfnamefont
  {K.}~\bibnamefont {{Sch{\"a}ffner}}}, \bibinfo {author} {\bibfnamefont
  {J.}~\bibnamefont {{Schieck}}}, \bibinfo {author} {\bibfnamefont
  {S.}~\bibnamefont {{Sch{\"o}nert}}}, \bibinfo {author} {\bibfnamefont
  {W.}~\bibnamefont {{Seidel}}}, \bibinfo {author} {\bibfnamefont
  {M.}~\bibnamefont {{Stahlberg}}}, \bibinfo {author} {\bibfnamefont
  {L.}~\bibnamefont {{Stodolsky}}}, \bibinfo {author} {\bibfnamefont
  {C.}~\bibnamefont {{Strandhagen}}}, \bibinfo {author} {\bibfnamefont
  {R.}~\bibnamefont {{Strauss}}}, \bibinfo {author} {\bibfnamefont
  {A.}~\bibnamefont {{Tanzke}}}, \bibinfo {author} {\bibfnamefont {H.~H.~T.}\
  \bibnamefont {{Thi}}}, \bibinfo {author} {\bibfnamefont {C.}~\bibnamefont
  {{T{\"u}rko{\v{g}}lu}}}, \bibinfo {author} {\bibfnamefont {M.}~\bibnamefont
  {{Uffinger}}}, \bibinfo {author} {\bibfnamefont {A.}~\bibnamefont
  {{Ulrich}}}, \bibinfo {author} {\bibfnamefont {I.}~\bibnamefont {{Usherov}}},
  \bibinfo {author} {\bibfnamefont {S.}~\bibnamefont {{Wawoczny}}}, \bibinfo
  {author} {\bibfnamefont {M.}~\bibnamefont {{Willers}}}, \bibinfo {author}
  {\bibfnamefont {M.}~\bibnamefont {{W{\"u}strich}}},\ and\ \bibinfo {author}
  {\bibfnamefont {A.}~\bibnamefont {{Z{\"o}ller}}},\ }\bibfield  {title}
  {\bibinfo {title} {{Results on MeV-scale dark matter from a gram-scale
  cryogenic calorimeter operated above ground}},\ }\href
  {https://doi.org/10.1140/epjc/s10052-017-5223-9} {\bibfield  {journal}
  {\bibinfo  {journal} {European Physical Journal C}\ }\textbf {\bibinfo
  {volume} {77}},\ \bibinfo {eid} {637} (\bibinfo {year} {2017})},\ \Eprint
  {https://arxiv.org/abs/1707.06749} {arXiv:1707.06749 [astro-ph.CO]}
  \BibitemShut {NoStop}%
\bibitem [{\citenamefont {{Armengaud}}\ \emph {et~al.}(2019)\citenamefont
  {{Armengaud}}, \citenamefont {{Augier}}, \citenamefont {{Beno{\^\i}t}},
  \citenamefont {{Benoit}}, \citenamefont {{Berg{\'e}}}, \citenamefont
  {{Billard}}, \citenamefont {{Broniatowski}}, \citenamefont {{Camus}},
  \citenamefont {{Cazes}}, \citenamefont {{Chapellier}}, \citenamefont
  {{Charlieux}}, \citenamefont {{Ducimeti{\`e}re}}, \citenamefont {{Dumoulin}},
  \citenamefont {{Eitel}}, \citenamefont {{Filosofov}}, \citenamefont
  {{Gascon}}, \citenamefont {{Giuliani}}, \citenamefont {{Gros}}, \citenamefont
  {{de J{\'e}sus}}, \citenamefont {{Jin}}, \citenamefont {{Juillard}},
  \citenamefont {{Kleifges}}, \citenamefont {{Maisonobe}}, \citenamefont
  {{Marnieros}}, \citenamefont {{Misiak}}, \citenamefont {{Navick}},
  \citenamefont {{Nones}}, \citenamefont {{Olivieri}}, \citenamefont {{Oriol}},
  \citenamefont {{Pari}}, \citenamefont {{Paul}}, \citenamefont {{Poda}},
  \citenamefont {{Queguiner}}, \citenamefont {{Rozov}}, \citenamefont
  {{Sanglard}}, \citenamefont {{Siebenborn}}, \citenamefont {{Vagneron}},
  \citenamefont {{Weber}}, \citenamefont {{Yakushev}}, \citenamefont
  {{Zolotarova}}, \citenamefont {{Kavanagh}},\ and\ \citenamefont {{EDELWEISS
  Collaboration}}}]{2019PhRvD..99h2003A}%
  \BibitemOpen
  \bibfield  {author} {\bibinfo {author} {\bibfnamefont {E.}~\bibnamefont
  {{Armengaud}}}, \bibinfo {author} {\bibfnamefont {C.}~\bibnamefont
  {{Augier}}}, \bibinfo {author} {\bibfnamefont {A.}~\bibnamefont
  {{Beno{\^\i}t}}}, \bibinfo {author} {\bibfnamefont {A.}~\bibnamefont
  {{Benoit}}}, \bibinfo {author} {\bibfnamefont {L.}~\bibnamefont
  {{Berg{\'e}}}}, \bibinfo {author} {\bibfnamefont {J.}~\bibnamefont
  {{Billard}}}, \bibinfo {author} {\bibfnamefont {A.}~\bibnamefont
  {{Broniatowski}}}, \bibinfo {author} {\bibfnamefont {P.}~\bibnamefont
  {{Camus}}}, \bibinfo {author} {\bibfnamefont {A.}~\bibnamefont {{Cazes}}},
  \bibinfo {author} {\bibfnamefont {M.}~\bibnamefont {{Chapellier}}}, \bibinfo
  {author} {\bibfnamefont {F.}~\bibnamefont {{Charlieux}}}, \bibinfo {author}
  {\bibfnamefont {D.}~\bibnamefont {{Ducimeti{\`e}re}}}, \bibinfo {author}
  {\bibfnamefont {L.}~\bibnamefont {{Dumoulin}}}, \bibinfo {author}
  {\bibfnamefont {K.}~\bibnamefont {{Eitel}}}, \bibinfo {author} {\bibfnamefont
  {D.}~\bibnamefont {{Filosofov}}}, \bibinfo {author} {\bibfnamefont
  {J.}~\bibnamefont {{Gascon}}}, \bibinfo {author} {\bibfnamefont
  {A.}~\bibnamefont {{Giuliani}}}, \bibinfo {author} {\bibfnamefont
  {M.}~\bibnamefont {{Gros}}}, \bibinfo {author} {\bibfnamefont
  {M.}~\bibnamefont {{de J{\'e}sus}}}, \bibinfo {author} {\bibfnamefont
  {Y.}~\bibnamefont {{Jin}}}, \bibinfo {author} {\bibfnamefont
  {A.}~\bibnamefont {{Juillard}}}, \bibinfo {author} {\bibfnamefont
  {M.}~\bibnamefont {{Kleifges}}}, \bibinfo {author} {\bibfnamefont
  {R.}~\bibnamefont {{Maisonobe}}}, \bibinfo {author} {\bibfnamefont
  {S.}~\bibnamefont {{Marnieros}}}, \bibinfo {author} {\bibfnamefont
  {D.}~\bibnamefont {{Misiak}}}, \bibinfo {author} {\bibfnamefont {X.~F.}\
  \bibnamefont {{Navick}}}, \bibinfo {author} {\bibfnamefont {C.}~\bibnamefont
  {{Nones}}}, \bibinfo {author} {\bibfnamefont {E.}~\bibnamefont {{Olivieri}}},
  \bibinfo {author} {\bibfnamefont {C.}~\bibnamefont {{Oriol}}}, \bibinfo
  {author} {\bibfnamefont {P.}~\bibnamefont {{Pari}}}, \bibinfo {author}
  {\bibfnamefont {B.}~\bibnamefont {{Paul}}}, \bibinfo {author} {\bibfnamefont
  {D.}~\bibnamefont {{Poda}}}, \bibinfo {author} {\bibfnamefont
  {E.}~\bibnamefont {{Queguiner}}}, \bibinfo {author} {\bibfnamefont
  {S.}~\bibnamefont {{Rozov}}}, \bibinfo {author} {\bibfnamefont
  {V.}~\bibnamefont {{Sanglard}}}, \bibinfo {author} {\bibfnamefont
  {B.}~\bibnamefont {{Siebenborn}}}, \bibinfo {author} {\bibfnamefont
  {L.}~\bibnamefont {{Vagneron}}}, \bibinfo {author} {\bibfnamefont
  {M.}~\bibnamefont {{Weber}}}, \bibinfo {author} {\bibfnamefont
  {E.}~\bibnamefont {{Yakushev}}}, \bibinfo {author} {\bibfnamefont
  {A.}~\bibnamefont {{Zolotarova}}}, \bibinfo {author} {\bibfnamefont {B.~J.}\
  \bibnamefont {{Kavanagh}}},\ and\ \bibinfo {author} {\bibnamefont {{EDELWEISS
  Collaboration}}},\ }\bibfield  {title} {\bibinfo {title} {{Searching for
  low-mass dark matter particles with a massive Ge bolometer operated above
  ground}},\ }\href {https://doi.org/10.1103/PhysRevD.99.082003} {\bibfield
  {journal} {\bibinfo  {journal} {\prd}\ }\textbf {\bibinfo {volume} {99}},\
  \bibinfo {eid} {082003} (\bibinfo {year} {2019})},\ \Eprint
  {https://arxiv.org/abs/1901.03588} {arXiv:1901.03588 [astro-ph.GA]}
  \BibitemShut {NoStop}%
\bibitem [{\citenamefont {{Mahdawi}}\ and\ \citenamefont
  {{Farrar}}(2018)}]{2018JCAP...10..007M}%
  \BibitemOpen
  \bibfield  {author} {\bibinfo {author} {\bibfnamefont {M.~S.}\ \bibnamefont
  {{Mahdawi}}}\ and\ \bibinfo {author} {\bibfnamefont {G.~R.}\ \bibnamefont
  {{Farrar}}},\ }\bibfield  {title} {\bibinfo {title} {{Constraints on Dark
  Matter with a moderately large and velocity-dependent DM-nucleon
  cross-section}},\ }\href {https://doi.org/10.1088/1475-7516/2018/10/007}
  {\bibfield  {journal} {\bibinfo  {journal} {\jcap}\ }\textbf {\bibinfo
  {volume} {2018}},\ \bibinfo {eid} {007} (\bibinfo {year} {2018})},\ \Eprint
  {https://arxiv.org/abs/1804.03073} {arXiv:1804.03073 [hep-ph]} \BibitemShut
  {NoStop}%
\bibitem [{\citenamefont {{Walther}}\ \emph {et~al.}(2019)\citenamefont
  {{Walther}}, \citenamefont {{O{\~n}orbe}}, \citenamefont {{Hennawi}},\ and\
  \citenamefont {{Luki{\'c}}}}]{2019ApJ...872...13W}%
  \BibitemOpen
  \bibfield  {author} {\bibinfo {author} {\bibfnamefont {M.}~\bibnamefont
  {{Walther}}}, \bibinfo {author} {\bibfnamefont {J.}~\bibnamefont
  {{O{\~n}orbe}}}, \bibinfo {author} {\bibfnamefont {J.~F.}\ \bibnamefont
  {{Hennawi}}},\ and\ \bibinfo {author} {\bibfnamefont {Z.}~\bibnamefont
  {{Luki{\'c}}}},\ }\bibfield  {title} {\bibinfo {title} {{New Constraints on
  IGM Thermal Evolution from the Ly{\ensuremath{\alpha}} Forest Power
  Spectrum}},\ }\href {https://doi.org/10.3847/1538-4357/aafad1} {\bibfield
  {journal} {\bibinfo  {journal} {\apj}\ }\textbf {\bibinfo {volume} {872}},\
  \bibinfo {eid} {13} (\bibinfo {year} {2019})},\ \Eprint
  {https://arxiv.org/abs/1808.04367} {arXiv:1808.04367 [astro-ph.CO]}
  \BibitemShut {NoStop}%
\bibitem [{\citenamefont {{Villasenor}}\ \emph {et~al.}(2021)\citenamefont
  {{Villasenor}}, \citenamefont {{Robertson}}, \citenamefont {{Madau}},\ and\
  \citenamefont {{Schneider}}}]{2021arXiv211100019V}%
  \BibitemOpen
  \bibfield  {author} {\bibinfo {author} {\bibfnamefont {B.}~\bibnamefont
  {{Villasenor}}}, \bibinfo {author} {\bibfnamefont {B.}~\bibnamefont
  {{Robertson}}}, \bibinfo {author} {\bibfnamefont {P.}~\bibnamefont
  {{Madau}}},\ and\ \bibinfo {author} {\bibfnamefont {E.}~\bibnamefont
  {{Schneider}}},\ }\bibfield  {title} {\bibinfo {title} {{Inferring the
  Thermal History of the Intergalactic Medium from the Properties of the
  Hydrogen and Helium Lyman-alpha Forest}},\ }\href@noop {} {\bibfield
  {journal} {\bibinfo  {journal} {arXiv e-prints}\ ,\ \bibinfo {eid}
  {arXiv:2111.00019}} (\bibinfo {year} {2021})},\ \Eprint
  {https://arxiv.org/abs/2111.00019} {arXiv:2111.00019 [astro-ph.CO]}
  \BibitemShut {NoStop}%
\bibitem [{\citenamefont {{Puchwein}}\ \emph {et~al.}(2019)\citenamefont
  {{Puchwein}}, \citenamefont {{Haardt}}, \citenamefont {{Haehnelt}},\ and\
  \citenamefont {{Madau}}}]{2019MNRAS.485...47P}%
  \BibitemOpen
  \bibfield  {author} {\bibinfo {author} {\bibfnamefont {E.}~\bibnamefont
  {{Puchwein}}}, \bibinfo {author} {\bibfnamefont {F.}~\bibnamefont
  {{Haardt}}}, \bibinfo {author} {\bibfnamefont {M.~G.}\ \bibnamefont
  {{Haehnelt}}},\ and\ \bibinfo {author} {\bibfnamefont {P.}~\bibnamefont
  {{Madau}}},\ }\bibfield  {title} {\bibinfo {title} {{Consistent modelling of
  the meta-galactic UV background and the thermal/ionization history of the
  intergalactic medium}},\ }\href {https://doi.org/10.1093/mnras/stz222}
  {\bibfield  {journal} {\bibinfo  {journal} {\mnras}\ }\textbf {\bibinfo
  {volume} {485}},\ \bibinfo {pages} {47} (\bibinfo {year} {2019})},\ \Eprint
  {https://arxiv.org/abs/1801.04931} {arXiv:1801.04931 [astro-ph.GA]}
  \BibitemShut {NoStop}%
\bibitem [{\citenamefont {{Buen-Abad}}\ \emph {et~al.}(2021)\citenamefont
  {{Buen-Abad}}, \citenamefont {{Essig}}, \citenamefont {{McKeen}},\ and\
  \citenamefont {{Zhong}}}]{2021arXiv210712377B}%
  \BibitemOpen
  \bibfield  {author} {\bibinfo {author} {\bibfnamefont {M.~A.}\ \bibnamefont
  {{Buen-Abad}}}, \bibinfo {author} {\bibfnamefont {R.}~\bibnamefont
  {{Essig}}}, \bibinfo {author} {\bibfnamefont {D.}~\bibnamefont {{McKeen}}},\
  and\ \bibinfo {author} {\bibfnamefont {Y.-M.}\ \bibnamefont {{Zhong}}},\
  }\bibfield  {title} {\bibinfo {title} {{Cosmological Constraints on Dark
  Matter Interactions with Ordinary Matter}},\ }\href@noop {} {\bibfield
  {journal} {\bibinfo  {journal} {arXiv e-prints}\ ,\ \bibinfo {eid}
  {arXiv:2107.12377}} (\bibinfo {year} {2021})},\ \Eprint
  {https://arxiv.org/abs/2107.12377} {arXiv:2107.12377 [astro-ph.CO]}
  \BibitemShut {NoStop}%
\bibitem [{\citenamefont {{Bringmann}}\ and\ \citenamefont
  {{Pospelov}}(2019)}]{2019PhRvL.122q1801B}%
  \BibitemOpen
  \bibfield  {author} {\bibinfo {author} {\bibfnamefont {T.}~\bibnamefont
  {{Bringmann}}}\ and\ \bibinfo {author} {\bibfnamefont {M.}~\bibnamefont
  {{Pospelov}}},\ }\bibfield  {title} {\bibinfo {title} {{Novel Direct
  Detection Constraints on Light Dark Matter}},\ }\href
  {https://doi.org/10.1103/PhysRevLett.122.171801} {\bibfield  {journal}
  {\bibinfo  {journal} {\prl}\ }\textbf {\bibinfo {volume} {122}},\ \bibinfo
  {eid} {171801} (\bibinfo {year} {2019})},\ \Eprint
  {https://arxiv.org/abs/1810.10543} {arXiv:1810.10543 [hep-ph]} \BibitemShut
  {NoStop}%
\bibitem [{\citenamefont {{Cappiello}}\ and\ \citenamefont
  {{Beacom}}(2019)}]{2019PhRvD.100j3011C}%
  \BibitemOpen
  \bibfield  {author} {\bibinfo {author} {\bibfnamefont {C.~V.}\ \bibnamefont
  {{Cappiello}}}\ and\ \bibinfo {author} {\bibfnamefont {J.~F.}\ \bibnamefont
  {{Beacom}}},\ }\bibfield  {title} {\bibinfo {title} {{Strong new limits on
  light dark matter from neutrino experiments}},\ }\href
  {https://doi.org/10.1103/PhysRevD.100.103011} {\bibfield  {journal} {\bibinfo
   {journal} {\prd}\ }\textbf {\bibinfo {volume} {100}},\ \bibinfo {eid}
  {103011} (\bibinfo {year} {2019})},\ \Eprint
  {https://arxiv.org/abs/1906.11283} {arXiv:1906.11283 [hep-ph]} \BibitemShut
  {NoStop}%
\bibitem [{\citenamefont {{Gilman}}\ \emph {et~al.}(2020)\citenamefont
  {{Gilman}}, \citenamefont {{Birrer}}, \citenamefont {{Nierenberg}},
  \citenamefont {{Treu}}, \citenamefont {{Du}},\ and\ \citenamefont
  {{Benson}}}]{2020MNRAS.491.6077G}%
  \BibitemOpen
  \bibfield  {author} {\bibinfo {author} {\bibfnamefont {D.}~\bibnamefont
  {{Gilman}}}, \bibinfo {author} {\bibfnamefont {S.}~\bibnamefont {{Birrer}}},
  \bibinfo {author} {\bibfnamefont {A.}~\bibnamefont {{Nierenberg}}}, \bibinfo
  {author} {\bibfnamefont {T.}~\bibnamefont {{Treu}}}, \bibinfo {author}
  {\bibfnamefont {X.}~\bibnamefont {{Du}}},\ and\ \bibinfo {author}
  {\bibfnamefont {A.}~\bibnamefont {{Benson}}},\ }\bibfield  {title} {\bibinfo
  {title} {{Warm dark matter chills out: constraints on the halo mass function
  and the free-streaming length of dark matter with eight quadruple-image
  strong gravitational lenses}},\ }\href
  {https://doi.org/10.1093/mnras/stz3480} {\bibfield  {journal} {\bibinfo
  {journal} {\mnras}\ }\textbf {\bibinfo {volume} {491}},\ \bibinfo {pages}
  {6077} (\bibinfo {year} {2020})},\ \Eprint {https://arxiv.org/abs/1908.06983}
  {arXiv:1908.06983 [astro-ph.CO]} \BibitemShut {NoStop}%
\bibitem [{\citenamefont {{Nadler}}\ \emph
  {et~al.}(2021{\natexlab{b}})\citenamefont {{Nadler}}, \citenamefont
  {{Birrer}}, \citenamefont {{Gilman}}, \citenamefont {{Wechsler}},
  \citenamefont {{Du}}, \citenamefont {{Benson}}, \citenamefont
  {{Nierenberg}},\ and\ \citenamefont {{Treu}}}]{2021ApJ...917....7N}%
  \BibitemOpen
  \bibfield  {author} {\bibinfo {author} {\bibfnamefont {E.~O.}\ \bibnamefont
  {{Nadler}}}, \bibinfo {author} {\bibfnamefont {S.}~\bibnamefont {{Birrer}}},
  \bibinfo {author} {\bibfnamefont {D.}~\bibnamefont {{Gilman}}}, \bibinfo
  {author} {\bibfnamefont {R.~H.}\ \bibnamefont {{Wechsler}}}, \bibinfo
  {author} {\bibfnamefont {X.}~\bibnamefont {{Du}}}, \bibinfo {author}
  {\bibfnamefont {A.}~\bibnamefont {{Benson}}}, \bibinfo {author}
  {\bibfnamefont {A.~M.}\ \bibnamefont {{Nierenberg}}},\ and\ \bibinfo {author}
  {\bibfnamefont {T.}~\bibnamefont {{Treu}}},\ }\bibfield  {title} {\bibinfo
  {title} {{Dark Matter Constraints from a Unified Analysis of Strong
  Gravitational Lenses and Milky Way Satellite Galaxies}},\ }\href
  {https://doi.org/10.3847/1538-4357/abf9a3} {\bibfield  {journal} {\bibinfo
  {journal} {\apj}\ }\textbf {\bibinfo {volume} {917}},\ \bibinfo {eid} {7}
  (\bibinfo {year} {2021}{\natexlab{b}})},\ \Eprint
  {https://arxiv.org/abs/2101.07810} {arXiv:2101.07810 [astro-ph.CO]}
  \BibitemShut {NoStop}%
\bibitem [{\citenamefont {{Viel}}\ \emph {et~al.}(2005)\citenamefont {{Viel}},
  \citenamefont {{Lesgourgues}}, \citenamefont {{Haehnelt}}, \citenamefont
  {{Matarrese}},\ and\ \citenamefont {{Riotto}}}]{2005PhRvD..71f3534V}%
  \BibitemOpen
  \bibfield  {author} {\bibinfo {author} {\bibfnamefont {M.}~\bibnamefont
  {{Viel}}}, \bibinfo {author} {\bibfnamefont {J.}~\bibnamefont
  {{Lesgourgues}}}, \bibinfo {author} {\bibfnamefont {M.~G.}\ \bibnamefont
  {{Haehnelt}}}, \bibinfo {author} {\bibfnamefont {S.}~\bibnamefont
  {{Matarrese}}},\ and\ \bibinfo {author} {\bibfnamefont {A.}~\bibnamefont
  {{Riotto}}},\ }\bibfield  {title} {\bibinfo {title} {{Constraining warm dark
  matter candidates including sterile neutrinos and light gravitinos with WMAP
  and the Lyman-{\ensuremath{\alpha}} forest}},\ }\href
  {https://doi.org/10.1103/PhysRevD.71.063534} {\bibfield  {journal} {\bibinfo
  {journal} {\prd}\ }\textbf {\bibinfo {volume} {71}},\ \bibinfo {eid} {063534}
  (\bibinfo {year} {2005})},\ \Eprint {https://arxiv.org/abs/astro-ph/0501562}
  {arXiv:astro-ph/0501562 [astro-ph]} \BibitemShut {NoStop}%
\bibitem [{\citenamefont {Enqvist}\ \emph {et~al.}(1990)\citenamefont
  {Enqvist}, \citenamefont {Kainulainen},\ and\ \citenamefont
  {Maalampi}}]{ENQVIST1990531}%
  \BibitemOpen
  \bibfield  {author} {\bibinfo {author} {\bibfnamefont {K.}~\bibnamefont
  {Enqvist}}, \bibinfo {author} {\bibfnamefont {K.}~\bibnamefont
  {Kainulainen}},\ and\ \bibinfo {author} {\bibfnamefont {J.}~\bibnamefont
  {Maalampi}},\ }\bibfield  {title} {\bibinfo {title} {Resonant neutrino
  transitions and nucleosynthesis},\ }\href
  {https://doi.org/https://doi.org/10.1016/0370-2693(90)91030-F} {\bibfield
  {journal} {\bibinfo  {journal} {Physics Letters B}\ }\textbf {\bibinfo
  {volume} {249}},\ \bibinfo {pages} {531 } (\bibinfo {year}
  {1990})}\BibitemShut {NoStop}%
\bibitem [{\citenamefont {{K{\"o}nig}}\ \emph {et~al.}(2016)\citenamefont
  {{K{\"o}nig}}, \citenamefont {{Merle}},\ and\ \citenamefont
  {{Totzauer}}}]{2016JCAP...11..038K}%
  \BibitemOpen
  \bibfield  {author} {\bibinfo {author} {\bibfnamefont {J.}~\bibnamefont
  {{K{\"o}nig}}}, \bibinfo {author} {\bibfnamefont {A.}~\bibnamefont
  {{Merle}}},\ and\ \bibinfo {author} {\bibfnamefont {M.}~\bibnamefont
  {{Totzauer}}},\ }\bibfield  {title} {\bibinfo {title} {{keV sterile neutrino
  dark matter from singlet scalar decays: the most general case}},\ }\href
  {https://doi.org/10.1088/1475-7516/2016/11/038} {\bibfield  {journal}
  {\bibinfo  {journal} {\jcap}\ }\textbf {\bibinfo {volume} {2016}},\ \bibinfo
  {eid} {038} (\bibinfo {year} {2016})},\ \Eprint
  {https://arxiv.org/abs/1609.01289} {arXiv:1609.01289 [hep-ph]} \BibitemShut
  {NoStop}%
\bibitem [{\citenamefont {{Boyarsky}}\ \emph {et~al.}(2009)\citenamefont
  {{Boyarsky}}, \citenamefont {{Lesgourgues}}, \citenamefont {{Ruchayskiy}},\
  and\ \citenamefont {{Viel}}}]{2009JCAP...05..012B}%
  \BibitemOpen
  \bibfield  {author} {\bibinfo {author} {\bibfnamefont {A.}~\bibnamefont
  {{Boyarsky}}}, \bibinfo {author} {\bibfnamefont {J.}~\bibnamefont
  {{Lesgourgues}}}, \bibinfo {author} {\bibfnamefont {O.}~\bibnamefont
  {{Ruchayskiy}}},\ and\ \bibinfo {author} {\bibfnamefont {M.}~\bibnamefont
  {{Viel}}},\ }\bibfield  {title} {\bibinfo {title}
  {{Lyman-{\ensuremath{\alpha}} constraints on warm and on warm-plus-cold dark
  matter models}},\ }\href {https://doi.org/10.1088/1475-7516/2009/05/012}
  {\bibfield  {journal} {\bibinfo  {journal} {\jcap}\ }\textbf {\bibinfo
  {volume} {2009}},\ \bibinfo {eid} {012} (\bibinfo {year} {2009})},\ \Eprint
  {https://arxiv.org/abs/0812.0010} {arXiv:0812.0010 [astro-ph]} \BibitemShut
  {NoStop}%
\bibitem [{\citenamefont {{Dvorkin}}\ \emph {et~al.}(2019)\citenamefont
  {{Dvorkin}}, \citenamefont {{Lin}},\ and\ \citenamefont
  {{Schutz}}}]{2019PhRvD..99k5009D}%
  \BibitemOpen
  \bibfield  {author} {\bibinfo {author} {\bibfnamefont {C.}~\bibnamefont
  {{Dvorkin}}}, \bibinfo {author} {\bibfnamefont {T.}~\bibnamefont {{Lin}}},\
  and\ \bibinfo {author} {\bibfnamefont {K.}~\bibnamefont {{Schutz}}},\
  }\bibfield  {title} {\bibinfo {title} {{Making dark matter out of light:
  Freeze-in from plasma effects}},\ }\href
  {https://doi.org/10.1103/PhysRevD.99.115009} {\bibfield  {journal} {\bibinfo
  {journal} {\prd}\ }\textbf {\bibinfo {volume} {99}},\ \bibinfo {eid} {115009}
  (\bibinfo {year} {2019})},\ \Eprint {https://arxiv.org/abs/1902.08623}
  {arXiv:1902.08623 [hep-ph]} \BibitemShut {NoStop}%
\bibitem [{\citenamefont {{Dvorkin}}\ \emph {et~al.}(2021)\citenamefont
  {{Dvorkin}}, \citenamefont {{Lin}},\ and\ \citenamefont
  {{Schutz}}}]{2021PhRvL.127k1301D}%
  \BibitemOpen
  \bibfield  {author} {\bibinfo {author} {\bibfnamefont {C.}~\bibnamefont
  {{Dvorkin}}}, \bibinfo {author} {\bibfnamefont {T.}~\bibnamefont {{Lin}}},\
  and\ \bibinfo {author} {\bibfnamefont {K.}~\bibnamefont {{Schutz}}},\
  }\bibfield  {title} {\bibinfo {title} {{Cosmology of Sub-MeV Dark Matter
  Freeze-In}},\ }\href {https://doi.org/10.1103/PhysRevLett.127.111301}
  {\bibfield  {journal} {\bibinfo  {journal} {\prl}\ }\textbf {\bibinfo
  {volume} {127}},\ \bibinfo {eid} {111301} (\bibinfo {year} {2021})},\ \Eprint
  {https://arxiv.org/abs/2011.08186} {arXiv:2011.08186 [astro-ph.CO]}
  \BibitemShut {NoStop}%
\bibitem [{\citenamefont {{Fan}}\ \emph {et~al.}(2006)\citenamefont {{Fan}},
  \citenamefont {{Strauss}}, \citenamefont {{Becker}}, \citenamefont {{White}},
  \citenamefont {{Gunn}}, \citenamefont {{Knapp}}, \citenamefont {{Richards}},
  \citenamefont {{Schneider}}, \citenamefont {{Brinkmann}},\ and\ \citenamefont
  {{Fukugita}}}]{2006AJ....132..117F}%
  \BibitemOpen
  \bibfield  {author} {\bibinfo {author} {\bibfnamefont {X.}~\bibnamefont
  {{Fan}}}, \bibinfo {author} {\bibfnamefont {M.~A.}\ \bibnamefont
  {{Strauss}}}, \bibinfo {author} {\bibfnamefont {R.~H.}\ \bibnamefont
  {{Becker}}}, \bibinfo {author} {\bibfnamefont {R.~L.}\ \bibnamefont
  {{White}}}, \bibinfo {author} {\bibfnamefont {J.~E.}\ \bibnamefont {{Gunn}}},
  \bibinfo {author} {\bibfnamefont {G.~R.}\ \bibnamefont {{Knapp}}}, \bibinfo
  {author} {\bibfnamefont {G.~T.}\ \bibnamefont {{Richards}}}, \bibinfo
  {author} {\bibfnamefont {D.~P.}\ \bibnamefont {{Schneider}}}, \bibinfo
  {author} {\bibfnamefont {J.}~\bibnamefont {{Brinkmann}}},\ and\ \bibinfo
  {author} {\bibfnamefont {M.}~\bibnamefont {{Fukugita}}},\ }\bibfield  {title}
  {\bibinfo {title} {{Constraining the Evolution of the Ionizing Background and
  the Epoch of Reionization with z\raisebox{-0.5ex}\textasciitilde6 Quasars.
  II. A Sample of 19 Quasars}},\ }\href {https://doi.org/10.1086/504836}
  {\bibfield  {journal} {\bibinfo  {journal} {\aj}\ }\textbf {\bibinfo {volume}
  {132}},\ \bibinfo {pages} {117} (\bibinfo {year} {2006})},\ \Eprint
  {https://arxiv.org/abs/astro-ph/0512082} {arXiv:astro-ph/0512082 [astro-ph]}
  \BibitemShut {NoStop}%
\bibitem [{\citenamefont {{Zhu}}\ \emph {et~al.}(2021)\citenamefont {{Zhu}},
  \citenamefont {{Becker}}, \citenamefont {{Bosman}}, \citenamefont
  {{Keating}}, \citenamefont {{Christenson}}, \citenamefont {{Ba{\~n}ados}},
  \citenamefont {{Bian}}, \citenamefont {{Davies}}, \citenamefont
  {{D'Odorico}}, \citenamefont {{Eilers}}, \citenamefont {{Fan}}, \citenamefont
  {{Haehnelt}}, \citenamefont {{Kulkarni}}, \citenamefont {{Pallottini}},
  \citenamefont {{Qin}}, \citenamefont {{Wang}},\ and\ \citenamefont
  {{Yang}}}]{2021arXiv210906295Z}%
  \BibitemOpen
  \bibfield  {author} {\bibinfo {author} {\bibfnamefont {Y.}~\bibnamefont
  {{Zhu}}}, \bibinfo {author} {\bibfnamefont {G.~D.}\ \bibnamefont {{Becker}}},
  \bibinfo {author} {\bibfnamefont {S.~E.~I.}\ \bibnamefont {{Bosman}}},
  \bibinfo {author} {\bibfnamefont {L.~C.}\ \bibnamefont {{Keating}}}, \bibinfo
  {author} {\bibfnamefont {H.~M.}\ \bibnamefont {{Christenson}}}, \bibinfo
  {author} {\bibfnamefont {E.}~\bibnamefont {{Ba{\~n}ados}}}, \bibinfo {author}
  {\bibfnamefont {F.}~\bibnamefont {{Bian}}}, \bibinfo {author} {\bibfnamefont
  {F.~B.}\ \bibnamefont {{Davies}}}, \bibinfo {author} {\bibfnamefont
  {V.}~\bibnamefont {{D'Odorico}}}, \bibinfo {author} {\bibfnamefont {A.-C.}\
  \bibnamefont {{Eilers}}}, \bibinfo {author} {\bibfnamefont {X.}~\bibnamefont
  {{Fan}}}, \bibinfo {author} {\bibfnamefont {M.~G.}\ \bibnamefont
  {{Haehnelt}}}, \bibinfo {author} {\bibfnamefont {G.}~\bibnamefont
  {{Kulkarni}}}, \bibinfo {author} {\bibfnamefont {A.}~\bibnamefont
  {{Pallottini}}}, \bibinfo {author} {\bibfnamefont {Y.}~\bibnamefont {{Qin}}},
  \bibinfo {author} {\bibfnamefont {F.}~\bibnamefont {{Wang}}},\ and\ \bibinfo
  {author} {\bibfnamefont {J.}~\bibnamefont {{Yang}}},\ }\bibfield  {title}
  {\bibinfo {title} {{Chasing the Tail of Cosmic Reionization with Dark Gap
  Statistics in the Ly$\alpha$ Forest over $5 < z < 6$}},\ }\href@noop {}
  {\bibfield  {journal} {\bibinfo  {journal} {arXiv e-prints}\ ,\ \bibinfo
  {eid} {arXiv:2109.06295}} (\bibinfo {year} {2021})},\ \Eprint
  {https://arxiv.org/abs/2109.06295} {arXiv:2109.06295 [astro-ph.CO]}
  \BibitemShut {NoStop}%
\bibitem [{\citenamefont {{DESI Collaboration}}\ and\ \citenamefont
  {{others}}(2016{\natexlab{a}})}]{2016arXiv161100036D}%
  \BibitemOpen
  \bibfield  {author} {\bibinfo {author} {\bibnamefont {{DESI Collaboration}}}\
  and\ \bibinfo {author} {\bibnamefont {{others}}},\ }\bibfield  {title}
  {\bibinfo {title} {{The DESI Experiment Part I: Science,Targeting, and Survey
  Design}},\ }\href@noop {} {\bibfield  {journal} {\bibinfo  {journal} {arXiv
  e-prints}\ } (\bibinfo {year} {2016}{\natexlab{a}})},\ \Eprint
  {https://arxiv.org/abs/1611.00036} {arXiv:1611.00036 [astro-ph.IM]}
  \BibitemShut {NoStop}%
\bibitem [{\citenamefont {{DESI Collaboration}}\ and\ \citenamefont
  {{others}}(2016{\natexlab{b}})}]{2016arXiv161100037D}%
  \BibitemOpen
  \bibfield  {author} {\bibinfo {author} {\bibnamefont {{DESI Collaboration}}}\
  and\ \bibinfo {author} {\bibnamefont {{others}}},\ }\bibfield  {title}
  {\bibinfo {title} {{The DESI Experiment Part II: Instrument Design}},\
  }\href@noop {} {\bibfield  {journal} {\bibinfo  {journal} {arXiv e-prints}\ }
  (\bibinfo {year} {2016}{\natexlab{b}})},\ \Eprint
  {https://arxiv.org/abs/1611.00037} {arXiv:1611.00037 [astro-ph.IM]}
  \BibitemShut {NoStop}%
\bibitem [{\citenamefont {{Banik}}\ \emph {et~al.}(2018)\citenamefont
  {{Banik}}, \citenamefont {{Bertone}}, \citenamefont {{Bovy}},\ and\
  \citenamefont {{Bozorgnia}}}]{2018JCAP...07..061B}%
  \BibitemOpen
  \bibfield  {author} {\bibinfo {author} {\bibfnamefont {N.}~\bibnamefont
  {{Banik}}}, \bibinfo {author} {\bibfnamefont {G.}~\bibnamefont {{Bertone}}},
  \bibinfo {author} {\bibfnamefont {J.}~\bibnamefont {{Bovy}}},\ and\ \bibinfo
  {author} {\bibfnamefont {N.}~\bibnamefont {{Bozorgnia}}},\ }\bibfield
  {title} {\bibinfo {title} {{Probing the nature of dark matter particles with
  stellar streams}},\ }\href {https://doi.org/10.1088/1475-7516/2018/07/061}
  {\bibfield  {journal} {\bibinfo  {journal} {\jcap}\ }\textbf {\bibinfo
  {volume} {2018}},\ \bibinfo {eid} {061} (\bibinfo {year} {2018})},\ \Eprint
  {https://arxiv.org/abs/1804.04384} {arXiv:1804.04384 [astro-ph.CO]}
  \BibitemShut {NoStop}%
\end{thebibliography}%
\end{document}